\documentclass[10pt,aps,pra,twocolumn,showpacs,superscriptaddress,nobalancelastpage,notitlepage,longbibliography]{revtex4-2}

\usepackage{graphicx,color,mathbbol,amsthm,amsmath,amsfonts,amssymb,bm,braket,footmisc,multirow,latexsym,times,dsfont}

\usepackage{braket,slashed}
\usepackage[mathscr]{euscript}
\usepackage[colorlinks=true, breaklinks=true, linkcolor=red, citecolor=blue, urlcolor=blue]{hyperref} 
\usepackage{hyperref}
\usepackage{subfigure}
\usepackage{xfrac}
\usepackage{kantlipsum}
\usepackage{enumerate}
\usepackage{framed}
\usepackage{subfigure}
\usepackage{cleveref}
\usepackage{csquotes}
\usepackage[dvipsnames]{xcolor}

\allowdisplaybreaks[1]

\usepackage{centernot}
\usepackage{xifthen}
\usepackage[normalem]{ulem}

\usepackage{mathtools}
\usepackage{physics}  
\usepackage{bbold}    

\DeclareMathSymbol{\shortminus}{\mathbin}{AMSa}{"39}
\usepackage{amsthm}

\DeclareFontFamily{U}{wncy}{}
\DeclareFontShape{U}{wncy}{m}{n}{<->wncyr10}{}
\DeclareSymbolFont{mcy}{U}{wncy}{m}{n}
\DeclareMathSymbol{\Sh}{\mathord}{mcy}{"58}

\newcommand{\lrt}{t_K}            
\newcommand{\ts}{t_s}             
\newcommand{\taug}{\tau_G}        
\newcommand{\mst}{\tau_\textrm{0}}

\newcommand{\LFstrength}{\Delta_0}
\newcommand{\LFwidth}{\gamma_0} 
\newcommand{\HFstrength}{\Delta_1}
\newcommand{\HFwidth}{\gamma_1}
\newcommand{\HFdetuning}{\delta \bar{\omega}_{1}}

\newcommand{\reset}{\mathcal{R}}

\begin{document}
\title{Limitations to Dynamical Error Suppression and Gate-Error Virtualization \\ from Temporally Correlated Nonclassical Noise}

\author{Michiel Burgelman}
\thanks{E-mail: \href{mailto:michiel.a.burgelman@dartmouth.edu}{michiel.a.burgelman@dartmouth.edu}}
\affiliation{Department of Physics and Astronomy, Dartmouth College, Hanover, New Hampshire 03755, USA}

\author{Nattaphong Wonglakhon}
\affiliation{Centre for Quantum Computation and Communication Technology (Australian Research Council), Centre for Quantum Dynamics, Griffith University, Brisbane, Queensland 4111, Australia}

\author{\mbox{Diego N. Bernal-Garc\'\i a}}
\affiliation{Centre for Quantum Computation and Communication Technology (Australian Research Council), Centre for Quantum Dynamics, Griffith University, Brisbane, Queensland 4111, Australia}

\author{Gerardo A. Paz-Silva}
\affiliation{Centre for Quantum Computation and Communication Technology (Australian Research Council), Centre for Quantum Dynamics, Griffith University, Brisbane, Queensland 4111, Australia}
  
\author{Lorenza Viola}
\thanks{E-mail: \href{mailto:lorenza.viola@dartmouth.edu}{lorenza.viola@dartmouth.edu}}
\affiliation{Department of Physics and Astronomy, Dartmouth College, Hanover, New Hampshire 03755, USA}

\begin{abstract}

Realistic multi-qubit noise processes often result in error mechanisms that are not captured by the probabilistic, Markovian error models commonly employed in circuit-level analyses of quantum fault-tolerance. By working within an open-quantum system Hamiltonian formulation, we revisit the validity of the notion of a constant gate error in the presence of noise that is both \emph{temporally correlated and nonclassical}, and whose impact is mitigated through perfect instantaneous dynamical decoupling subject to finite timing constraints.
We study a minimal exactly solvable single-qubit model under Gaussian quantum dephasing noise, showing that the fidelity of a dynamically protected idling gate can depend strongly on its location in the circuit and the history of applied control, even when the system-side error propagation is fully removed through perfect reset operations.
For digital periodic control, we prove that, under mild conditions on the low-frequency behavior of the nonclassical noise spectrum, the \emph{gate fidelity saturates} at a value that is strictly smaller than the one attainable in the absence of control history; the presence of high-frequency noise peaks is also found as especially harmful, due to the possible onset of \emph{control-induced resonances}.
We explicitly relate these features to the evolution of the bath statistics during the computation, which has not been fully accounted for in existing treatments.
We find that only if decoupling can keep the qubit highly pure over a timescale larger than the correlation time of the noise, the bath approximately converges to its original statistics and a stable-in-time control performance is recovered.
{Our work highlights the significance of the full bath evolution between circuit locations, and suggests that additional tradeoffs and design constraints may arise in layered quantum fault-tolerant architectures from the need to appropriately re-equilibrate the quantum bath statistics, particularly when Markovian noise is present alongside temporally correlated  noise.
}
\end{abstract}

\date{\today}
\maketitle

\section{Introduction}
\label{sec:intro}

\subsection{Context and motivation}

Noise resulting from unwanted interactions between a quantum system of interest and its environment poses a major challenge towards implementing coherence-enabled quantum technologies.
In particular, the possibility to realize the potential of quantum computing with realistically noisy devices rests on the validity of cornerstone fault-tolerance threshold theorems~\cite{manny1998,dorit1998,preskill1998,dorit2008,preskill2013}, according to which arbitrarily accurate quantum computation of arbitrary size can be achieved with bounded resources, provided that the noise is sufficiently uncorrelated in time and space, and sufficiently weak.
{This then allows} for arbitrary operations to be executed with an ``error per gate'' (EPG) below a threshold value.
At the circuit-level description, the impact of noise is typically modeled in terms of local Markovian Pauli (or, more generally, local Markovian) channels, whereby errors are treated as local and stochastic.
{Under these assumptions, the EPG can be naturally defined and computed} in terms of (classical) error probabilities and gate infidelities, or otherwise bounded in terms of appropriate distance measures \cite{Sanders_2016}.

While the exact nature of the relevant environment varies depending on the experimental platform at hand, many realistic noise processes that plague current-day quantum devices are known to be inherently correlated to varying degrees{: temporal or spatial correlations may be present, often together,} as directly revealed by quantum noise spectroscopy {(QNS)} and characterization experiments -- see for instance \cite{bylander, andrea, pappas, quintana,zhang2022predicting, lara2021, Pal2022relaxation, Tarucha1, Tarucha2, Qasim} for representative examples. On the one hand, strong temporal noise correlations, as manifested in ``colored'' noise spectra, result in complex non-Markovian error dynamics \cite{Markov}, 
which presents new challenges for developing scalable characterization techniques \cite{ball2016effect,figueroa2021randomized,White}.
On the other hand, as long as these noise correlations are sufficiently slow, they make it possible for dynamical error suppression techniques (DES) to be extremely effective in countering the effects of low-frequency noise.
{In its simplest form}, DES includes general-purpose dynamical decoupling (DD) protocols for quantum memory \cite{viola1999dynamical,khodjasteh2005fault,limits} or DD-protected quantum gates \cite{viola1999universal,khodjasteh2008rigorous,west2010high} based on instantaneous (unbounded) control.
{More} sophisticated protocols allow for executing dynamically corrected and geometric gates solely using bounded controls \cite{edd,DCG,khodjasteh2009dynamical,khodjasteh2010arbitrarily,Bluhm,Barnes_2022,nelson2023designing}, as well as for optimizing the performance in specific noise environments, upon explicitly incorporating noise spectral information \cite{10min}.
Remarkably, the use of DD has proved instrumental in enabling recent landmark achievements, ranging from the demonstration of a provable algorithmic quantum speedup \cite{speedup} to initial realizations of {multi-round} quantum error correction (QEC) {in repetition \cite{GoogleQEC}, color \cite{HayesQEC}, surface codes \cite{Blais,GoogleQEC}, and surface code hybrids~\cite{Takita}}.
Ultimately, DES is expected to play a crucial role in reducing the physical-layer noise strength to a sufficiently small level for fault-tolerant QEC to become possible with acceptable overheads \cite{NgCombining,paz2013optimally}.

The intuition behind this successful marriage of DES and QEC is that DES techniques can, as noted, efficiently suppress temporally correlated, ``low-frequency'' components of the noise, while QEC is ideally suited to efficiently correct for the Markovian errors resulting from the  ``high-frequency'' remainder. This idea is formalized in layered architectures for fault-tolerant quantum computation \cite{CodyJones}, where one defines new information primitives via quantum firmware,  in terms of virtual qubits and gates that are inherently ``dressed'' with and protected by DES \cite{BallPT}. By design, the virtual layer experiences a lower effective noise strength, suitable for QEC to produce, in turn, low-error logical qubits and gates in the next layer. Clearly, the foundational premise of the virtual layer is that one can define a meaningful EPG, which is lowered by the DES and, importantly, may be considered to be \emph{constant}, independent of the spatio-temporal location at which the gate occurs in the circuit.
In other words, DES techniques should remain equally effective throughout the duration of the computation, in order for fault-tolerant quantum computation to be even in principle possible in the presence of temporally correlated noise. 

The question we set out to answer in this work, is the extent to which the above picture remains valid when the system is coupled to a noise environment (``bath'') that is not only temporally correlated, but, crucially, also genuinely {\em nonclassical}, that is, not describable in terms of a classical (commuting) stochastic process.
Aside from filling a fundamental gap, addressing this question is also timely and practically important, as larger and more performant quantum processors are rapidly becoming available, and new or previously unimportant error mechanisms may need to be accounted for, in order to engineering a control system that can reach its performance limits~\cite{Klimov}.
Indeed, since a quantum-mechanical bath has a nontrivial dynamics of its own, it may be expected to evolve in response to the control that is exerted on the system it is coupled to.
As a result, the statistical properties of the noise experienced by the system will themselves change over time~\footnote{Note that this is beyond standard definitions of ``weak coupling'', where the statistical properties of the bath are explicitly required not to ``appreciably'' change \cite{petruccione}. While such an approximation may be accurate for baths with a white noise spectrum and Markovian reduced dynamics, it becomes delicate in the presence of strong temporal correlation and ``long'' memory time scales. See also \cite{bath_update_paper} for related discussion}.
In such a scenario, one would envision that EPG notions may not only be non-constant, but also explicitly dependent on the control applied on the system at earlier times -- undermining the possibility of consistently ``virtualizing'' the underlying imperfect hardware.
{Notably, nonclassical noise spectra have already shown to be relevant in superconducting-qubit platforms~\cite{quintana} and able to be detected by QNS methods \cite{Fei2018,Uwe2020, Qasim}; further to that, the possibility that emergent, {\em a priori} nonclassical, error behavior may arise from the presence of a bath of ``spectator qubits'' has been recently pointed out and experimentally validated \cite{Luke}. Thus, we expect that the effects we are calling attention to may emerge for a variety of hardware platforms, either in near-term implementations, or as currently dominant Markovian or classical noise processes will be increasingly suppressed through improved hardware design or fabrication.}

{From a theoretical standpoint,} understanding the impact of such general -- correlated and nonclassical -- noise processes clearly necessitates moving beyond the simplified probabilistic error models we mentioned earlier on, to a description where information about the relevant bath degrees of freedom and their evolution is included. Of course, this realization is, in itself, not new.
The fact that quantum-correlated environments would introduce extra challenges and render relevant notions of EPG or effective noise strength time-dependent was acknowledged early on \cite{Klesse,novais2006decoherence} ,
and the particular importance of the asymptotically long-time regime was shown in~\cite{alicki2002dynamical}. By working within a fully Hamiltonian formulation, useful accuracy threshold theorems could still be established for quantum (Gaussian) ``non-Markovian'' noise models with sufficiently well-behaved correlations \cite{Terhal,dorit2006,NgGaussian}.
In particular, detailed error analyses of the impact of noise due to a bosonic bath on a system running QEC was carried out in a series of works by leveraging tools from statistical mechanics \cite{novais2007,novais2010}, with special emphasis on surface codes \cite{novais2013surface, Mucciolo2013BosonicBath, novais2017}. While, interestingly, recent work has argued that current proof techniques fail to establish the existence of a nontrivial threshold for surface codes exposed to general nonclassical noise \cite{chai2022fault}, existing analyses have not captured the full implications of quantum bath dynamics \emph{in the presence of DES and QEC simultaneously}, which is our focus here. Most relevant in this context is Ref.\,\cite{NgCombining}, which rigorously establishes how, by incorporating DD-protected gates, fault-tolerant quantum circuits can tolerate stronger noise and have a lower overhead cost than circuits built out of unprotected gates {have}.
In order to obtain a notion of EPG that is independent upon the bath state, however, the bath state is therein discarded at the end of each circuit location, and refreshed to its initial state at the start of the next one. 
{In turn,} in order for this assumption to be relaxed and the threshold result to be extended to a less restricted setting, the need to access properties of the noise correlations in ``conditional'' states the bath may evolve into is explicitly acknowledged, but not further investigated. 

Our main goal in this work is to make headway in overcoming these assumptions and, by accounting for the full interplay between applied control and bath evolution in determining the time- and control-history dependence of gate errors, to quantify the extent to which temporally correlated nonclassical noise impacts the effectiveness of DES to begin with.
The upshot, as we will see, is a mechanism of error propagation that is \emph{entirely attributable to the bath dynamics}, even under the (idealized) assumption that perfect control and resetting capabilities can completely block error propagation on the system side. 

\subsection{Outline and summary of main results} 

The minimal model we consider consists of a single physical qubit experiencing quantum Gaussian dephasing noise -- characterized by a ``classical'' (symmetrized) and ``quantum'' (anti-symmetrized) noise spectra, $S^{+}(\omega)$ and $S^{-}(\omega)$ \cite{ClerkRMP, Paz2017multiqubit} -- and undergoing a form of idealized error-corrected computation.
The joint system-bath state is assumed to be factorized, with the qubit initialized in a fixed initial state to be protected, and a sequence of DD-protected idling (``no-op'') gates is implemented via perfect and instantaneous dephasing-preserving $\pi$-pulse control, subject to a finite minimum-switching time constraint \cite{limits,khodjasteh2013designing}.
At designated times, the qubit is perfectly and instantaneously refreshed to its initial state via a reset operation $\mathcal{R}$, providing a mock-up for rounds of QEC.
The fidelity $\mathcal{F}_{\textrm{G}}(\ts)$ of a target DD-protected identity gate ($G=I$), performed after a last reset operation, is then studied as a function of its starting time $\ts$ and the control history between $0 \leq t  < t_s$, under the condition that such a DD-protected gate is high-fidelity when applied at $\ts = 0$.

We derive exact expressions for the gate fidelity $\mathcal{F}_{\textrm{G}}(\ts)$, valid for an arbitrary control history, and arbitrary noise spectra $S^{\pm}(\omega)$.
To obtain a concrete quantitative study, however, we consider noise spectra deriving from a dephasing spin-boson model that may include both a low-frequency (LF) contribution, as well as a high-frequency (HF) contribution in the spectral density $J(\omega)$, as is common for phononic baths \cite{nemati2022coupling}.
The control histories that we consider are, in turn, one of three possible control scenarios (see also Fig.\,\ref{fig:general_protocol}):
\begin{enumerate}
    \item[(${\bm {c_1}}$)] A sequence comprising multiple (say, $M$) repetitions of one and the same DD-protected identity gate,
          followed by a reset operation -- before the final DD-protected gate is executed.
    \item[(${\bm {c_2}}$)] A similar periodic sequence, but with multiple (say, $K$) rounds of resets interspersed every $M$ control repetitions, for a total of $KM$ repetitions -- before the final DD-protected gate is executed.
    \item[(${\bm {c_3}}$)] Non-periodic blocks of evolution where high-order DD-schemes, meant to protect the qubit over a longer idling time, are interspersed with DD-protected gates and resets -- before the final DD-protected gate is executed.
\end{enumerate}

After the relevant setup is made more precise in Sec.\,\ref{sec:setup}, the control-history dependence of the gate fidelity for these three scenarios is studied in the presence of both LF and HF-noise in Secs.\,\ref{sec:single_reset} and \ref{sec:multiple_resets}.
Specifically, in Sec.\,\ref{sec:single_reset}, we focus on the simplest scenario $({\bm {c_1}})$, involving a single reset operation.
We start by establishing exact analytical formulas for the gate fidelity, revealing that, for purely classical stationary noise, the gate performance does not depend on the control history.
For nonclassical noise, a non-trivial dependence upon the control history is instead encoded in a \emph{bath-induced phase contribution}, $\theta_{\textrm{q}}$, that enters the fidelity expression in addition to the usual classical decay factor $\chi_{\textrm{c}}$.
In the limit of  an asymptotically large number of control repetitions, $M \to\infty$, we rigorously show that such a phase contribution causes the gate fidelity to converge to a fixed value $\mathcal{F}_{\textrm{G}}(\infty)$ that is always (possibly substantially) lower than $\mathcal{F}_{\textrm{G}}(0)$.
Thus, a \emph{performance plateau} for DES is engendered, even when the classical decay is effectively suppressed.
Explicit conditions on the LF-properties of the noise and the filtering order of the control are established, which reveals that this asymptotic gate performance saturation is generic.

Complementary to these exact results, the dependence on both LF- and HF- noise properties is quantitatively studied for a (zero-temperature) bosonic bath, showing that the asymptotic gate fidelity can be substantially lower than the original gate fidelity.
For LF-noise, we attribute this to a difference in the filtering orders to which the control suppresses the classical (respectively, quantum) noise spectra at low frequencies.
We show how a good quantitative understanding of the relative importance of the decay and phase contributions, $\chi_{\textrm{c}}$ and $\theta_{\textrm{q}}$, can be obtained based on the competition between two distinct characteristic timescales: one determined by the inherent noise coupling strength, and the other indicating how fast the protected gate is applied w.r.t. the LF-noise correlation time.
Notably, we further show that the presence of narrow HF noise peaks can be especially harmful in combination with periodic control, due to the possible occurrence of \emph{control-induced resonances} between the control periodicity and the center frequency of the HF-peak.
As it turns out, these HF-resonance effects are not linked exclusively to periodic repetition of control.
{Rather}, they can also occur for prolonged non-periodic high-order DD schemes (e.g., concatenated DD (CDD) \cite{khodjasteh2005fault,khodjasteh2007performance}), which cancel LF-noise up to high accuracy -- reinforcing the highly detrimental role of HF noise. 

The more general case of multiple reset operations is studied in Sec.~\ref{sec:multiple_resets}, again by first focusing on the equivalent scenario $({\bm {c_2}})$ of periodic control repetitions.
Exact analytical expressions for the gate fidelity show that multiple quantum phase contributions, one for each reset, $\theta_{\textrm{q},[k]}$, now encode the control history in-between all previous resets.
Importantly, this contribution would vanish even for a quantum bath, if its state could be refreshed to its initial ($t=0$) state after every qubit reset, as assumed in simplified treatments \cite{NgCombining,Mucciolo2013BosonicBath}.
This demonstrates how the bath dynamics result in temporal correlations across multiple cycles and effectively in a control-dependent, non-stationary noise process. For an asymptotically large number of resets, $K\to\infty$, we find that most of the intuition gained from the single-reset case carries over: a saturation of the gate performance still emerges. {Notably}, additional unpredictability arises from the intricate interplay between the quantum effect and the number of repetitions $M$ between consecutive resets. Lastly, in scenario $({\bm {c_3}})$, we consider idling the qubit for a prolonged period of time under higher-order DD sequences -- now, for simplicity, in the presence of only LF-noise. We observe that the gate fidelity after a sufficiently-long idling time during which the qubit is kept approximately pure, and a subsequent reset operation, the gate fidelity approximately returns to its initial value. While this effectively removes the intervening history dependence, it comes at the cost of long evolution times, during which temporally uncorrelated (white) noise may become important.

In Sec.\,\ref{sec:bath_pov}, we provide a physical interpretation of the behavior of the gate fidelity we observed in Secs.\,\ref{sec:single_reset}-\ref{sec:multiple_resets} in terms of the evolution of the quantum bath itself.
After introducing the necessary tools for describing the exact statistics of the updated bath state for a general control history, we show that the observed error propagation between resets, as encoded by the quantum phases $\theta_{\textrm{q}, [k]}$, is explicitly due to an \emph{error propagation on the bath side}, as reflected in changing statistical properties.
In turn, the approximate reversal of the control history dependence for prolonged high-fidelity idling is shown to be due to an effective approximate re-equilibration of the bath statistics to their original values.
We argue that such prolonged high-fidelity idling provides a systematic and robust way to re-obtain the original control performance, and therefore also approximately restore an error virtualization notion.
We end in Sec.\,\ref{sec:discussion} with 
{a summary of our key findings and a discussion of their significance and implications in the context of   
QEC theory and realistic fault-tolerant architectures.
}

\section{A minimal model for assessing gate-error virtualization}
\label{sec:setup}

\subsection{Open system and control setting}\label{ssec:controlled_noise_model}

We consider a single qubit with quantization axis set by $\sigma_z$, with internal Hamiltonian  $H_S\equiv \tfrac{\omega_q}{2} \sigma_z$ (in units $\hbar=1$), and experiencing purely dephasing noise that is temporally-correlated and nonclassical.
Within an open quantum systems setting, we model this noise as a coupling to an uncontrollable (possibly infinite-dimensional) bath with a free Hamiltonian $H_B$.
In the interaction picture w.r.t.\ $H_S+H_B$, the total Hamiltonian in the absence of control may be written as 
\begin{equation}
\label{eq:def_Bt_rotframe}
    H_{\text{free}} (t) \equiv 
    \sigma_z \otimes B(t), \quad B(t) = e^{i H_B t} B e^{- i H_B t}, 
\end{equation}
where $B$ is the lab-frame bath-side coupling operator.
We assume system and bath to be initially disentangled, in a separable joint state, $\rho_{\textrm{SB}}(0) = \rho_{\textrm{S}}(0) \otimes \rho_{\textrm{B}}(0).$ In particular, we work in a scenario where the bath is assumed to reach a steady state before coming in contact with the qubit, described by a thermal state w.r.t.\ its free Hamiltonian $H_B$
\footnote{In a situation where the qubit is prepared in its ground state and the environment is allowed to equilibrate while being in contact with it, the Hamiltonian $H_{\text{free}}(t)$ would be better described by a rank-1 coupling model as considered in~\cite{Paz2017multiqubit}.
The resulting, additional phase evolution has a formal connection with the quantum phase we identify here in the case where no control is applied, but is physically distinct.}
\begin{equation}
    \rho_{\textrm{B}}(0) =   \bar{\rho}_B \equiv \frac{e^{- \beta H_B}}{\Tr(e^{- \beta H_B})},
\end{equation}
where $\beta$ is the inverse temperature of the bath in units of time (so that we also have $k_B = 1$). 

Besides interacting with the bath, the qubit is subject to external time-dependent control that preserves the dephasing nature of the noise in the toggling frame w.r.t.\ the applied control Hamiltonian \cite{Paz2017multiqubit}.
We will work in the idealized limit where one can perform perfectly-accurate and instantaneous $\pi_x$-pulses on the qubit at designated times.
Under these assumptions, the total Hamiltonian reads $H(t) \equiv y(t)  \sigma_z \otimes B(t)$, and one obtains the following exact equations of motion in the toggling frame:
\begin{equation}
\label{eq:general_EOM_toggling_frame}
    \dot{\rho}_{\textrm{SB}} (t) = - i y(t) \comm{\sigma_z \otimes B(t)}{\rho_{\textrm{SB}} (t)},
\end{equation}
where the control switching function $y(t) \in \{1, -1\}$ and switches sign at times where a $\pi_x$-pulse is applied.
To impose finite (realistic) timing constraints, we demand that the instants at which $y(t)$ switches sign obey a finite minimum switching time constraint \cite{limits,khodjasteh2013designing}, that is, any two consecutive pulses be separated by a minimum interval $\tau_{\text{0}}>0$.

The effect of the control can equivalently be described in the frequency domain, within the transfer filter-function (FF) formalism~\cite{Kofman,Nave, paz2014general,Soare,Paz2017multiqubit}.
In particular, the first-order fundamental FF is obtained by considering the finite Fourier transform of the switching function $y(t)$ over the relevant time-interval \cite{paz2014general}:
\begin{equation*}
    F(\omega ; t) \equiv \int_{0}^{t} \dd s \, y(s) e^{i \omega s}.
\end{equation*}
A quantity central to the capability of the control to suppress LF-noise is the (fundamental) filtering order (FO) $\alpha_{\textrm{p}}$ around $\omega=0$, defined such that
\begin{equation}
\abs{F(\omega ; t)} \sim \abs{\omega}^{\alpha_{\textrm{p}}}  , \quad \alpha_{\textrm{p}} \in {\mathbb N},
\label{eq:def_classical_filtering_order}
\end{equation}
at low frequencies.
Since, in the context of Gaussian dephasing noise on a single qubit of this work, the above first-order FF is the only one of interest, we henceforth refer to $F(\omega ; t)$ simply as the FF of the control. 

The joint unitary propagator of system and bath is consequently given by 
\begin{equation}
\label{eq:def_joint_unitary}
    U(t_1, t_2) \equiv \mathcal{T}_+ \exp(\!- i \sigma_z \otimes \int_{t_1}^{t_2} \!y(s) B(s) \dd s),
\end{equation}
where $\mathcal{T}_+$ stands for positive time-ordering.
Associating the qubit ground and excited states to eigenstates in the $z$ basis, that is, $\ket{g} \equiv \ket{z=-1}$ and $\ket{e}\equiv \ket{z=+1}$, with $\sigma_z \ket{z=\mp 1}= \mp \ket{z=\mp 1}$, respectively, note that their populations, $\ketbra{z}$ are constant throughout the evolution -- consistent with the assumed dephasing-preserving nature of the controlled dynamics.

The statistical properties of the noise imparted on the qubit can be described by the multi-time cumulant-averages of the bath variables, $C(B(t_j), \ldots, B(t_1))$, which can be expressed in terms of statistical moments, namely, multi-time correlation functions of the form
\begin{eqnarray*}
\expval{B(t_j) \cdots B(t_1)   }_{\bar{\rho}_B} 
 = \Tr[ B(t_j) \cdots B(t_1) \bar{\rho}_B] ,
\end{eqnarray*}
where expectation values are taken with respect to the initial thermal bath state.
We also simply speak of the cumulants or statistics of the (generally non-commutative) noise process $(B(t) ; \bar{\rho}_B)$.
Note that, by virtue of the stationarity condition, $\comm{H_B}{\bar{\rho}_B} = 0$, any multi-time cumulant-average depends only on pairwise relative differences of the times $t_j$.
We additionally assume the noise process $(B(t) ; \bar{\rho}_B)$ to be Gaussian, implying that all $j$-time cumulants with $j \geq 3$ vanish, and zero-mean, in the sense that 
\[C(B(t)) = \expval{B(t)}_{\bar{\rho}_B} \equiv 0,\quad \forall t.\]
In this case, the two-time cumulant average $C(B(t_2),B(t_1))$, now coinciding with the traditional two-time (connected, auto-)correlation function,
fully captures the statistical properties of the noise.
It is useful to separate $C(B(t_2),B(t_1))$ into the classical (symmetrized) correlation function
\begin{equation}
\label{eq:def_classical_correlation_function}
    C^{+}_{\bar{\rho}_B}(t_2, t_1) \equiv \expval{\acomm{B(t_2)}{B(t_1)}}_{\bar{\rho}_B},
\end{equation}
and the quantum (anti-symmetrized) correlation function
\begin{equation}
\label{eq:def_quantum_correlation_function}
    C^{-}_{\bar{\rho}_B}(t_2, t_1) \equiv \expval{\comm{B(t_2)}{B(t_1)}}_{\bar{\rho}_B},
\end{equation}
with $\acomm{\cdot}{\cdot}$ (resp.\ $\comm{\cdot}{\cdot}$) denoting the anti-commutator (resp.\ commutator) of two operators. 
Again, owing to stationarity, we have that $C^{\pm}$ are time-translation invariant, depending only on the time lag $ \tau = t_2 - t_1$ -- which allows us to write, with a slight abuse of notation,
\begin{equation}
\label{eq:def_stationary_correlation_functions}
C^{\pm}(\tau) \equiv C^{\pm}(\tau, 0) = C^{\pm}(t_0 + \tau, t_0), \quad \forall t_0 \in \mathbb{R}.
\end{equation}

The stationarity of the correlation functions allows describing the effect of the noise in frequency domain by considering their single-variable Fourier transform, yielding the classical and quantum noise spectra \cite{ClerkRMP,Paz2017multiqubit}: 
\begin{equation}
\label{eq:general_def_noise_spectra}
    S^{\pm}(\omega) \equiv \int_{-\infty}^{\infty} \dd \tau \, e^{- i \omega \tau} C^{\pm}(\tau).
\end{equation}
Importantly, $C^{-}(\tau)$ and $S^{-}(\omega)$ are non-zero when the noise is nonclassical, since otherwise the commutator $\comm{B(\tau)}{B(0)}$ vanishes identically.
Mathematically, one can see that $C^{+}(\tau)$ is a real and even function of $\tau$, while $C^{-}(\tau)$ is purely imaginary and odd in $\tau$.
Consequently, $S^{+}(\omega)$ is even, while $S{^-}(\omega)$ is odd in $\omega$, and both are real.

A purely dephasing spin-boson model provides a ubiquitous setting where all the stated assumptions on the noise are satisfied and exact analysis is possible, making it an important benchmark for non-perturbative treatments \cite{petruccione,PRA98,Ischi,Paz2017multiqubit,NgGaussian,Riberi_2022,nemati2022coupling}.
In this case, the bath comprises a collection of non-interacting harmonic oscillators, with a free bath Hamiltonian $ H_B = \sum_{k} \Omega_k b_k^\dagger b_k,$ with $\Omega_k > 0,$ and the bath coupling operator in Eq.\,\eqref{eq:def_Bt_rotframe} has a linear form, 
\begin{equation*}
    B = \sum_{k} ( g_k b_k^\dagger + g_k^* b_k),
\end{equation*}
in terms of creation and annihilation operators $b_k^\dagger$ and $b_k$ that satisfy canonical commutation relations, $[b_k, b_k^\dagger] = 1$, for all $k$.
The initial, thermal bath state is then given explicitly by
\begin{equation}
\label{eq:def_thermal_state_spin_boson}
    \bar{\rho}_B  = \bigotimes_k \frac{e^{- \beta \Omega_k b^\dagger_k b_k}}{ ({1 - e^{- \beta \Omega_k})}^{-1}}.
\end{equation}
The classical and quantum spectra both follow from the spectral density of the modes, defined as
\begin{equation}
\label{eq:microscopic_def_spectral_density}
    J(\omega) \equiv \frac{1}{2 \pi} \sum_{k} {\abs{g_k}}^2 \big(\delta(\omega - \Omega_k) + \delta(\omega + \Omega_k)\big).
\end{equation}
Explicitly, direct calculation yields:
\begin{subequations}
\begin{align}
S^{+}(\omega) &= \coth(\frac{\beta \abs{\omega}}{2}) J(\omega), 
 \label{eq:relation_Sp_spectral_density} \\
 S^{-}(\omega)& = \mathrm{sgn}(\omega) J(\omega).
\label{eq:relation_Sm_spectral_density}
\end{align}
\end{subequations}
We consider the typical case of a large number of modes, for which $J(\omega)$ can be considered continuous.
On physical grounds, we impose two additional constraints.
First, at low frequencies, $J(\omega)$ is taken to vanish according to a power law,
\begin{equation}
\label{eq:def_ohmicity_parameter_J}
    J(\omega) \sim {\abs{\omega}}^s \qq*{,} s > 0,
\end{equation}
where $s$ is called the Ohmicity parameter \cite{petruccione}.
That is, we require a vanishing number of modes per frequency unit are found at $\omega = 0$.
Second, $J(\omega)$ is assumed to decay to zero for $\omega \rightarrow \infty$, in a way that we take to be super-polynomial in this work.
This latter condition reflects the existence of a maximal frequency scale, or physical ``cut-off'' frequency above which modes are sparse or have a negligible coupling rate.
Relevant cases of exponential decay for $J(\omega)$ can be found e.g., in Appendix B of~\cite{mascherpa2017open}.

We stress that the spin-boson model satisfies the Gaussianity assumption for a completely general $\omega$-dependence of the spectral density $J(\omega)$.
While the resulting dephasing noise is often considered LF-dominated, realistic (e.g., phononic) environments can also display nontrivial HF spectral features, resulting from a spectral density composed of peaks centered at different frequencies with different widths~\cite{nemati2022coupling}.
To capture the effect of both LF and HF noise, here $J(\omega)$ is chosen to be composed of two separate peaks which, for analytical convenience, we assume to have Gaussian lineshapes.
That is,
\begin{align}
    J(\omega) &\equiv \frac{1}{\Gamma\qty(\frac{1 + s}{2})} \frac{\Delta_0}{\gamma_0} {\qty(\frac{\abs{\omega}}{\gamma_0})}^{\!s}
                    e^{-\omega^2/\gamma_0^2} 
                    \label{eq:def_spectral_density_gaussian_peaks}\\
              &+ \frac{\Delta_1}{\sqrt{\pi} \gamma_1} \qty[
	e^{-(\omega - \bar{\omega}_1)^2/ \gamma_1^2} +  e^{-(\omega + \bar{\omega}_1)^2/ \gamma_1^2}], 
              \nonumber
\end{align}
where $\gamma_j$ represents the linewidth of the $j$-th peak centered at $\bar{\omega}_j$ ($\bar{\omega}_{0} = 0$), with corresponding surface area $\Delta_j$.
We also refer to $\Delta_j$ as the noise strength of the corresponding peak, since the quantity $\pi \qty(\Delta_0 + 2 \Delta_1) = \sum_k {\abs{g_k}}^2$ corresponds to an overall coupling rate (squared).

We stress that, while most of Sec.\,\ref{sec:single_reset}-\ref{sec:multiple_resets} deals explicitly with a spin-boson model with a spectral density of the above form (and at zero temperature, $\beta = \infty$), the general analytical results for the gate fidelity we will establish in Sec.\,\ref{sec:single_reset}-\ref{sec:multiple_resets}, and all of the results in Sec.\,\ref{sec:bath_pov} are valid for arbitrary quantum Gaussian noise.
We then impart the same regularity requirements on the quantum spectrum, namely, $S^{-}(\omega)$ decays super-polynomially for $\omega \rightarrow \infty$, and is smooth everywhere except (possibly) at $\omega = 0$, where
\begin{equation}
\label{LF_ohmicity_Sm}
    S^{-}(\omega) \sim \mathrm{sgn}(\omega) {\abs{\omega}}^{s}.
\end{equation}
Here, with a slight abuse of notation, we use the same symbol $s$ as for the Ohmicity parameter in Eq.~\eqref{eq:def_ohmicity_parameter_J} for a spin-boson model, where the two notions clearly coincide. 
More generally, in the important case where the noise environment satisfies a Kubo-Martin-Schwinger condition, as it happens if $\bar{\rho}_B$ is a thermal equilibrium state, the classical and quantum components of $S(\omega)$ are constrained by the fluctuation-dissipation theorem \cite{ClerkRMP,Lukas}, resulting in $S^+(\omega) = \coth(\beta \omega/2) \, S^-(\omega)$, for $\omega>0$.
The above {regularity} requirements then translate directly to the symmetrized spectrum, where they are commonly assumed for purely classical (stochastic) noise.

\subsection{Protocol and performance metric}\label{ssec:protocol}

As stated in the introduction, we will assess the performance of DD-protected gates at the physical layer, under the noise and control assumptions   outlined in the previous subsection.
While several approaches have been put forward and experimentally demonstrated for combining DES with computation~\cite{viola1999universal,NgCombining,DCG,west2010high,Slava,Suter,MikeDCG,Barnes_2022}, in order to isolate the essential features of our argument, we focus here on the simplest ``decouple-then-compute'' approach.
{Here}, the desired gates are assumed to be applied instantaneously, in-between full cycles of DD that suppress temporally correlated noise over timescales long compared to the base cycle.
Further to that, in order to realize the exact dephasing-preserving dynamics described Eq.\,\eqref{eq:general_EOM_toggling_frame}, we restrict ourselves to considering only DD-protected identity gates, which then coincide with DD-schemes, meant to preserve arbitrary superpositions between $\ket{g}$ and $\ket{e}$ states.
Focusing on identity gates arguably illustrates the control-history dependence in the most fundamental way, since basic identity operations are an ubiquitous primitive on any quantum device. 

Concretely, a DD-protected identity gate of duration $\taug$, applied at the initial time of the computation, $t_s=0$, is characterized by its switching function $y_G(t)$, $t \in [0, \taug]$, where $y_G$ switches sign an even number of times (e.g., $G = I$ could correspond to Hahn-echo DD with $\tau_{\textrm{G}} \equiv 2 \tau_{\textrm{p}}$, and the pulse separation $\tau_p > \mst$.)
While, for stochastic error models, different metrics exist for quantifying the error probability of a quantum gate in an average or worst-case sense \cite{Sanders_2016, Flammia}, joint system-bath Hamiltonian evolution as in Eq.~\eqref{eq:def_Bt_rotframe} results in coherent error-amplitude propagation, and care is needed in isolating the effects of the bath in terms of an ``error action operator''~\cite{DCG,khodjasteh2010arbitrarily,paz2014general} or ``deviation maps'' \cite{NgCombining,chai2022fault}, and upper-bounding the overall error amplitude.
In our setting, the situation is greatly simplified by the fact that no error propagation happens on the system side; furthermore, the dephasing-preserving assumption we work under makes it natural to consider, for $G=I$, a ``worst-case gate fidelity'' defined by 
\begin{eqnarray}
\mathcal{F}_{\text{G}}(0)\equiv \underset{|\psi(0)\rangle}{\text{min}} \langle \psi (0) | \rho_S(\taug) | \psi (0)\rangle  =\! \mel{+}{\rho_S(\taug)}{+},\quad
\label{gate_fid}
\end{eqnarray}   
where as usual $\ket{\pm} \equiv \tfrac{1}{\sqrt{2}}\big(\ket{z=-1} \pm \ket{z=+1}\big)$ (equivalently, $\sigma_x \ket{\pm} = \pm \ket{\pm}$) and $\rho_S(\taug)\equiv  \text{Tr}_B [\rho_{SB}(\taug)]$ is understood as the reduced state the qubit reaches at time $\taug$ starting from initial condition $|\psi(0)\rangle\langle\psi(0)| \otimes \bar{\rho}_B$ \footnote{Since, under the given noise model, the populations of $\ketbra{z}$ are preserved, the worst-case gate fidelity is attained by choosing any pure state in the equatorial plane of the Bloch sphere as the input state at $t_s$.
Here, we choose $\rho_S(\ts) = \ketbra{+}$ without loss of generality.}.
The above fidelity is related to an ``average gate fidelity'' $\overline{\mathcal{F}}_{\text{G}}$, which quantifies the probability that the gate induces an error on a random input state, via $\overline{\mathcal{F}}_{\text{G}}= [1+2 \mathcal{F}_{\text{G}}]/3$ \cite{Fortunato_2002}.
While $\overline{\mathcal{F}}_{\text{G}}$ can be directly measured by randomized benchmarking \cite{Flammia}, $\mathcal{F}_{\text{G}}$ and the corresponding worst-case infidelity or error rate, $1-{\mathcal{F}}_{\text{G}}$, provide a more stringent performance guarantee in the context of fault tolerance.

\begin{figure*}[ht!]
\includegraphics[width=0.9\linewidth]{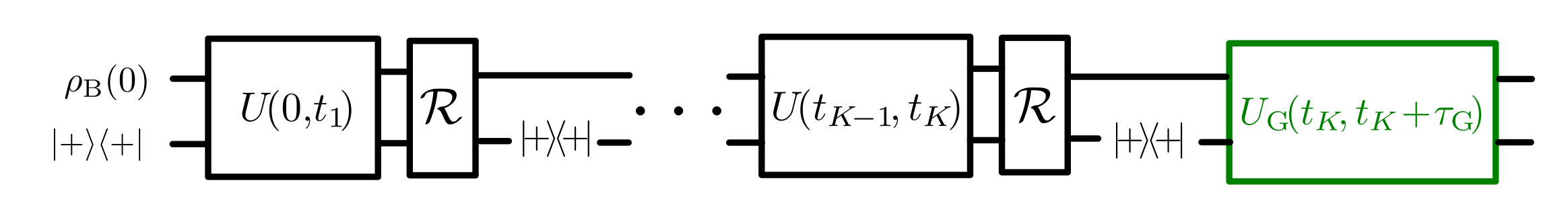}
\vspace*{-2mm}
\caption{Schematics of the most general protocol for assessing the validity of error virtualization in effecting a target DD-protected idling gate $U_G$ (green box).
A total of $K = 0, 1, \ldots$ blocks of joint system-bath evolution are allowed, during which DD is applied to the qubit, each followed by an instantaneous reset operation that perfectly disentangles system and bath, and returns the qubit to the $\ketbra{+}$-state.}
\label{fig:general_protocol}
\end{figure*}

Note that an unprotected gate of shorter duration (say, $\taug = \mst$) can be outperformed by a DD-protected gate that takes a longer time, as this is the basic premise of DES~\cite{viola1999dynamical,DCG}: suppression of temporally-correlated noise.
This work considers the case where in order to obtain an identity gate that achieves high-fidelity, say, $\mathcal{F}_{\textrm{G}}(0) > \mathcal{F}_{\textrm{G}}^{\text{\,min}}$, one must in fact resort to DES -- in line with the notion of a ``virtualized error'' in layered architectures for fault-tolerant QEC \cite{CodyJones}.
Throughout the following analysis, we will uphold the condition $\mathcal{F}_{\textrm{G}}^{\text{\,min}} =0.98$ as a general figure of merit.

Consider now a general control history applied before the starting time $\ts$ of a same DD-protected identity gate, characterized by a switching function $y_{\textrm{hist}}(t), t \in [0, \ts]$ (see also Fig.\,\ref{fig:general_protocol}). 
In order to understand how the bath dynamics influence the dependence of the gate performance upon the control history for different scenarios and starting times $\ts$, we introduce a reset-operation $\reset$ that fully disentangles the system and the bath, and re-prepares the system in the $\ketbra{+}$-state at designated times.
That is:
\begin{equation}
\label{eq:general_def_reset}
    \reset(\rho_{\textrm{SB}}) \equiv \ketbra{+} \otimes \Tr_S(\rho_{\textrm{SB}}).
\end{equation}
One possible implementation of such a reset-procedure would be to perform a perfect projective measurement in the $\sigma_x$-basis, and to apply a corrective $\pi_z$-pulse when measuring $\ketbra{-}$ as an outcome. {Or, it could be implemented in an  unconditional fashion, in the spirit of driven reset schemes \cite{Reset1,Reset2}. Regardless, a Kraus representation}
reads
\begin{equation}
\label{eq:krausmap_closedloop_reset}
    \reset(\rho_{\textrm{SB}}) = \ketbra{+} \rho_{\textrm{SB}} \ketbra{+} + \sigma_z \ketbra{-} \rho_{\textrm{SB}} \ketbra{-} \sigma_z.
\end{equation}
We stress, however, that the particular implementation of $\mathcal{R}$ has no importance, as Eq.\,\eqref{eq:general_def_reset} and Eq.\,\eqref{eq:krausmap_closedloop_reset} are two equivalent representations of the same operation.
As for the control pulses, we assume $\reset$ to be applied perfectly and instantaneously. 
Let us introduce the notation $\ts^-$ ($\ts^+$) to denote the instant immediately before (after) the reset.
If the gate $G=I$ is applied at a start time $\ts\equiv \ts^+ > 0$, after the given control history and a reset operation, the joint state is given by 
\begin{equation*}
    \rho_{\textrm{SB}}(t_s^+) = \reset(\rho_{\textrm{SB}}(t_s^-)) =  \ketbra{+} \otimes \Tr_S \big[\rho_{\textrm{SB}}(t_s^-)\big], 
\end{equation*}
and the fidelity given in Eq.\,\eqref{gate_fid} thus becomes 
\begin{eqnarray}
\label{eq:def_gate_fidelity_single_reset}
    \mathcal{F}_{\textrm{G}}(\ts)  
    = \mel{+}{\rho_S(\ts + \taug)}{+} .
\end{eqnarray}
Apart from providing a fair comparison of the performance of one same gate applied in different temporal environments, $\reset$ serves as a mock-up of a QEC operation.
Indeed, $\reset$ disentangles the system and bath, while reducing to the identity operation were there no noise (and hence entanglement between system and bath \cite{Roszak}), given that the qubit is properly initialized to $\ketbra{+}$.
Of course, this analogy is only partial, since in our single-qubit setting the full state of the physical system is preserved, as opposed to the state of a logical subsystem in a ``true'' QEC setting.

The above description applies to single-reset scenarios, which we referred to as $({\bm {c_1}}$) in the introduction.
However, the more general protocols we consider, $({\bm {c_2}})$ and $({\bm {c_3}}$), involve the application of $K$ reset operations $\reset$ at designated times $t_k, k = 1, \ldots, K$, interspersed in-between a general control history characterized by $y_{\textrm{hist}}(t), t \in [0, t_{K})$, before the application of the DD-protected gate of (fixed) duration $\taug$ (depicted in green in Fig.\,\ref{fig:general_protocol}).
The figure of merit is still the gate fidelity given in Eq.\,\eqref{eq:def_gate_fidelity_single_reset}, except that now the starting time of the gate is $t_s = t_K$, and the joint post-reset states are given by 
\begin{align*}
    \rho_{\textrm{SB}}(t_{k}^{+}) = \reset\qty(\rho_{\textrm{SB}}(t_{k}^{-})), \quad k = 1, \ldots, K.
\end{align*}
Consistent with the timing constraints we impose on unitary control, we analogously demand a minimal separation in time between two consecutive reset operations, which we take to be the same as the minimum switching time $\mst$ for convenience; thus, $t_{k} - t_{k-1} \geq \mst,$ for all $k = 1, \ldots, K$.

As a last remark on our problem setup, we note that for a true finite-distance QEC code, there is a finite probability of correcting to the wrong logical state, whereas our mock-up reset operation is assumed to be perfect.
This has for a consequence that there is no (coherent) error propagation on the system side in-between different reset operations.
In this way, we can reasonably expect the gate fidelities we study to upper-bound an analogous performance for a true QEC situation with encoded qubits. 
We will see, however, that there is propagation of errors between different resets, which then stems entirely from error propagation on the bath side, and is encoded entirely in the updated state of the bath following reset operations:
\begin{equation}
\rho_B(t_{k}^+) \equiv \Tr_S[\rho_{\textrm{SB}}(t_{k}^-)],\quad k = 1, \ldots, K.
\end{equation}
It is exactly this separate phenomenon we focus on in this work, independently of known and studied error propagation on the system side.

\section{
Gate performance after a single error-correction cycle}
\label{sec:single_reset}

In this section, we zoom in on the case of a single reset $\mathcal{R}$ in studying the control-history dependence of a DD-protected identity gate of fixed duration $\tau_G$.
While this study will reveal a distinct set of possible phenomena, it will additionally set the stage for exploring the general case in Sec.\,\ref{sec:multiple_resets}.

\subsection{Exact results: Asymptotic gate-performance saturation}

\label{ssec:exact_asymptotics_single_reset}

\subsubsection{General fidelity expressions}

A detailed derivation provided in Appendix~\ref{app:sec:deriv_x_exp_vals} yields the following exact formula for the gate fidelity defined in Eq.~\eqref{eq:def_gate_fidelity_single_reset}, for the case of a single reset operation ($\ts\equiv t_1$):
\begin{equation}
\label{eq:general_fidelity_single_reset}
    \mathcal{F}_{\textrm{G}}(\ts) = \frac{1}{2}\Big(1 + e^{- \chi_{\textrm{c}} }\cos(\theta_{\textrm{q}})\Big),
\end{equation}
with
\begin{equation}
\label{eq:timedom_def_chi}
    \chi_{\textrm{c}} \equiv \int_{0}^{\taug} \dd \!\tau_2 \int_{0}^{\taug} \dd \tau_1 \;
                                    y_{\textrm{G}}(\tau_1) y_{\textrm{G}}(\tau_2) \, C^+(\tau_2 - \tau_1),
\end{equation}
and
\begin{equation}
\label{eq:timedom_def_single_theta}
    \theta_{\textrm{q}} \equiv - 2 i \int_{0}^{\taug} \! \! \dd \tau \int_{0}^{t_s} \! \! \dd t
                                    \, y_{\textrm{G}}(\tau) \, y_{\textrm{hist}}(t) \, C^-(\tau - t + t_s),
\end{equation}
respectively. Moving to the frequency domain yields the alternative representations
\begin{subequations}
\begin{align}
    \chi_{c}   &\equiv \frac{1}{2 \pi} \int_{- \infty}^{\infty} {\abs{F_{\textrm{G}}(\omega ; \taug)}}^2 S^{+}(\omega) \dd \omega,
    \label{eq:freqdom_def_chi} \\
    \theta_{q} &\equiv  \frac{1}{\pi} \!\! \int_{- \infty}^{\infty} \!\!\!\!\!\!
                       \mathrm{Im}\qty[e^{i \omega \ts} F_{\textrm{G}}(\omega ; \taug) F^*_{\textrm{hist}}(\omega ; \ts) ] S^{-}(\omega) \dd \omega \label{eq:freqdom_def_single_theta},
\end{align}
\end{subequations}
where $\mathrm{Im}$ stands for taking the imaginary part, and the FFs
\begin{align}
F_{\textrm{G}}(\omega; \taug) \equiv & \int_{0}^{\taug} y_{\textrm{G}}(\tau) e^{i \omega \tau} \dd \tau, \label{ff1}\\
F_{\textrm{hist}}(\omega; \ts) \equiv & \int_{0}^{\ts} y_{\textrm{hist}}(t) e^{i \omega t} \dd t.
\label{ff2}
\end{align}
Note that the loss of fidelity in Eq.\,\eqref{eq:general_fidelity_single_reset} stems from a genuine (dephasing) decoherence process, in the sense that no part of the effect on the qubit can be described by a unitary rotation around the $\sigma_z$-axis of the Bloch sphere.
This is evidenced by the fact that we have $$\mel{z}{\rho_{\textrm{S}}(\ts + \taug)}{z'} = e^{-\chi_{\textrm{c}} } \cos(\theta_{\textrm{q}}) \mel{z}{\rho_{\textrm{S}}(0)}{z'}, \quad z\neq z',$$ for the $z$-coherence element itself.

We can draw a few immediate conclusions from the above general formulas.
First, $\chi_{\textrm{c}}$ depends only on the control applied \emph{after} the reset, and is due only to the classical part of the noise, captured by $C^{+}$.
We hence refer to $\chi_{\textrm{c}}$ as the classical decay factor.
The quantum phase angle $\theta_{\textrm{q}}$, on the other hand, depends also on the control applied \emph{before} the reset, and takes contribution solely from $C^{-}$.
This highlights the necessity for the noise to be genuinely nonclassical for a control-history dependence to arise in the gate fidelity: if the noise was classical ($C^{-} \equiv 0$), we would have that $\theta_{\textrm{q}} = 0$, and one would recover the same gate fidelity as for $\ts = 0$, for any control history.
Indeed, when $\ts = 0$ (hence no reset is involved), Eqs.\,\eqref{eq:general_fidelity_single_reset}-\eqref{eq:timedom_def_chi} reduce to the well-known result for Gaussian dephasing noise under purely dephasing-preserving control $-$ see Appendix A of~\cite{Paz2017multiqubit} for a derivation of the system-only dynamics, employing similar techniques as ours.
To the best of our knowledge, the present derivation of exact expectation values for our protocol is novel.

As a further remark, when there is genuine nonclassical noise, note that the control history dependence engendered by $\theta_{\textrm{q}}$ can only lower the fidelity of a given implementation of a DD-protected gate, regardless of the applied control history.
It will prove instructive to quantify the extent to which the quantum phase worsens the EPG compared to the case of vanishing nonclassical noise, or, equivalently, to the case where the gate is applied at $\ts = 0$.
This motivates defining a relative quantum error (RQE) as follows:
\begin{equation}
\label{eq:def_RQE}
    \mathrm{RQE} \equiv \frac{1 - \mathcal{F}_{\textrm{G}}(\ts)}{1 - \mathcal{F}_{\textrm{G}}(0)}
                 = \frac{1 - e^{- \chi_{\textrm{c}}} \cos(\theta_{\textrm{q}})}{1 - e^{- \chi_{\textrm{c}}}} .
\end{equation}
Upholding the fact that $\chi_{\textrm{c}} \ll 1$ as discussed, and realizing that when $\cos(\theta_{\textrm{q}})$ cannot be expanded up to quadratic terms, the gate fidelity will have dropped below any possible threshold, we can approximate
\begin{equation}
\label{eq:approximate_RQE}
    \mathrm{RQE} \approx 1 + \frac{{\theta_{\textrm{q}}}^2}{2 \chi_{\textrm{c}}},
\end{equation}
up to the leading order.
While parallel analysis indicates that terms linear in $\theta_{\textrm{q}}$ are relevant for a different class of control protocols~\cite{free_evolution_paper}, in our context $\theta_{\textrm{q}}$ only enters up to quadratic order. 
The quantitative consequences of this will be discussed in Sec.\,\ref{ssec:case_studies_single_reset}, in reference to the concrete spin-boson setting. 

\subsubsection{Asymptotics for periodic control}

By maintaining the analysis at the general level, consider now scenario $({\bm {c_1}}$), whereby the same gate sequence is repeated $M$ times before the reset, and once more after the reset.
Formally, we have $\ts = M \taug$ and \[y_{\textrm{hist}}(t) = y_{\textrm{G}}(t \; \mathrm{mod} \, \taug) \qq*{,} t \in [0, M \taug).\] 
It is easy to verify from the definition of the FFs in Eqs.\,\eqref{ff1}-\eqref{ff2} that, in this case, \[F_{\textrm{hist}}(\omega ; M \taug) = e^{i (M - 1) \frac{\omega \taug}{2}} \,\frac{\sin(M \frac{\omega \taug}{2})}{\sin(\frac{\omega \taug}{2})} F_{\textrm{G}}(\omega ; \taug).\]
The quantum phase angle can consequently be written as
\begin{equation}
\label{single_theta_periodic_repetition}
    \theta_{\textrm{q}}^{(M)} = \frac{1}{2 \pi} \int_{- \infty}^{\infty} \xi^{(M)}(\omega \taug) {\abs{F_{\textrm{G}}(\omega ; \taug)}}^2 S^{-}(\omega) \dd \omega,
\end{equation}
where the kernel capturing the periodic repetition reads
\begin{equation}
\label{eq:def_xi_kernel}
    \xi^{(M)}(\omega \taug) \equiv \cot(\frac{\omega \taug}{2}) \big(1 - \cos( M \omega \taug)\big) + \sin(M \omega \taug).
\end{equation}

The remainder of this section studies the limiting case of an asymptotically-large number of repetitions $M$.
While $\theta_{\textrm{q}}^{(M)}$ could \emph{a priori} have general $M$-dependence, we will show that the sequence asymptotically converges to a finite limit value, under very general conditions on the remaining term ${\abs{F_{\textrm{G}}(\omega ; \taug)}}^2 S^{-}(\omega)$ entering the integral in Eq.\,\eqref{single_theta_periodic_repetition}.
Formally, we have the following:

\smallskip

{\bf Theorem 1. $\!$[Single-reset asymptotic quantum phase].} 
Assume that the control FF for the DD-protected gate $G$ [Eq.\,\eqref{ff1}] has a LF-expansion of the form ${\abs{F_{\textrm{G}}(\omega ; \taug)}}^2 \sim \omega^{2 \alpha_{\textrm{p}}},$ in terms of its FO [Eq.\,\eqref{eq:def_classical_filtering_order}], and that 
\begin{equation}
s_{\textrm{p}} \equiv s + 2 \alpha_{\textrm{p}} >0, 
\label{eq:plateau_condition_single_reset}
\end{equation}
where the Ohmicity parameter $s$ characterizes the LF power-law behavior of the quantum noise spectrum. Then the quantum phase in Eq.\,\eqref{single_theta_periodic_repetition} converges to a finite value,
\begin{equation*}
\lim_{M \rightarrow \infty} \theta_{\textrm{q}}^{(M)}  \equiv \theta_{\textrm{q}}^{(\infty)}  .
\end{equation*}

{\bf Proof.} The desired result can be obtained from the properties of the kernel $\xi^{(M)}(\omega \taug)$. First, let us rewrite $\xi^{(M)}(\omega \taug)$ in the alternative representation
\begin{equation}
\label{eq:fourier_representation_xi_M}
    \xi^{(M)}(\omega \taug) = 2 \sum_{n = 1}^{M} \sin(n \omega \taug),
\end{equation}
which allows us to write
\begin{equation}
\label{eq:partial_sum_thetas_single_reset}
    \theta_{\textrm{q}}^{(M)} = \sum_{n = 1}^{M} \theta_{\textrm{q}, n},
\end{equation}
where the elementary phase contribution reads
\begin{align}
\theta_{\textrm{q},n} 
= - \frac{i}{\pi} \int_{- \infty}^{\infty} e^{i n \taug \omega} {\abs{F_{\textrm{G}}(\omega ; \taug)}}^2 S^{-}(\omega) \, \dd \omega. 
\label{eq:theta_delta_m_fourier_tf}
\end{align}
Since $\theta_{\textrm{q},n}$ is the (inverse) Fourier transform of the function $- 2 i {\abs{F_{\textrm{G}}(\omega ; \taug)}}^2 S^{-}(\omega)$, evaluated at time $n  \taug$, we can utilize known results \cite{saichev2018distributions} that relate the regularity (or smoothness) of a function (here, ${\abs{F_{\textrm{G}}(\omega ; \taug)}}^2 S^{-}(\omega)$) to the long-time asymptotic decay of its (inverse) Fourier transform (i.e., how fast $\theta_{\textrm{q},n} \rightarrow 0$ for $n \rightarrow \infty$).
Excluding a finite number of terms in Eq.~\eqref{eq:partial_sum_thetas_single_reset} for $M \gg 1$, we can then estimate the magnitude of the tail of the sum, to prove convergence.

By using Eq.~\eqref{eq:def_classical_filtering_order}, together with Eq.~\eqref{LF_ohmicity_Sm}, we have 
\[ {\abs{F_{\textrm{G}}(\omega ; \taug)}}^2 S^{-}(\omega) \sim \mathrm{sgn}(\omega) {\abs{\omega}}^{s_{\textrm{p}}},\]
and hence the level of smoothness at $\omega = 0$ is fully determined by $s_{\textrm{p}}$.
As stated in Sec.\,\ref{ssec:controlled_noise_model}, $S^{-}(\omega)$ is further assumed to be $\mathcal{C}^{\infty}$ everywhere except at $\omega = 0$,
and to decay super-polynomially for $\omega \rightarrow \infty$.
One can then see that: 

i) For $s_{\textrm{p}} \in \mathbb{N}$ and odd (or, equivalently, $\frac{s_{\textrm{p}} - 1}{2} \in \mathbb{N}$), we have ${\abs{F_{\textrm{G}}}}^2 S^{-} \in \mathcal{C}^{\infty}(\mathbb{R})$; 

ii) For $s_{\textrm{p}} \in \mathbb{N}$ and even (or, $\frac{s_{\textrm{p}}}{2} \in \mathbb{N}$), ${\abs{F_{\textrm{G}}}}^2 S^{-} \in \mathcal{C}^{s_{\textrm{p}} - 1}(\mathbb{R})$, while its derivative of order $s_{\textrm{p}}$ is bounded but discontinuous. 

From the results in Chapter 4 of~\cite{saichev2018distributions}, it then follows that $\theta_{\textrm{q},n} \rightarrow 0$ super-polynomially for $n \rightarrow \infty$ in case (i), while $\abs{\theta_{\textrm{q},n}} \sim 1 / n^{1 + s_{\textrm{p}}}$ in case (ii) \footnote{The more general case $s_{\textrm{p}} \in \mathbb{R} \setminus \mathbb{N}$ can be treated in a natural way utilizing the theory of homogeneous distributions.}.
For intermediate values of $s_{\textrm{p}}$, one obtains more generally that
\begin{equation}
\label{eq:asymptotics_theta_delta_m}
    \abs{\theta_{\textrm{q},n}} \sim \Gamma\qty(s_{\textrm{p}} + 1)
        \frac{\abs{\cos(\frac{\pi s_{\textrm{p}}}{2})}}{n^{1 + s_{\textrm{p}}}},\quad s_{\textrm{p}} > - 1,  \frac{s_{\textrm{p}} - 1}{2} \notin \mathbb{N}.
\end{equation}
We conclude that $\theta_{\textrm{q}}^{(M)}$ converges to a finite value for $M \rightarrow \infty$ whenever $s_{\textrm{p}}>0$, as assumed.
Moreover, when $s_{\textrm{p}}$ is not an odd integer, we have 
\begin{align*}
    \abs{ \theta_{\textrm{q}}^{(M)} - \theta_{\textrm{q}}^{(\infty)} } \; \; \stackrel{M \rightarrow \infty}{\sim}&
         \; \;\Gamma\qty(s_{\textrm{p}} + 1) \abs{\cos(\frac{\pi s_{\textrm{p}}}{2})} \nonumber \\
        \! \! &\times \! \! \bigg(\zeta(1 + s_{\textrm{p}}) - \sum_{n = 1}^{M - 1} \frac{1}{n^{1 + s_{\textrm{p}}}}\bigg),
\end{align*}
where $\zeta$ is the Riemann zeta function. \qed

\begin{figure}[t!]
\includegraphics[width=0.98\linewidth]{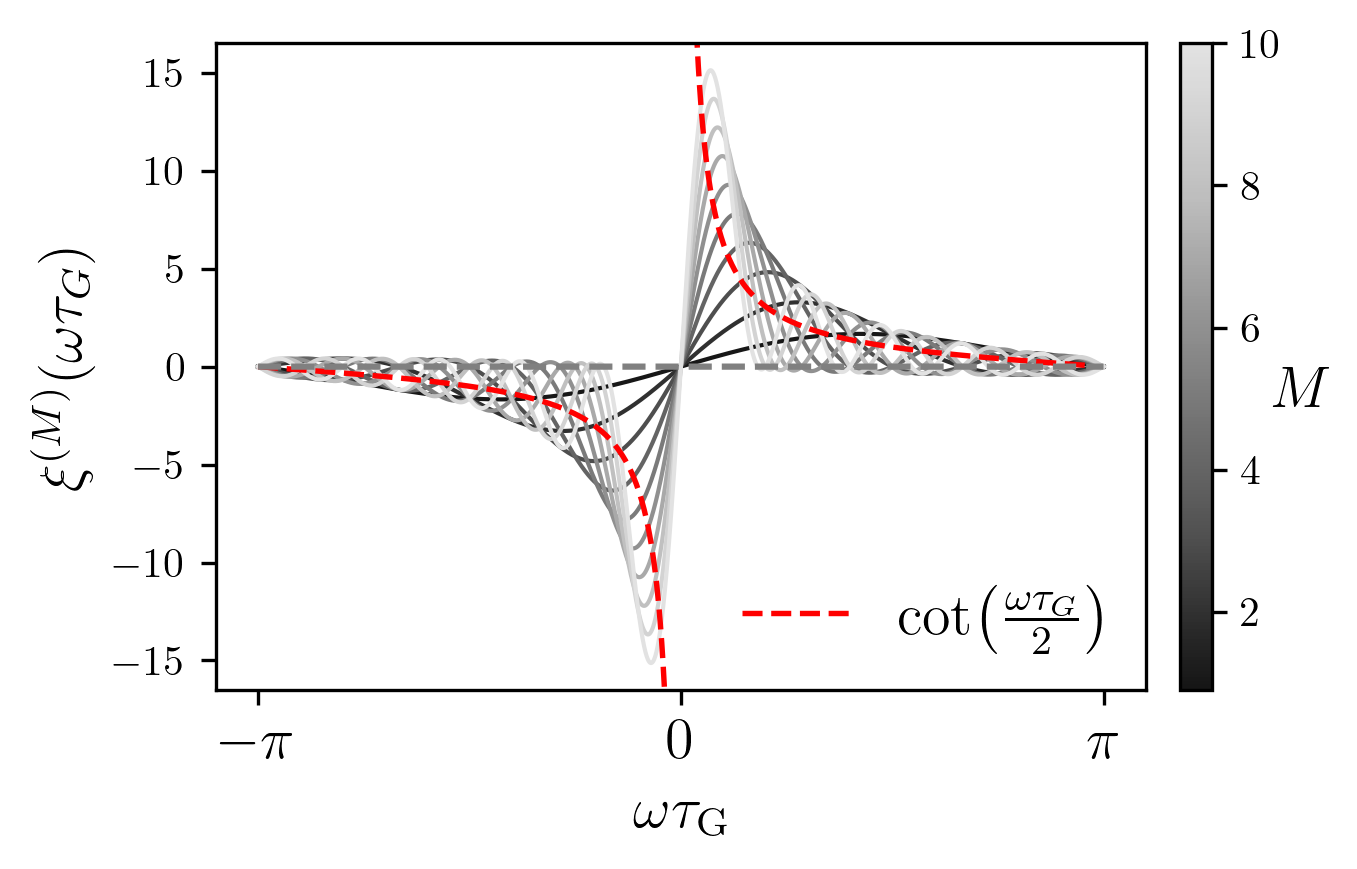}
\vspace*{-3mm}
\caption{Illustration of the repetition kernel $\xi^{(M)}(\omega \taug)$ defined in Eq.\,\eqref{eq:def_xi_kernel} for increasing values of $M$, alongside its limit in a distributional sense, Eq.\,\eqref{eq:convergence_xi_M_distributional_sense}.}
\label{fig:kernel_convergence}
\end{figure}

\smallskip

A last question to be answered in this section is which specific value  the quantum phase $\theta_{\textrm{q}}^{(\infty)}$ converges to.
We can answer this by considering the kernel $\xi^{(M)}(\omega \taug)$ in a distributional sense, acting on the appropriate space of test functions ${\abs{F_{\textrm{G}}}}^2 S^{-}$.
First, note that $\theta_{\textrm{q}}^{(M)}$ is well-defined whenever $s_{\textrm{p}} > -2$, as then $\xi^{(M)} {\abs{F_{\textrm{G}}}}^2 S^{-}$ is locally integrable (around $\omega = 0$), for all finite $M \in \mathbb{N}$.
While, for finite $M$, considering the space of test functions characterized by $s_{\textrm{p}} > -2$ thus suffices, to take the limit $M \rightarrow \infty$, one needs $s_{\textrm{p}} > 0$, in which case it can be shown that
\begin{equation}
\label{eq:convergence_xi_M_distributional_sense}
    \xi^{(M)}(\omega \taug) \stackrel{M \rightarrow \infty}{\rightarrow} \mathcal{P}v_{\cot(\frac{\omega \taug}{2})},
\end{equation}
in a distributional sense.
Note that the continuity condition $s_{\textrm{p}} > 0$ indeed matches the condition in Eq.~\eqref{eq:plateau_condition_single_reset} for the convergence of $\theta_{\textrm{q}}^{(M)}$.
A graphical illustration of Eq.~\eqref{eq:convergence_xi_M_distributional_sense} can be seen in Fig.\,\ref{fig:kernel_convergence}.
Summarizing, we have that
\begin{equation*}
    \theta_{\textrm{q}}^{(\infty)} = \frac{1}{2 \pi} \mathcal{P}v \! \int_{- \infty}^{\infty} \! \!
        \cot(\frac{\omega \taug}{2}) {\abs{F_{\textrm{G}}(\omega ; \taug)}}^2 S^{-}(\omega) \, \dd \omega, 
\end{equation*}
under the conditions in the above theorem, where $\mathcal{P}v$ denotes that the integration limits around the singular points $\omega_\ell \equiv {2 \ell \pi}/{\taug}$, $\ell  \in \mathbb{Z}$, need to be taken symmetrical.

For the purpose of clarity, we stress that the results established here have no bearing on the coherence plateaus shown to exist for periodic DD in a purely dephasing-preserving setting in~\cite{khodjasteh2013designing}, or vice versa.
On the level of the protocol, the repetition index $M$ here represents a possibly family of experiments where a DD scheme is repeated $M$ times before a reset operation, after which the same scheme is applied one more time.
In~\cite{khodjasteh2013designing}, all operations are dephasing-preserving -- which the reset operation is manifestly not.
Hence, while condition Eq.~\eqref{eq:plateau_condition_single_reset} resembles the first condition in Equation (3) of~\cite{khodjasteh2013designing}, the two results are not directly related. 

In summary, for the elementary case of periodic control repetition followed by a single reset operation, we have shown that for an asymptotically-large number of repetitions $M$, the fidelity of the DD-protected gate saturates to a value
\begin{equation*}
    \mathcal{F}_{\textrm{G}}(M \taug \to \infty) = \frac{1}{2}\Big(1 + e^{- \chi_{\textrm{c}}} \cos( \theta_{\textrm{q}}^{(\infty)}) \Big) <  \mathcal{F}_{\textrm{G}}(0) ,
\end{equation*}
that is, strictly smaller than the fidelity of the same gate were it applied at time $\ts = 0$, in the absence of any prior history.
This gate performance saturation constitutes our first main result, and will be further elaborated on in the next section.

\subsection{Impact of noise-spectral properties and control}
\label{ssec:case_studies_single_reset}

In this section, we consider a spin-boson model at zero temperature ($\beta = \infty$), with a spectral density of the form  Eq.~\eqref{eq:def_spectral_density_gaussian_peaks}, capturing both LF- as well as HF-noise.
Anticipating the results to follow, combined with a typical timescale of the applied control, namely, the gate duration $\taug$, this noise model allows for the magnitude of both $\chi_{\textrm{c}}$ and $\theta_{\textrm{q}}$ to be described as a function of the following $5$ dimensionless parameters:
\begin{equation}
\LFstrength \taug^2, \quad \HFstrength \taug^2,\quad \LFwidth \taug, \quad \HFwidth \taug, \quad \bar{\omega}_{1} \taug.
\label{parameters}
\end{equation}
Here, $\LFstrength \taug^2$ and $\HFstrength \taug^2$ are to be interpreted as the LF- and, respectively, HF-noise strengths, given the relevant timescale of evolution, $\taug$.
The parameters $\LFwidth \taug$ and $\HFwidth \taug$ capture how fast the gate is performed with respect to the typical correlation time of the LF- resp.\ HF-components of the noise.
Lastly, $\bar{\omega}_{1} \taug$ designates where the HF-noise peak is centered w.r.t.\ the characteristic frequency scale $\omega_G\equiv 2 \pi / \taug$ introduced by the applied periodic control.

We will uphold two general physically-motivated constraints on the noise.
First, we assume LF-noise to be always present, and to \emph{a priori} dominate the HF noise, in the sense that $\LFstrength$ is at least an order of magnitude (OOM) larger than $\HFstrength$.
Second, the width of the HF-noise peak is assumed to be much smaller than that of the LF-noise peak, $\HFwidth \ll \LFwidth$.
Both these assumptions are typical for dephasing noise specifically.
In addition, not only do we assume the gate fidelity for $\ts = 0$ to be high -- say, $\mathcal{F}_{\textrm{G}}(0) > 0.98$, as stated in Sec.\,\ref{ssec:protocol} -- but we demand that this high-fidelity regime is achieved thanks to DD-protection of the gate.
Concretely, this means that we assume DD to be implemented deep in a fast control limit, whereby $\gamma_0 \mst \ll 1$ (recall that $\gamma_0 \mst \ll 2\pi $ is a necessary condition for LF-noise to be perturbatively suppressed by DD \cite{limits}).
In addition, since in periodic DD protocols $\taug$ provides the duration of the base identity-gate sequence that is being repeated, we will also assume the short gate limit (SGL), for which \(\LFwidth \taug \ll 1,\) ensuring that the ``resonance'' frequency $\omega_G$ introduced by the control is much higher than the cutoff of the LF noise, determined by $\LFwidth$~\cite{Irene,khodjasteh2013designing}.

\subsubsection{Transition in effective filtering order for low-frequency noise}
\label{ssec:transition_filtering_order}

\begin{figure*}[ht!]
\includegraphics[width=0.95\linewidth]{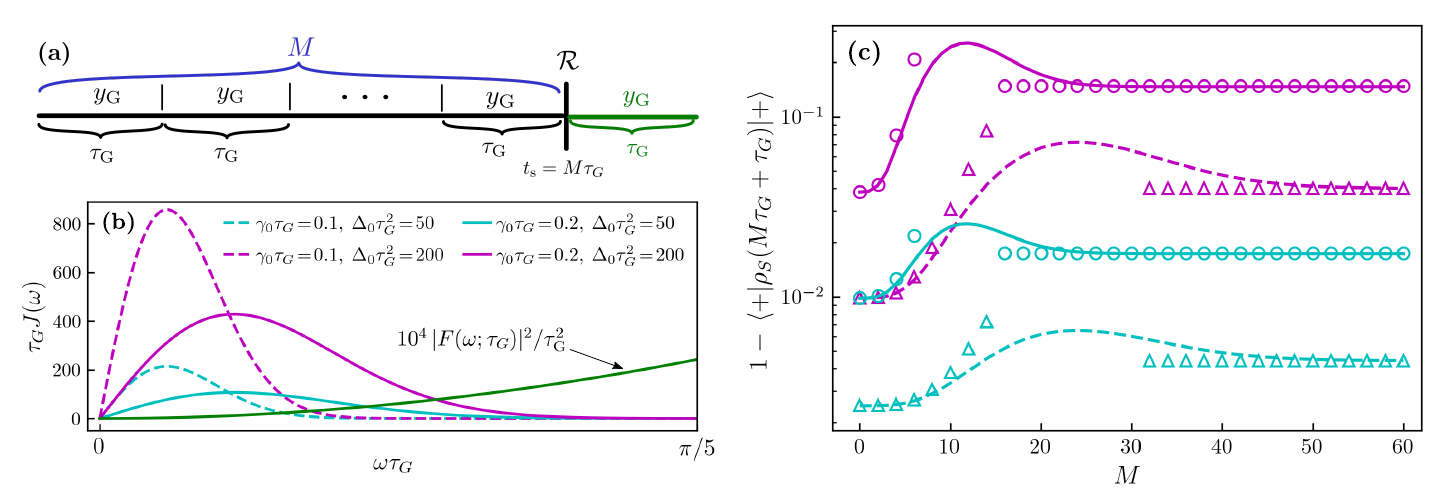}
\vspace*{-6mm}
\caption{Performance of an identity gate protected by Hahn-echo DD after $M$ echo repetitions and a single reset, for the case of only LF-noise ($\HFstrength = 0$), and Ohmic ($s=1$), zero-temperature spin-boson dephasing. 
{\bf (a)} Graphical illustration of the applied control protocol. 
{\bf (b)} FF in the LF limit, and spectral densities for two different values of $\LFstrength, \LFwidth$.
{\bf (c)} Gate infidelity as a function of the number of repetitions $M$ of Hahn-echo DD ($\taug =2\tau_0, \alpha_{\textrm{p}}=1$) before the reset operation.
The open markers (triangles resp. circles depending on the peak width) depict the predicted value for the infidelity from the approximate formulas in Eqs.\,\eqref{eq:OOM_formula_chi_LF_noise}-\eqref{eq:OOM_formulas_theta_LF_noise}, both in the short-time (left), as well as asymptotic regime (right).
For clarity, only even values of $M$ are plotted.
One can see that the convergence towards the asymptotic gate performance plateau is twice as fast for the noise-peak that is twice as broad.}
\label{fig:infidelities_spectra_single_reset_LF_noise}
\end{figure*}

The behavior of the gate fidelity for the case where the repeated base sequence is a Hahn echo of duration $\taug =2 \tau_p=\mst$ is illustrated in Fig.\,\ref{fig:infidelities_spectra_single_reset_LF_noise}(c) (solid circles resp.\ triangles), for four different spectral densities that include only LF-noise ($\HFstrength = 0$) and are depicted in Fig.\,\ref{fig:infidelities_spectra_single_reset_LF_noise}(b).
As predicted, we see that, for asymptotically large values of $M$, the gate infidelity converges to a plateau error value, which is strictly higher than the error the gate incurs when applied at $\ts = 0$ (i.e., the first point $M = 0$).
The transient time in $M$ towards the plateau is found to be inversely proportional to the width $\LFwidth$ of the LF-peak.
The infidelity for fixed $M$ is seen to depend intricately on both $\LFstrength$ and $\LFwidth$, and is increasing for both stronger noise ($\LFstrength \nearrow$), and larger  
peak width ($\LFwidth \nearrow$).
This latter fact can be attributed to a worsened ability of DD to cancel the noise, since it is less supported on low frequencies.
As a function of $M$, however, we observe that the asymptotic gate error can be almost a whole OOM larger than for small values of $M$, and even more so for transient values of $M$, as seen for the pink curves.
Clearly, a large portion of the DD-suppression capabilities of LF-noise after the reset can be lost in the asymptotic regime, while for small values of $M$ there seems to be a much more limited control-history dependence of the gate infidelity.

To gain insight, it is useful to think of the ability of a given DD-scheme to suppress LF-noise in terms of a filtering order of the relevant FFs, which, for the family of CDD-schemes~\cite{khodjasteh2005fault,khodjasteh2007performance} (including CDD1 $\equiv$ Hahn echo) additionally coincides with the ``cancellation'' or decoupling order~\cite{paz2014general}.
Specifically, we employ the filtering order defined in Eq.\,\eqref{eq:def_classical_filtering_order} to show to which OOM the gate infidelity is suppressed by the DD, starting from the exact expressions for the decay and the phase we previously derived, and considering the two limiting scenarios of small values of $M$ or asymptotically-large values of $M$, respectively.
By using the explicit form of $J(\omega)$ in Eqs.\,\eqref{eq:freqdom_def_chi} and Eq.~\eqref{single_theta_periodic_repetition}, we have: 
\begin{eqnarray}
\chi_{\textrm{c}}        
   & = &\frac{1}{2 \pi \Gamma\qty(\frac{1 + s}{2})} \frac{\LFstrength}{\LFwidth} 
\label{eq:chi_LF_noise} \\
&\times & \int_{- \infty}^{\infty}                         {\abs{F_{\textrm{G}}(\omega ; \taug)}}^2
                                 {\qty(\frac{\abs{\omega}}{\LFwidth})}^{\!s} e^{ - {\omega}^2/\LFwidth^2}  \,\dd \omega , \nonumber\\  
    \theta_{\textrm{q}}^{(M)} &= &\frac{1}{2 \pi \Gamma\qty(\frac{1 + s}{2})} \frac{\LFstrength}{\LFwidth} 
    \label{eq:single_theta_LF_noise}\\
                   & \times & \int_{- \infty}^{\infty} \xi^{(M)}(\omega \taug) {\abs{F_{\textrm{G}}(\omega ; \taug)}}^2
                                 {\qty(\frac{\abs{\omega}}{\LFwidth})}^{\!s}  e^{ - {\omega}^2/\LFwidth^2}  \,
                                 \dd \omega. \nonumber
\end{eqnarray}
It is straightforward to verify that, for a fixed (-order) DD scheme, the filter $F_{\textrm{G}}$ is such that we can write
\[{\abs{F_{\textrm{G}}(\omega ; \taug)}}^{2} = \taug^2 {|\tilde{F}_{\textrm{G}}(\omega \taug)|}^{2},\]
where $\tilde{F}_{\textrm{G}}$ depends only on the product $\omega \taug$.
Given that, in Eq.~\eqref{eq:chi_LF_noise} and Eq.~\eqref{eq:single_theta_LF_noise}, the integral predominantly has support on frequencies that are at most of the order of $\LFwidth$, due to the Gaussian roll-off of the noise, we can expand $\tilde{F}_{\textrm{G}}$ around $\omega \taug = 0$ up to leading order.
That is, 
\begin{equation}
\label{eq:LF_expansion_classical_filter}
    {|\tilde{F}_{\textrm{G}}(\omega \taug)|}^{2} \simeq \tilde{F}^{2}_{\textrm{G},0} \, {(\omega \taug)}^{2 \alpha_{\textrm{p}}},
\end{equation}
with $\tilde{F}^{2}_{\textrm{G},0} > 0$ being the appropriate leading-order coefficient, which is valid in the SGL we assumed.

From Eq.\,\eqref{eq:fourier_representation_xi_M}, however, we observe that a lowest-order Taylor expansion of the kernel $\xi^{(M)}(\omega \taug)$ for fixed $M$ is only valid over a frequency scale that goes like $2 \pi / M \taug$ (see Fig.\,\ref{fig:kernel_convergence} for an illustration).
Hence, we cannot approximate $\xi^{(M)}$ uniformly in $M$, for a fixed SGL parameter $\LFwidth \taug$. 
For $M \LFwidth \taug \ll 1$, we take
\[\xi^{(M)}(\omega \taug) \simeq M (M + 1) \, \omega \taug,\]
whereas for $M \LFwidth \taug \gg 1$, one can consider the limit
\begin{align*}
    \xi^{(M)}(\omega \taug) \simeq \mathcal{P}v \cot(\frac{\omega \taug}{2}) 
                          \simeq \mathcal{P}v_{\frac{2}{\omega \taug}}.
\end{align*}
Performing these approximations in Eqs.\,\eqref{eq:chi_LF_noise}-\eqref{eq:single_theta_LF_noise}, and introducing the shorthand notations
\begin{subequations}
\begin{align*}
    d_{\textrm{c}}                 &\equiv \frac{\Gamma\qty(\frac{s + 2 \alpha_{\textrm{p}} + 1}{2})}{2 \pi \Gamma\qty(\frac{s + 1}{2})} \tilde{F}^{2}_{\textrm{G},0}, \\
    d_{\textrm{q}}^{\textrm{\,sh}} &\equiv \frac{\Gamma\qty(\frac{s + 2 \alpha_{\textrm{p}} + 2}{2})}{2 \pi \Gamma\qty(\frac{s + 1}{2})} \tilde{F}^{2}_{\textrm{G},0},\quad
    d_{\textrm{q}}^{\textrm{\,as}} \equiv \frac{\Gamma\qty(\frac{s + 2 \alpha_{\textrm{p}}    }{2})}{  \pi \Gamma\qty(\frac{s + 1}{2})} \tilde{F}^{2}_{\textrm{G},0}, 
\end{align*}
\end{subequations}
we obtain
\begin{align}
\chi_{\textrm{c}} &\simeq 
                      d_{\textrm{c}}  \qty(\LFstrength \taug^2) {(\LFwidth \taug)}^{2 \alpha_{\textrm{p}}} 
\label{eq:OOM_formula_chi_LF_noise}.
\end{align}
In a completely analogous manner, we find 
\begin{equation}
\label{eq:OOM_formulas_theta_LF_noise}
    \theta_{\textrm{q}}^{(M)} \simeq \begin{cases}
                                      \!  M (M + 1)             d_{\textrm{q}}^{\,\textrm{sh}} \qty(\LFstrength \taug^2) {(\LFwidth \taug)}^{2 \alpha_{\textrm{p}} + 1}
\!\! , \,M \LFwidth \taug \ll 1,\\
                                      \! \hphantom{M (M + 1)}  d_{\textrm{q}}^{\,\textrm{as}} \qty(\LFstrength \taug^2) {(\LFwidth \taug)}^{2 \alpha_{\textrm{p}} - 1}
\!\! , \, M \LFwidth \taug \gg 1.
                                     \end{cases}
\end{equation}
We can thus conclude that the effective filtering order for the quantum phase $\theta_{\textrm{q}}^{(M)}$ is $2 \alpha_{\textrm{p}} + 1$ for $M \LFwidth \taug \ll 1$ and $2 \alpha_{\textrm{p}} - 1$ for $M \LFwidth \taug \gg 1$. 
The respective corrections to these LF-approximations of $\chi_{\textrm{c}}$ and $\theta_{\textrm{q}}^{(M)}$ when including the next-order Taylor term in Eq.~\eqref{eq:LF_expansion_classical_filter} all come with an extra factor ${(\LFwidth \taug)}^{2}$, so the leading-order expansion in Eq.~\eqref{eq:LF_expansion_classical_filter} was indeed justified in the SGL. 
The gate infidelity as predicted by Eqs.\,\eqref{eq:OOM_formula_chi_LF_noise}-\eqref{eq:OOM_formulas_theta_LF_noise} can be seen in the hollow markers in Fig.\,\ref{fig:infidelities_spectra_single_reset_LF_noise}(c), showing excellent agreement in their respective limiting regime.

From Eq.\,\eqref{eq:OOM_formulas_theta_LF_noise}, we see that the quantum phase $\theta_{\textrm{q}}^{(M)}$ grows two OOMs in ${(\LFwidth \taug)}^{-1}$ for $M$ increasing from $1$ to $\infty$, and 
\[\theta_{\textrm{q}}^{(1)} \ll \chi_{\textrm{c}} \ll \theta_{\textrm{q}}^{(\infty)},\]
where each is separated by one OOM in ${(\LFwidth \taug)}^{-1}$.
This allows us to draw the following conclusions about the control-history dependence of the gate:

{\bf i)} The gate fidelity (and so the EPG) can appear constant for all practical purposes for a low number of repetitions, whereas the quantum effect is only felt after a relatively large number of repetitions, depending on how deep we are in the SGL. 

{\bf ii)} If one attempts to suppress the quantum effect with more powerful control resources, e.g., by removing the timing-constraint and approaching the Zeno limit $\taug \rightarrow 0$, all three quantities $\chi_{\textrm{c}}$, $\theta_{\textrm{q}}^{(1)}$ and $\theta_{\textrm{q}}^{(\infty)}$ decay to $0$ following a power law.
However, the asymptotic quantum phase $\theta_{\textrm{q}}^{(\infty)}$ does so slower than the classical decay factor $\chi_{\textrm{c}}$ does.
Hence, it is not \emph{a priori} given that applying maximal control resources is sufficient to obtain a constant EPG.

{\bf iii)} Related to the above, given a fixed achievable OOM for $\chi_{\textrm{c}}$, a narrower LF-noise peak creates a much larger asymptotic quantum effect than a broader one.
This might seem counter-intuitive at first, since LF-noise with a long correlation time is more amenable to be well-suppressed by DD. On the other hand, for same $\chi_{\textrm{c}}$ a longer correlation time also signifies, from the point of view of the quantum bath, the possibility of a stronger response.
This is explicitly illustrated by the middle two curves (cyan circles and magenta triangles) of Fig.\,\ref{fig:infidelities_spectra_single_reset_LF_noise}.
The fact that these two scenarios are not equivalent, even for equal values of $\mathcal{F}_{\textrm{G}}(0) = e^{-\chi_{\textrm{c}}}$, is an important feature of the control-history dependence of the gate fidelity.

Using the OOM in Eqs.\,\eqref{eq:OOM_formula_chi_LF_noise} and Eq.~\eqref{eq:OOM_formulas_theta_LF_noise}, and dropping the constants, we conclude that a large RQE, as defined in Eq.\,\eqref{eq:approximate_RQE}, is engendered when
\[\LFstrength \taug^2 {(\LFwidth \taug)}^{2(\alpha_{\textrm{p}} - 1)} \gg 1.\]
While the ratio $\chi_{\textrm{c}} / \theta_{\textrm{q}}^{(\infty)} \sim \LFwidth \taug$ depends only on how deep we are in the SGL, we see that the RQE explicitly depends also on $\alpha_{\textrm{p}}$ and $\LFstrength$.
For a Hahn-echo scheme, where $\alpha_{\textrm{p}} = 1$, one requires that $\LFstrength \taug^2 \gg 1$, which implies a low (gate) fidelity for a round of free evolution of length $\taug$ applied at $\ts = 0$, as then $\chi_{\textrm{c}} \simeq \LFstrength \taug^2$.
For a Carr-Purcell-Meiboom-Gill (CPMG) scheme, where $\alpha_{\textrm{p}} = 2$, one in turn requires that $\LFstrength \taug^2 {(\LFwidth \taug)}^{2} \gg 1$, which implies  a low fidelity for a gate protected by Hahn-echo DD with the same cycle time, applied at time $\ts = 0$.
Thus, for a fixed gate time $\taug$ and effecting the final DD-protected identity gate with increasingly higher FO $\alpha_{\textrm{p}}$, a DD scheme of correspondingly high FO $\alpha_{\textrm{p}}-1$ would have to perform poorly when applied at time $t_s = 0$ in order for a large RQE to emerge -- also meaning that the noise strength would have to be increasingly large.
Note, however, that there is a limit to the filtering order, since the interpulse time cannot be shorter than $\tau_{0}$.

In summary, the results of this section point to a first scenario where caution is warranted regarding gate-error virtualization.
Namely, while comparatively more amenable to DD suppression, LF noise with strong temporal correlations can engender a large (negative) control-history dependent quantum effect on the gate fidelity, provided it is sufficiently strong.
In the next section, we see a dual scenario, where relatively weak HF-noise can have detrimental effects, if a control-induced resonance condition is satisfied.

\subsubsection{Control-induced resonance effects for high-frequency noise under periodic protocols}
\label{ssec:integer_resonances}

\begin{figure}[t!]
\includegraphics[width=0.95\linewidth]{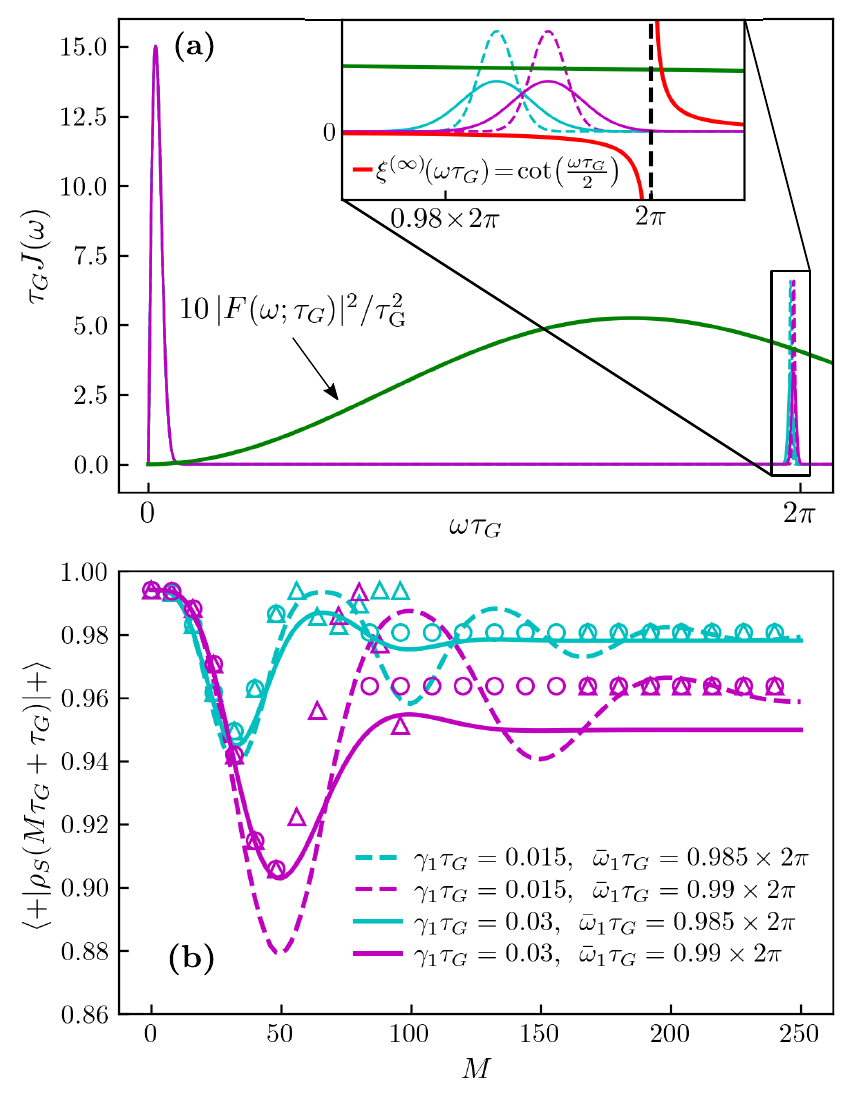}
\vspace*{-3mm}
\caption{Fidelity of an identity gate protected by Hahn-echo DD, following the same control history of Fig.\,\ref{fig:infidelities_spectra_single_reset_LF_noise}, in the presence of both LF- as well as HF-noise of fixed strength. The LF-parameters read $s = 1$, $\LFstrength \taug^2 = 3.5$ and $\LFwidth \taug = 0.1$, while $\HFstrength \taug^2 = 0.175$.
{\bf (a)} FF of the protected gate, and spectral density for two different values of $\HFstrength, \HFwidth$.
{\bf (b)} Gate fidelity as a function of $M$. Given the large number of discrete points, continuous lines are plotted for clarity.
}
\label{fig:infidelities_spectra_single_reset_HF_noise}
\end{figure}

Illustrative gate-fidelity results for the same control scenario where both the repeated control sequence and the final identity gate are based on Hahn-echo DD is given in Fig.\,\ref{fig:infidelities_spectra_single_reset_HF_noise}(b), for four different spectral densities that, in addition to LF-noise, include a narrow HF-peak, depicted in Fig.\,\ref{fig:infidelities_spectra_single_reset_HF_noise}(a).
The case where $\bar{\omega}_{1} \simeq \omega_{\textrm{G}} = 2 \pi / \taug$ was chosen, which highlights a possible highly detrimental history-dependence of the gate fidelity.
The gate fidelity is seen to drop almost an OOM for both transient and asymptotic values of $M$, compared to the case $M=0$ (in which case $\mathcal{F}_{\textrm{G}} (0) \simeq 0.995$).
This effect is more dramatic for the HF-peak with the center frequency closer to $2 \pi / \taug$.
Further to that, the gate fidelity also undergoes multiple oscillations in the transient regime, of a periodicity which varies with the detuning of $\bar{\omega}_{1}$ away from $\omega_{\textrm{G}}$.
{Notably}, the asymptotic regime is reached only after a much larger number of repetitions than for LF-noise only, adding to the apparent unpredictability of the gate fidelity from a practical point of view in an experiment.

To further understand these observations, we can in first instance focus on the HF-noise only: even though LF-noise is assumed intrinsically stronger ($\LFstrength \gg \HFstrength$), it is efficiently suppressed by DD, yielding the typical OOMs detailed in the previous section.
Intuitively speaking, the quantum phase $\theta_{\textrm{q}}$ caused by a HF-noise peak supported mainly around the integer resonance frequencies (IRFs) of the periodic control, 
\[ \omega_{\textrm{G},\ell} \equiv \frac{2 \ell \pi}{\taug}, \quad \ell \in \mathbb{N} \setminus \{0\}, \]
undergoes much the same increasing transition in OOM for $M$ going from $1$ to $\infty$, which can be explained by an analysis of effective filtering orders similar to before.
Indeed, firstly, the analogous SGL $\HFwidth \taug \ll 1$ is satisfied, since $\HFwidth \ll \LFwidth$.
Secondly, the kernel $\xi^{(M)}(\omega \taug)$ has a periodicity of $\omega_{\text{G}}$ for all $M$ and, analogously to the LF-noise case we discussed before, it transitions from adding a filtering order locally around $\omega_{\textrm{G},\ell}$ for $M \HFwidth \taug \ll 1$, to subtracting one for $M \HFwidth \taug \gg 1$.

Yet, there are two main differences w.r.t.\ the result established for LF-noise in the SGL.
First, the (repeated) DD-sequence is assumed to have zero filtering order locally around the HF-peak \footnote{Besides the fact that DD is designed to filter out LF-noise specifically, there is a more fundamental reason to this fact, namely, a DD scheme comprising repetitions of a base sequence of cycle time $\taug$ cannot be expected to cancel at all IRFs.
We have observed this for CDD$n$-sequences of order $n \geq 1$, and conjecture this to be the case for all \emph{digital} DD sequences.}.
Consequently, we can assume ${\abs{F_{\textrm{G}}(\omega ; \taug)}}^2$ to take the constant value
\begin{align*}
    {\abs{F_{\textrm{G}}(\omega ; \taug)}}^2 \simeq \taug^2 {|\tilde{F}_{\textrm{G}}(\bar{\omega}_{1} \taug)|}^2 
                                             \simeq \taug^2 {|\tilde{F}_{\textrm{G}}(2 \ell \pi)|}^2,
\end{align*}
since $\HFdetuning \approx 2 \ell \pi / \taug$. 
A second major difference is that, while the quantum noise spectrum $S^{-}(\omega)$ is anti-symmetric around $\omega = 0$, maximizing the overlap with the anti-symmetric kernel $\xi^{(M)}(\omega \taug)$, $S^{-}(\omega)$ has a fixed (positive) sign for $\omega > 0$.
As a consequence, the precise position of the center frequency $\bar{\omega}_1$ w.r.t.\ the IRF can have a large effect on the quantum phase. 

Quantitatively, upon defining the detuning w.r.t.\ the IRF as
\begin{equation*}
    \HFdetuning \equiv \bar{\omega}_{1} - \omega_{\textrm{G},\ell}, \quad 
    \delta \bar{\omega}_{1} \ll \omega_{\textrm{G}}, 
\end{equation*}
and the prefactor 
$$d_{\textrm{HF},\ell} \equiv \frac{1}{\pi} {{|\tilde{F}_{\textrm{G}}(2 \ell \pi)|}^2},$$
we obtain
\begin{equation}
\label{eq:OOM_chi_HF_single_reset}
    \chi_{\textrm{c,HF}} \simeq \frac{d_{\textrm{HF},\ell}}{2} \HFstrength \taug^2,
\end{equation}
and, uniformly in $M$,
\begin{align}
\label{eq:OOM_theta_HF_single_reset}
    \theta_{\textrm{q,HF}} \simeq \frac{ d_{\textrm{HF},\ell} }{2 \sqrt{\pi}} 
      \HFstrength \taug^2  \int_{- \infty}^{\infty}  \!  \xi^{(M)}_{\textrm{HF}}
    \big(\omega' \taug \big)
        e^{- {{(\omega' - \HFdetuning)}^2} \! / {\HFwidth^2}} \, \frac{\dd \omega'}{\HFwidth},
\end{align}
where the relevant repetition kernel has the form
\begin{equation*}
{\xi^{(M)}_{\textrm{HF}} (\omega' \taug) \equiv 2 \,\frac{1 - \cos(M \omega' \taug)}{\omega' \taug} + \sin(M \omega' \taug).}
\end{equation*}
In the limit $\HFdetuning \ll \HFwidth$, the integral in Eq.~\eqref{eq:OOM_theta_HF_single_reset} vanishes, because of the  anti-symmetry of the integrand.
The largest quantum effect can be found in the opposite limit, when $\HFwidth \ll \HFdetuning$.
Again, when $M \HFwidth \taug \ll 1$, we can expand $\xi^{(M)}_{\textrm{HF}} \big(\omega' \taug)$ up to leading order around $\omega' = \HFdetuning$, which for $\HFdetuning \gg \HFwidth$ simply corresponds to putting $\omega' = \HFdetuning$.
For $M \HFwidth \taug \ll 1$, one can again neglect the oscillating sine and cosine functions in $\xi^{(M)}_{{\textrm{HF}}}$.
Altogether, up to leading order, we then obtain
\begin{equation}
\label{eq:OOM_formulas_theta_HF_noise}
    \theta_{\textrm{q, HF}}^{(M)} \! \simeq \! \begin{cases}
                                          d_{\textrm{HF},\ell}\frac{\HFstrength \taug}{\HFdetuning} \big((1 \! - \cos(M \HFdetuning \taug)\big),
                                      & M \HFwidth \taug \ll 1,\\
                                          d_{\textrm{HF},\ell}\, \frac{\HFstrength \taug}{\HFdetuning}, 
                                        & M \HFwidth \taug \gg 1,
                                     \end{cases}                                                                        
\end{equation}
where we repeat the assumed double timescale separation
\[\HFwidth \taug \ll \HFdetuning \taug \ll 1.\]
Under these conditions, we now have a simple approximate formula for the transient regime, namely, an oscillation of period $1 / \HFdetuning \taug$ in $M$, with corresponding amplitude of the order of $\HFstrength \taug / \HFdetuning$.
The approximation of the gate infidelity from Eqs.\,\eqref{eq:OOM_chi_HF_single_reset} and Eq.~\eqref{eq:OOM_formulas_theta_HF_noise} are shown in the hollow markers in Fig.\,\ref{fig:infidelities_spectra_single_reset_HF_noise}(b).
Note that, besides the expected range of validity in $M$, these formulas do not explicitly depend on $\HFwidth$, explaining why, for the relevant values of $M$, the triangles and circles of equal color coincide.
Also note that, for the parameters of the solid pink line, $\HFwidth \centernot{\ll} \HFdetuning$, so the pink circles do not approximate the exact infidelities up to good accuracy.

Analogously to the case of LF-noise, we have that
\[\theta_{\textrm{q, HF}}^{(1)} \ll \chi_{\textrm{c,HF}} \ll \theta_{\textrm{q,HF}}^{(\infty)}.\]
Now, however, the OOM separations are ${(\HFdetuning  \taug)}^{-1}$ and hence are determined by the detuning $\HFdetuning$ and not the width $\HFwidth$ of the noise peak.
This is a general consequence of the odd parity of the repetition kernel $\xi^{(M)}$ that enters the FF corresponding to the quantum phase.

We conclude that when a narrow HF-noise peak is closely detuned with the frequency of repetition of control, the gate fidelity can oscillate wildly as a function of $M$, and this happens regardless of the specific sequence being repeated.
Contrary to the case of strong LF-noise with a long correlation time, this resonant effect can be large for moderate or even small HF-noise strengths.
Notably, one could even have that $\chi_{\textrm{c}}$ is LF-dominated, while $\theta_{\textrm{q}}^{(\infty)}$ is HF-dominated, since the timescale separations $\HFdetuning \taug$ and $\HFwidth \taug$ are completely independent from the SGL parameter $\LFwidth \taug$.
On the other hand, the resonance condition of the HF-noise is more constraining for this phenomenon to occur, although it cannot be excluded \emph{a priori} without careful advance characterization of the noise -- for instance via QNS, in both the classical and quantum regime \cite{Qasim,Wang}.

\subsubsection{Control-induced resonance effects for high-frequency noise under non-periodic protocols}
\label{ssec:non_periodic_resonances}

\begin{figure}[t!]
\includegraphics[width=0.95\linewidth]{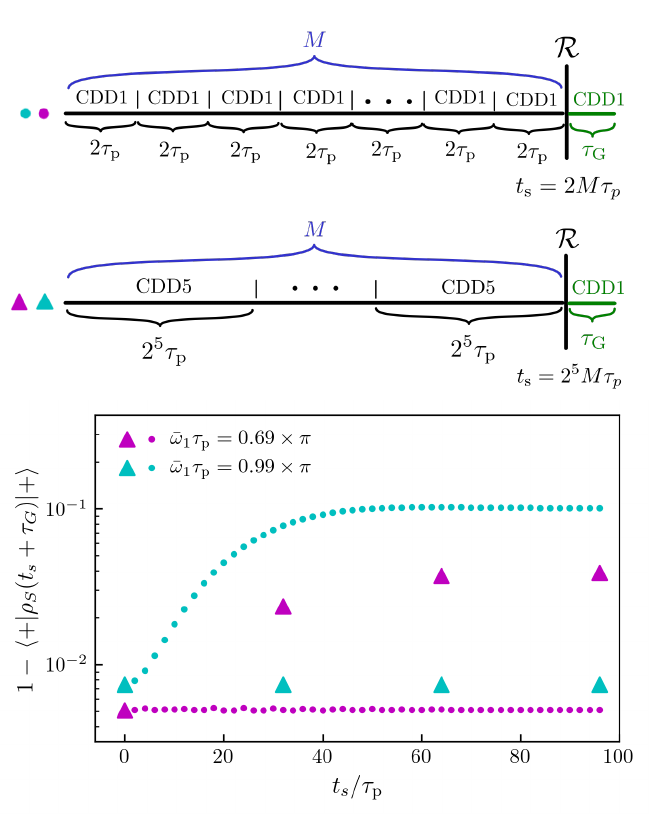}
\vspace*{-4mm}
\caption{
Gate infidelity of a Hahn-echo sequence with cycle time $\tau_{\textrm{G}} = 2 \tau_{0}$, after a variable number of repetitions of Hahn-echo DD with cycle time $2 \tau_{\textrm{p}}$ (circles, top diagram), and after a variable number of repetitions of CDD5 with cycle time $2^5 \tau_{\textrm{p}}$ (triangles, bottom diagram) respectively, with ${\tau_{\textrm{p}} / \tau_{0} =  (1 + \sqrt{5}) / 2}$.
The LF-noise parameters read $s = 1$, $\LFstrength \tau_{\textrm{p}}^2 = 2.5$, $\LFwidth \tau_{\textrm{p}} = 0.05$, while the strength and peak width of HF noise are $\HFstrength \tau_{\textrm{p}}^2 = 0.125$ and $\HFwidth \tau_{\textrm{p}} = 0.05$.
A resonance is found to occur at a frequency $\bar{\omega}_{1} \tau_{\textrm{p}} \approx 0.69 \pi$, for CDD5.
The common $x$-axis is the waiting time before the reset, $t_{s} = 2 M_{\textrm{CDD1}} \, \tau_{\textrm{p}} = 2^5 M_{\textrm{CDD5}} \, \tau_{\textrm{p}}$.}
\label{fig:CDD_single_reset}
\end{figure}

The effects of the previous subsection were explained based on the properties of the repetition kernel $\xi^{(M)}$, which came about when assuming that a same DD-sequence was repeated $M$ times before the reset operation, after which the very same DD (identity-gate-) sequence with the same cycle time was applied once more.
It is natural to ask how much of the observed behavior and the resulting large quantum effect depends specifically on these periodicity assumptions. 

As a first step to break periodicity, in Fig.\,\ref{fig:CDD_single_reset} (circles), we consider applying a Hahn-echo sequence with minimal gate time $\taug = 2 \mst$ as a gate sequence, after having repeated $M$ Hahn echo sequences with a different inter-pulse time $\tau_{\textrm{p}}$.
To show that no particular resonance relationship between the inter-pulse times before and after the reset is required in order for resonance effects to emerge, we take their ratio to be the golden ratio: $ \tau_{\textrm{p}} / \mst = (1 + \sqrt{5}) / 2$.
We can see the gate error increases by over one OOM for an asymptotically-large number of repetitions.
No transient oscillation is present, however, since here $\HFdetuning / \HFwidth = \pi / 5 \centernot{\gg} 1$.
While the details go beyond the scope of this paper, we mention that the fact that a large quantum effect is possible can be traced back to the fact that the relevant repetition kernel defining $\theta_{\textrm{q}}^{(M)}$ now does not adhere to the same anti-symmetry as $\xi^{(M)}$.
{Therefore}, it is not imperative that the HF-noise peak be narrow, and supported mainly on one side of the IRF, unlike in the fully periodic setting of Sec.~\ref{ssec:transition_filtering_order}-\ref{ssec:integer_resonances}.

As a second step, we consider a control history consisting of only a handful repetitions of a DD-sequence different from the (last) gate sequence.
Concretely, we choose a CDD5 sequence meant to idle the qubit.
CDD5 suppresses LF-noise extremely well, given the fast control limit \(\LFwidth \tau_{\textrm{p}} \ll 1,\) and the high FO $\alpha_{\textrm{p}} = 5$.
An account of the resulting gate infidelity is shown in Fig.\,\ref{fig:CDD_single_reset} (triangles).
We can see that the gate infidelity increases an OOM from the first repetition of the CDD5 sequence, and hence here the resonance effect is \emph{not} due to any periodic repetition of control.
Rather, the fact that the same resonance frequency, which we found numerically, causes a large quantum effect for an arbitrary (nonzero) number of repetitions is due to the digital nature of CDD$n$, namely, the fact that all pulse times are an integer multiple of the minimal inter-pulse time $\tau_{\textrm{p}}$.
The cycle time of CDD$n$ being increasingly long allows, in turn, for considerable ``build-up'' of the resonance over a single cycle.
Thus, while digital sequences offer important advantages, for instance in terms of hardware compatibility and minimum sequencing complexity \cite{Hayes,Haoyu,khodjasteh2013designing}, digital timing may be more prone to adverse effects from nonclassical noise.

\section{Gate performance after multiple error-correction cycles}
\label{sec:multiple_resets}

Analogously to Sec.\,\ref{sec:single_reset}, we first outline exact formulas for the gate fidelity, as well as exact asymptotic results for periodic control, after which a parameter study is performed for the concrete setting of spin-boson dephasing.

\subsection{Exact results on asymptotic gate-performance saturation}
\label{ssec:exact_asymptotics_multiple_resets}

\subsubsection{General fidelity expressions}

The gate fidelity for the general case of $K$ reset operations takes the following form:
\begin{equation}
\label{eq:general_fidelity_multiple_resets}
    \mathcal{F}_{G} (\ts) = \frac{1}{2}\Big(1 + e^{- \chi_{c}} \!\prod_{k = 1}^{K} \cos(\theta_{\textrm{q},[k]})\Big), 
\end{equation}
where the classical decay factor $\chi_{c}$ is still defined as in Eq.\,\eqref{eq:timedom_def_chi}, but now the quantum phase contribution reads
\begin{equation}
\label{eq:timedom_def_thetas}
    \theta_{\textrm{q},[k]} \equiv - 2 i \int_{0}^{\taug} \! \! \! \! \dd \tau \int_{t_{k-1}}^{t_{k}} \! \! \! \! \dd t
                                    \, y_{\textrm{G}}(\tau) \, y_{\textrm{hist}}(t) \, C^-(\tau - t + \lrt).
\end{equation}
A full derivation of these formulas is provided in Appendix~\ref{app:sec:deriv_x_exp_vals}.
Expressing $\theta_{\textrm{q},[k]}$ in the frequency domain, we obtain
\begin{equation}
\label{eq:freqdom_def_thetas}
    \theta_{\textrm{q},[k]} \equiv \frac{1}{\pi} \! \! \int\limits_{- \infty}^{\infty} \! \! \mathrm{Im}
                 \qty[e^{i \omega t_{K}} F_{\textrm{G}}(\omega ; \taug) F^*_{\textrm{hist}}(\omega ; t_{k-1}, t_{k})]
                 S^{-}\!(\omega) \, \dd \omega,
\end{equation}
where the FF of the control history is explicitly indexed by start- and end-times,
\[F_{\textrm{hist}}(\omega ; t_{k-1}, t_{k}) \equiv \int_{t_{k-1}}^{t_{k}} \!\!e^{i \omega s} y_{\textrm{hist}}(s) \, \dd s.\]
Analogously to the single-reset case ($K=1$), we see that the $\theta_{\textrm{q},[k]}$ in Eq.\,\eqref{eq:timedom_def_thetas} depend both on the gating control applied after the last reset, as well as the control history, and vanish when the noise is purely classical.
Interestingly, for $K>1$, $\theta_{\textrm{q},[k]}$ depends only on the control applied during the time interval in-between the $(k-1)$-th and $k$-th reset operation. 
In a way, the system-resets ``compartmentalize'' the quantum memory effects that are due to the different inter-reset evolutions.
We stress that these effects would vanish even for a quantum bath, if its state could be refreshed to the initial $\bar{\rho}_B$ after every qubit reset, as assumed in \cite{NgCombining}.
This can be seen more explicitly in Sec.\,\ref{ssec:decomposition_gaussian_components}, where an alternative expression for $\theta_{\textrm{q}, [k]}$ will be derived, directly in terms of the updated bath statistics.

We refrain from defining a RQE analogous to Eq.\,\eqref{eq:def_RQE}, since it is not \emph{a priori} clear if multiple dominant phases $\theta_{\textrm{q}, [k]}$ have a significant contribution, and if so, which those would be.
We will rather address these questions on a case-by-case basis in the rest of this section, by considering control-history scenarios similar to those we had in the single-reset case.

\subsubsection{Asymptotics for periodic control}

Consider a fixed DD sequence that is repeated $M$ times before every reset, with a total of $K$ reset operations.
Thus, the total number of repetitions before the last reset is $N = K M$, after which the same gate sequence is repeated once more to effect the target identity gate.
Formally, we have that $\ts = t_{K} = N \taug =  K M \taug$, and
\[y_{\textrm{hist}}(t) = y_{\textrm{G}}(t \; \mathrm{mod} \, \taug) \qq*{,} t \in [0, N \taug).\]
Introducing the index \(\tilde{k} \equiv K - k + 1,\) which ``counts back'' from the last ($K$-th) reset, we define
\[\theta_{\textrm{q},\tilde{k}}^{(M)} \equiv \theta_{\textrm{q}, [k = K - \tilde{k} + 1]}, \quad \tilde{k} = 1, \ldots, K,\]
and, from the definition of the FF, we also note that
\begin{align*}
      & e^ {i K M \omega \taug} F_{\textrm{hist}}^{*}(\omega ; (k-1) M \taug, k M \taug) = \\
     & \qquad e^{i \tilde{k}\, M \omega \taug} e^{-i (M - 1) \frac{\omega \taug}{2}}
\frac{\sin(M \frac{\omega \taug}{2})}{\sin(\frac{\omega \taug}{2})} F_{\textrm{G}}^{*}(\omega ; \taug). 
\end{align*}
We can then write the quantum phase as
\begin{equation}\label{eq:single_theta_periodic_repetition}
    \theta_{\textrm{q}, \tilde{k}}^{(M)} = \frac{1}{2 \pi} \int_{- \infty}^{\infty} \!\eta_{\tilde{k}}^{(M)}(\omega \taug)
                                          {\abs{F_{\textrm{G}}(\omega ; \taug)}}^2 S^{-}(\omega) \dd \omega,
\end{equation}
where the relevant repetition kernel now reads
\begin{align}
     \eta_{\tilde{k}}^{(M)}(\omega \taug)
             &\equiv \cos((\tilde{k} - 1) M \omega \taug) \; \xi^{(M)}(\omega \taug) \label{eq:rep_kernel_multiple_resets} \\
             &+ \, \sin((\tilde{k} - 1) M \omega \taug) \; \qty( D^{(M)}(\omega \taug) - 1),\nonumber
\end{align}
with $D^{(M)}$ being the Dirichlet kernel,
\begin{equation*}
    D^{(M)}(\omega \taug) = \frac{\sin((2 M + 1)\frac{\omega \taug}{2})}{\sin(\frac{\omega \taug}{2})}.
\end{equation*}

\begin{figure*}[t!]
\includegraphics[width=1.0\linewidth]{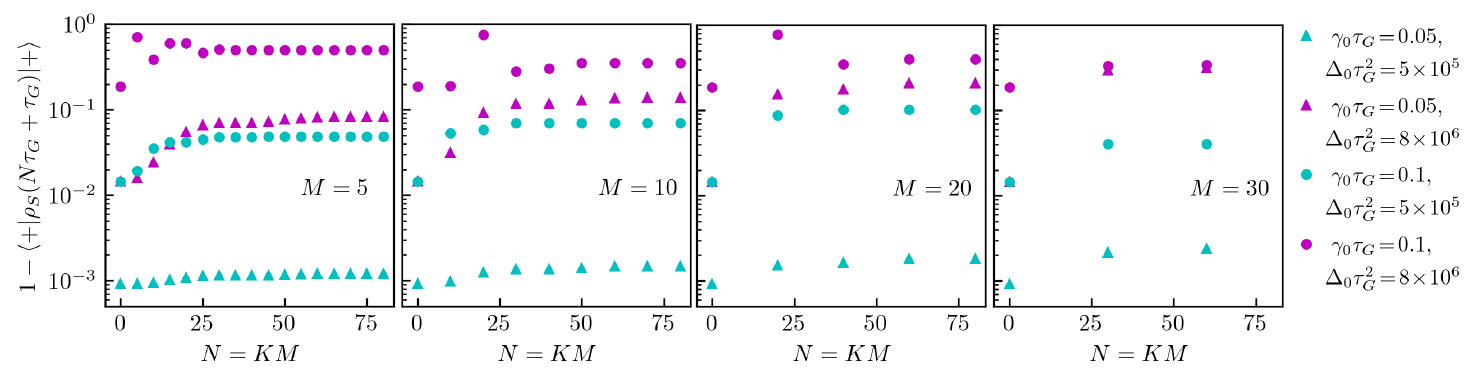}
\vspace*{-5mm}
\caption{Gate infidelity of a CPMG-sequence with cycle time $\tau_{\textrm{G}}$, after a variable number $K$ of resets, with $M$ repetitions of the CPMG sequence before every reset operation. Only LF-noise is considered, with Ohmicity parameter $s = 2$. Note that for the larger value $M = 30$ (right-most panel), we have $M \LFwidth \taug \centernot{\ll} 1,$ hence these parameters are not in a fast reset limit.}
\label{fig:LF_multiple_resets}
\end{figure*}

We can then establish the following analog to Theorem 1: 

\smallskip

{\bf Theorem 2 [Multiple-resets asymptotic quantum phase].} 
Assume, as before, that the control FF for the DD-protected gate $G$ has a LF-expansion of the form 
${\abs{F_{\textrm{G}}(\omega ; \taug)}}^2 \sim \omega^{2 \alpha_{\textrm{p}}}$. Then, if  
\begin{equation}
s_{\textrm{p}} = s + 2 \alpha_{\textrm{p}} > - \frac{1}{2} , 
\label{eq:plateau_condition_multiple_reset}
\end{equation}
the quantum phase contribution converges to a finite value,
\begin{equation*}
\lim_{K \rightarrow \infty} \prod_{\tilde{k} = 1}^{K} \cos(\theta_{\textrm{q}, \tilde{k}}^{(M)}) = \prod_{\tilde{k} = 1}^{\infty} \cos(\theta_{\textrm{q}, \tilde{k}}^{({M})}).
\end{equation*}

{\bf Proof.} We wish to show that, under the above assumptions, $\theta_{\textrm{q}, \tilde{k}}^{(M)} \stackrel{\tilde{k} \rightarrow \infty}{\longrightarrow} 0$ fast enough such that $\prod_{\tilde{k} = 1}^{\infty} \cos(\theta_{\textrm{q}, \tilde{k}}^{(M)})$ is finite.
To perform a similar analysis as in Sec.\,\ref{ssec:exact_asymptotics_single_reset}, we utilize the alternative representation
\begin{equation*}
    \eta_{\tilde{k}}^{(M)}(\omega \taug) = 2 \sum_{m = 1}^{M} \sin((m + (\tilde{k} - 1) M ) \omega \taug).
\end{equation*}
{This yields}
\begin{equation}
\label{eq:summation_theta_multiple_resets}
    \theta_{\textrm{q}, \tilde{k}}^{(M)} = \sum_{m = 1}^{M} \theta_{\textrm{q}, m + (\tilde{k} - 1) M},
\end{equation}
where each phase contribution $\theta_{\textrm{q}, n}$ has the form given in Eq.\,\eqref{eq:theta_delta_m_fourier_tf}.
Under the stated assumptions on the LF-behavior of noise and control we can invoke the asymptotic expression for $\theta_{\textrm{q}, n}$ given in Eq.\,\eqref{eq:asymptotics_theta_delta_m}.
Using the triangle inequality, we obtain from Eq.~\eqref{eq:summation_theta_multiple_resets},
\begin{align}
\label{eq:asymptotics_theta_multiple_reset}
   \abs{\theta_{\textrm{q}, \tilde{k}}^{(M)}} \sim \; & \Gamma(s_{\textrm{p}} + 1) \abs{\cos(\frac{\pi s_{\textrm{p}}}{2})}\nonumber\\
                            & \times \; \mathcal{O}\bigg(\sum_{m = 1}^{M} \frac{1}{{(m + (\tilde{k} - 1) M)}^{1 + s_{\textrm{p}}}} \bigg),
\end{align}
in the limit where $\tilde{k} \rightarrow \infty$. We can now quantify under what conditions and how fast the following product converges
\[\prod_{\tilde{k} = 1}^{K} \cos(\theta_{\textrm{q}, \tilde{k}}^{(M)})
\stackrel{K \rightarrow \infty}{\longrightarrow}
\prod_{\tilde{k} = 1}^{\infty} \cos(\theta_{\textrm{q}, \tilde{k}}^{(M)}),\]
by upper-bounding the relative difference as follows:
\begin{align*}
1 -   \frac{\prod\limits_{\tilde{k} = 1}^{\infty} \! \! \cos(\theta_{\textrm{q}, \tilde{k}}^{(M)})}{\prod\limits_{\tilde{k} = 1}^{K} \! \! \cos(\theta_{\textrm{q}, \tilde{k}}^{(M)})}
&   = 1 - \! \! \prod_{\tilde{k} = K + 1}^{\infty} \! \! \cos(\theta_{\textrm{q}, \tilde{k}}^{(M)}), \nonumber\\
&   = 1 - e^{  \sum_{\tilde{k} = K + 1}^{\infty}  \log \!\big[ \!\cos(\theta_{\textrm{q}, \tilde{k}}^{(M)}) \big] } 
\nonumber\\
&
\leq \sum_{\tilde{k} = K + 1}^{\infty} \! \! {\qty(\theta_{\textrm{q}, \tilde{k}}^{(M)})}^2,
\end{align*}
where we have assumed the $\theta_{\textrm{q}, \tilde{k}}^{(M)}$ to be small enough for a lowest-order Taylor expansion to hold. We can then insert the asymptotics Eq.~\eqref{eq:asymptotics_theta_multiple_reset} to obtain
\begin{align*}
    1 -   \frac{\prod\limits_{\tilde{k} = 1}^{\infty} \! \! \cos(\theta_{\textrm{q}, \tilde{k}}^{(M)})}{\prod\limits_{\tilde{k} = 1}^{K} \! \! \cos(\theta_{\textrm{q}, \tilde{k}}^{(M)})}
        & \leq \frac{\Gamma^2(s_{\textrm{p}} + 1) \cos^2\qty(\frac{\pi s_{\textrm{p}}}{2})}{M^{2 s_{\textrm{p}}}}\\
           & \times  \mathcal{O}\bigg(\zeta(2 + 2 s_{\textrm{p}}) - \sum_{\tilde{k} = 1}^{K - 1} \frac{1}{{\tilde{k}}^{2 + 2 s_{\textrm{p}}}} \bigg) \nonumber ,
\end{align*}
after some simple estimates.
Eq.\,\eqref{eq:plateau_condition_multiple_reset} provides a sufficient condition for convergence, thus the claim follows. \qed

\begin{figure*}[ht!]
\includegraphics[width=1.0\linewidth]{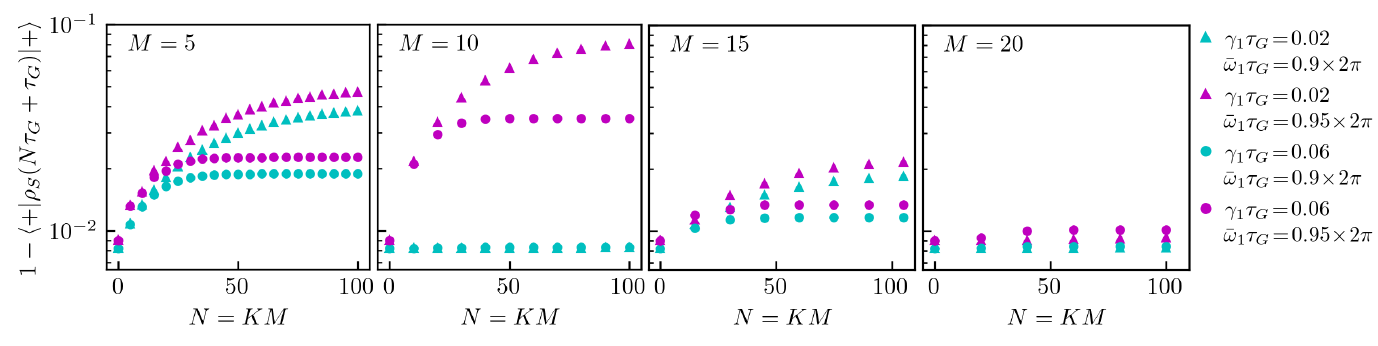}
\vspace*{-3mm}
\caption{Gate infidelity of a CPMG-sequence with cycle time $\tau_{\textrm{G}}$, after a variable number of resets, with $M$ repetitions of the CPMG sequence before every reset operation, in the presence of both LF- and HF-noise.
The LF-noise parameters read $s = 2$, $\LFstrength \taug^2 = 7$ and $\LFwidth \taug = 0.1$, while the HF-noise strength is $\HFstrength \taug^2 = 0.3$.
{For these noise parameters, the (in)fidelity is dominantly determined by the HF noise. For comparison, when only LF-noise is present as in Fig.\,\ref{fig:LF_multiple_resets} ($\LFstrength \taug^2 = 7$, $\Delta_{1} = 0$), the infidelity drops to approximately $2 \times 10^{-7}$.} }
\label{fig:HF_multiple_resets}
\end{figure*}

\smallskip

It is worth noting that the condition in Eq.\,\eqref{eq:plateau_condition_multiple_reset} is a weaker condition than the one in Eq.~\eqref{eq:plateau_condition_single_reset} for a single reset.
Even for the case of free evolution ($\alpha_{\textrm{p}} = 0$), for instance, a negative Ohmicity parameter would be allowable in the multiple-reset setting, as long as $s > - \frac{1}{2}$.
For fixed values of $s$ and $\alpha_{\textrm{p}}$, the convergence towards the asymptotic gate fidelity is faster for the multiple-reset case than it is for the single-reset case, in terms of the total number of repetitions 
$N$. Note that this \emph{a priori} has no implications on the value of the asymptotic gate fidelity, as we do not have explicit formulas for the case of multiple resets.
This question and other (transient) phenomena are studied for the spin-boson model, as a function of relevant parameters, in the remainder of this section.

\subsection{The range of possibilities}
\label{ssec:case_studies_multiple_resets}

In this section, we will work under the same assumptions outlined in Sec.\,\ref{ssec:case_studies_single_reset} [Eq.\,\eqref{parameters}]. 
However, we will not require a fast reset limit where the inter-reset time is constrained to be shorter than $1 / \LFwidth$, $M \taug \ll 1/ \LFwidth$. Rather, we shall generally allow that $M \LFwidth \taug \centernot{\ll} 1,$ in line with the fact that DES is relatively ``cheap'', whereas performing QEC operations is more costly. From this point of view, using relatively small values of $M$ would be considered non-optimal, in terms of leveraging maximal benefit from DD. Additionally, the general idea of combining DES with QEC is to perform error correction on a timescale longer than the correlation time of the LF noise, hence dealing with the approximately uncorrelated (white)
remainder of the noise, as outlined in the introduction.

\subsubsection{Asymptotic gate fidelity under low-frequency noise}
\label{ssec:break_even}

For the case of periodic repetition in the presence of LF-noise only, we can readily make two observations that are in line with the conclusions for the $K=1$ case:

{\bf i)} For a given fixed initial gate infidelity at $t_s=0$ a narrower LF-noise peak consistently performs worse in the asymptotic limit.
This can be seen in the cyan circles versus magenta triangles of Fig.\,\ref{fig:LF_multiple_resets}, starting at $N = KM= 30$.
The asymptotic gate fidelity is accordingly reached faster for the wider noise peak.
Notably, for an increasing number of repetitions $M$, the difference in the asymptotic gate infidelity becomes larger, as evidenced in the last plot, with $M = 30$.

{\bf ii)} While one can show that the ratio $\chi_{\textrm{c}} / \theta_{\textrm{q}, \tilde{k}}^{(M)}$ does not depend on $\Delta_0$, the history dependence of the gate infidelity is significantly weaker when both $\chi_{\textrm{c}}$ and $\theta_{\textrm{q}, \tilde{k}}^{(M)}$ become smaller, as can be seen for the cyan triangles in Fig.\,\ref{fig:LF_multiple_resets}.
The opposite effect can be seen for the magenta circles, where the infidelity rapidly reaches $1 / 2$, indicating a maximally-mixed qubit state at the final time $N \taug + \taug = t_s+\taug$.

In terms of the number of repetitions between resets, we can see that, when increasing $M$ from a small number (say, $M = 5$) to values that are more resource-efficient in terms of reset operations (e.g., $M = 10, 20$), the gate fidelity generally decreases.
This is not all too surprising, since the reset operation can in itself only help to preserve the system state.
This is in perfect analogy with a true QEC, where only if the rounds of error correction are performed frequently enough, the (logical) qubit state can be protected.
Nonetheless, when comparing $M = 20$ and $M = 30$ for the cyan circles, for instance, we see that the asymptotic gate fidelity increases for the larger value of $M$.
This shows an intricate dependence of the quantum effect on the inter-reset time, in a way that is \emph{a priori} unpredictable.

\subsubsection{Control-induced resonance effects for high-frequency noise}

In Fig.\,\ref{fig:HF_multiple_resets}, we show representative results for a case where a narrow HF-noise peak is present and, for simplicity, the LF-noise parameters are chosen so that LF-noise effects are negligible in first instance, given the DD-protection.
{This is to be contrasted to the previous scenario in Fig.\,\ref{fig:LF_multiple_resets}, where LF-noise alone causes for a substantial gate-infidelity.
In turn,} the HF-noise parameters are chosen in the same regime of Sec.\,\ref{ssec:integer_resonances}, characterized by the double timescale separation
\[\HFwidth \taug \ll \HFdetuning \taug \ll 1.\]
A first conclusion is that the case of a narrower peak (triangles) consistently leads to higher gate infidelities than the case of a broad peak (circles), no matter where exactly the noise-peak is centered.
This is to be contrasted to the single-reset case, where for large-enough detuning $\HFdetuning$ from the IRF's, the width $\HFwidth$ did not play a major role.
This may be explained by noticing that, for a small 
$M$, the relevant repetition kernel $\eta_{\tilde{k}}^{(M)}(\omega \taug)$ is not sharply peaked around the IRF's, making the precise location of the HF-noise peak comparatively less important.
Furthermore, as a function of $\tilde{k}$, the phases $\theta_{\textrm{q}, \tilde{k}}^{(M)}$  decay slower for a more narrow peak, as the individual $\theta_{\textrm{q}, n}$ (defined in Eq.~\eqref{eq:theta_delta_m_fourier_tf}) that enter $\theta_{q, \tilde{k}}^{(M)}$ (see Eq.~\eqref{eq:summation_theta_multiple_resets}) decay over a characteristic ``timescale'' of the order $n \sim 1 / (\HFwidth \taug)$ in practice, for HF-dominated noise.
A more narrow HF-noise peak will thus have a larger number of phases contributing significantly to the control-history dependence.

Turning our attention to the center frequency $\bar{\omega}_{1}$ of the peak, we find that the gate fidelity is (monotonically) decreasing in the fast reset limit where $M \HFdetuning \taug \ll 1$.
However, for the noise parameters of Fig.\,\ref{fig:HF_multiple_resets}, this would correspond only to $M = 1, 2$, which we consider to be resource-inefficient scenarios, and hence are not shown.
For increasing $M$, the gate fidelity exhibits a non-monotonic and highly-sensitive dependence upon $M$, both in the transient and asymptotic regime.
This behavior can be partially explained from the intuition we gained from the single-reset case.
Indeed, the noise parameters corresponding to the cyan plots are chosen to yield a local maximum of the quantum phase $\theta_{\textrm{q}}^{(M)}$ for $M = 5$ in the $K=1$ setting, and a local minimum for $M = 10$.
The two left-most panels of Fig.\,\ref{fig:HF_multiple_resets} clearly show that in fact, for $M = 10$, all the quantum phases $\theta_{\textrm{q}, \tilde{k}}^{(M = 10)}, \tilde{k} = 1, \ldots, K$ are correspondingly small (not just for $\tilde{k}= K = 1$), confirming that our intuition gained from the single-reset case carries over.
For the magenta curves we have an exactly analogous scenario, where now a local maximum for $\theta_{\textrm{q}}^{(M)}$ is chosen to occur at $M = 10$, while a local minimum happens at $M = 20$.

In summary, on top of the non-monotonic dependence on $M$ we identified for LF-noise, the presence of HF-noise can lead to an \emph{a priori} increasingly unpredictable control-history dependence of the gate fidelity.
In realistic scenarios where both LF- and HF-noise are simultaneously present, one can only expect the complexity of this multi-parameter problem to increase further, making the gate performance essentially unpredictable.

\subsubsection{High-order performant decoupling for idling}
\label{sub:idling}

\begin{figure}[ht!]
\includegraphics[width=1.0\linewidth]{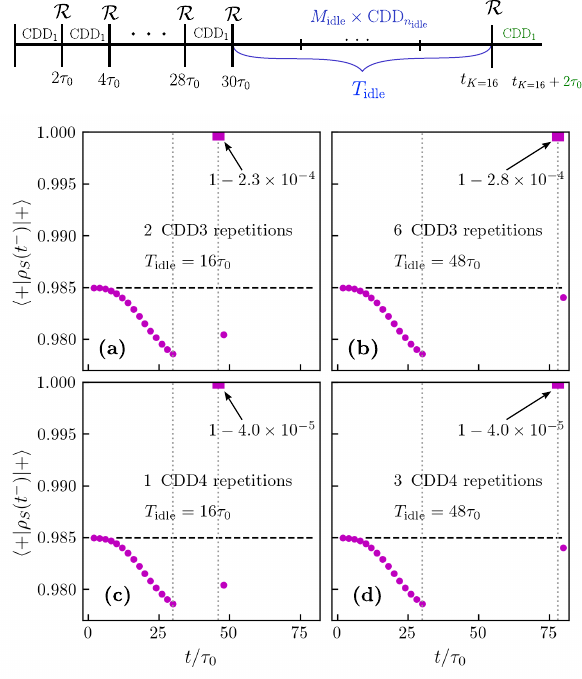}
\vspace*{-4mm}
\caption{Fidelity of the qubit at designated pre-reset times, {$t^-$, for} 
a family of protocols as depicted on top. A Hahn-echo (CDD{$_1$}) sequence (circles) with cycle time $2 \tau_{0}$ is repeated a fixed number of times, after which a higher-order CDD$_n$ idling sequence, with the same minimal switching time $\tau_{0}$, is repeated a variable number $M_{\textrm{idle}}$ of times (square), finally followed by one last application of the Hahn-echo sequence (last circle). 
A reset operation $\mathcal{R}$ is applied after every Hahn echo- and the (repeated) CDD$_n$-sequence. {The duration of the idling period is marked by dotted vertical lines}.
The fidelity of the final identity gate is seen to depend mostly on the total duration of the repeated idling sequence, provided that the idling sequence is {sufficiently} high-fidelity.
}
\label{fig:rethermalization}
\end{figure}

Throughout our analysis thus far, we have seen that the fidelity of a fixed DD-protected gate applied at $t_s$ is impacted negatively by the control history applied before $t_s$, in a way that generically only gets worse with increasing number of reset operations.
Furthermore, for a wide range of control histories under the established conditions, the gate error can increase by a whole order of magnitude.
One could argue that such a poor performance is due to the low-order nature of the DD schemes we considered before executing the DD-protected gate, which in our situation are key to allow for a fast application of subsequent gates.
Often during a computation, however, there may be time intervals  during which one does not require increasingly more performing operations on the system, in which case one can realistically apply higher-order DD schemes to achieve better protection.
Given the assumed constraint on a finite minimal switching time, however, this can only be accomplished at the cost of longer time durations -- which may make it challenging to combine them with the ``decouple, then compute" approach we consider in this work.

In Fig.\,\ref{fig:rethermalization}, we consider the effect of interjecting a period of such high-order DD sequence, idling the system, after several repetitions of a fixed low-order gate sequence.
Since the goal here is not to focus on possible resonance effects with HF-noise, here we consider only LF-noise. 
Furthermore, to maximally cut off the system-side error propagation, we perform a reset after every application of the low-order DD.
While the evolution of the fidelity of the repeated Hahn-echoes (circles) is qualitatively similar to that of Fig.\,\ref{fig:LF_multiple_resets}, a first important observation is that the fidelity of the idling sequence (squares), which precedes the last reset, is several OOMs higher.
By keeping the minimum switching time $\tau_0$ constant, we explore the dependence of the fidelity of the last DD sequence, effecting the target identity gate (right-most circle) on two parameters that characterize the idling: the order $n_{\textrm{idle}}$ of the CDD idling scheme, and its number of repetitions, $M_{\textrm{idle}}$, before the last reset.
The following observations may be made:

{\bf i)} For the higher value of $n_{\textrm{idle}}$ (bottom two panels, (c) and (d)), the fidelity of the idling scheme is always higher.
One could then expect that, since in such cases the qubit is less entangled with the bath prior to the reset and the implementation of the protected gate, the final gate fidelity would always be higher for larger $n_{\textrm{idle}}$. Instead, we do not see an explicit change with $n_{\textrm{idle}}$, since panels (c) and (a) (or, respectively, panels (d) and (b)) show very similar final gate fidelities.
We note that this approximate independence of $n_{\textrm{idle}}$ is true as long as the idling sequence is high-fidelity relative to that of the low-order DD.

{\bf ii)} The number of repetitions $M_{\textrm{idle}}$ of the idling sequence appears to play a key role, and the final gate fidelity improves drastically for the larger value of $M_{\textrm{idle}}$, as can be seen in the right two panels, (b) and (d).

{\bf iii)} As long as the idling fidelity remains high, we also see that the final gate fidelity is monotonically increasing with $M_{\textrm{idle}}$, and eventually \emph{approximately} restores the initial fidelity at time $t = \taug$, when the gate is applied at $t_s=0$.

Thus, the use of high-performant DD prior to execution of a DD protected gate effectively suppresses the control-history dependence that the quantum memory introduces in the gate fidelity and, with that, it allows for a notion of a constant EPG to be approximately recovered.
In the next section, we seek further physical insight into this behavior by directly examining the bath dynamics, and by characterizing the extent to which the emergence of an approximately constant control performance is related to an approximate re-equilibration of the quantum bath statistics towards their original values, after being perturbed by an intervening low-fidelity control history. 

\section{Physical interpretation from \\ quantum bath dynamics}
\label{sec:bath_pov}

In both the single- and multiple-reset scenarios of Secs.\,\ref{sec:single_reset} and \ref{sec:multiple_resets}, a key step has been to establish exact formulas that capture the control history of the intended DD-protected gate in one or more quantum phase contributions $\theta_{\textrm{q}}$, which would identically vanish if the noise were classical.
We now provide an alternative viewpoint in explaining the emergence and meaning of these nonclassical contributions, by explicitly considering the quantum state of the bath after the final reset operation. 
Indeed, as noted at the end of Sec.\,\ref{ssec:protocol}, since the reset operations $\mathcal{R}$ are considered to be perfectly disentangling, any relevant control history applied before the $k$-th reset must be encoded in the post-reset state $\rho_{\textrm{B}}(t_{k}^{+})$.
Explicitly, to capture the relevant noise properties of the bath during the DD-protected gate, we consider the bath multi-time correlation functions 
\begin{equation}
\label{eq:general_correlation_functions_updated_bath_state}
\expval{B(\lrt + \tau_{j}) \cdots B(\lrt + \tau_{1})}_{\rho_{\textrm{B}}(\lrt^{+})}, \quad j = 1, 2, 3 \ldots ,
\end{equation}
which together fully determine the qubit reduced dynamics after the $K$-th (last) reset.

First, we will work towards a decomposition of $\rho_{\textrm{B}}(\lrt^{+})$ into Gaussian components.
This will reveal a clear interpretation of the different Gaussian components as consecutive displacements of the bath state, conditional on the state of the qubit.
Additionally, in combination with the general strategy for updating quantum bath statistics of \cite{bath_update_paper}, this decomposition will allow us to obtain closed-form expressions for the correlation functions in Eq.\,\eqref{eq:general_correlation_functions_updated_bath_state}. 
Next, as alluded to above, we will re-interpret the results of Fig.\,\ref{fig:rethermalization} in the light of a re-equilibration process of the quantum bath statistics after a prolonged period of high-fidelity control.

\subsection{Exact bath-statistics update by decomposition \\into Gaussian components}
\label{ssec:decomposition_gaussian_components}

In order to develop the argument, consider first the case of a single reset operation, applied at time $t_{K = 1} = \ts$.
By using the fact that the joint system-bath unitary evolution is diagonal in the $\{\ket{z=\pm 1}\}$ basis and the reset is perfect, it is immediate to see that the post-reset bath state can be written as the (equal) convex combination
\begin{align}
    \rho_{\textrm{B}}(\ts^{+}) =  \rho_{\textrm{B}}(t_1^{+})
    = \frac{1}{2} \Big( \rho_{\textrm{B}, -1}(t_1) + \rho_{\textrm{B}, +1}(t_1)\Big),
\label{eq:sum_of_gaussians_single_reset}
\end{align}
where the two components are given by 
\begin{equation*}
\rho_{\textrm{B}, \mp 1}(t_1) 
     \equiv U_{\mp 1}(0, t_{1}) \,\bar{\rho}_{\textrm{B}}\,
U_{\mp 1}^\dagger(0,t_1),                                              
\end{equation*}
in terms of the bath unitaries 
\[U_{\mp 1 }(0, t_1) = \mathcal{T}_{+} \exp( \pm i 
\int_{0}^{t_1} \! \! y_{\textrm{hist}}(t) B(t) \dd t).\]
By linearity of the correlation functions in Eq.~\eqref{eq:general_correlation_functions_updated_bath_state} with respect to the bath state, we can focus on the correlation functions corresponding to the components $\rho_{\textrm{B}, \mp 1}(\ts)$ individually.

Because, by assumption, $\bar{\rho}_{\textrm{B}}$ is Gaussian, it follows that $\rho_{\textrm{B}, \mp 1}(t_1)$ are such that the noise (sub-)processes $\qty( B(t_1 + \tau) ; \rho_{\textrm{B}, \mp 1}(t_1) )$ are Gaussian.
In Appendix~\ref{app:sec:deriv_bath_update}, we derive the first two cumulants, thus completely characterizing their statistical properties.
The qubit-induced evolution of the means are exactly opposite, and read
\begin{equation}
\label{eq:def_opposite_means_single_reset}
\expval{B(t_1 + \tau)}_{\rho_{\textrm{B}, \mp 1}(t_1)} =  
\mp  \,  \mu(t_1; \tau),
\end{equation}
with
\begin{equation}
\label{eq:updated_mean_single_reset}
    \mu( t_1; \tau) \equiv - i  \int_{0}^{t_1} \! \! y_{\textrm{hist}}(t) \, C^{-}_{\bar{\rho}_B}(t_1 + \tau - t) \, \dd t.
\end{equation}
We see that the respective means depend on the control history, and the original quantum correlation function.
Notably, the explicit dependence upon the elapsed time $\tau$ makes the resulting noise statistics \emph{non-stationary}.
The subprocesses ($ B(t_1 + \tau) ; \rho_{\textrm{B},\mp 1 (t_1)}$) have the same centralized second-order moment (equivalently, second-order cumulant), as the original noise process at time $t_1=\ts = 0$, namely, 
\begin{align*}
        &\expval{ B(t_1  +  \tau_{2}) B(t_1  +  \tau_{1}) }_{\rho_{\textrm{B}, \mp 1}(t_1)} = 
        \expval{B(\tau_{2}) B(\tau_{1})}_{\bar{\rho}_{\textrm{B}}}\\
        &\hphantom{---} \;+ \expval{B(t_1 + \tau_{2})}_{\rho_{\textrm{B} \mp 1}(t_1)}
        \expval{B(t_1 +  \tau_{1})}_{\rho_{\textrm{B} \mp 1}(t_1)} , 
\end{align*}
where now non-stationarity manifests through the dependence on both $\tau_1$ and $\tau_2$.
One can check that, together with Gaussianity, the above two properties imply that each of the sub-processes $(B(\ts + \tau) ; \rho_{\textrm{B}, \mp 1}(t_1))$ is statistically indistinguishable from a mean-displaced version of the original zero-mean, stationary noise process, 
\[\big(B(\tau) + \expval{B(t_1 + \tau)}_{\rho_{\textrm{B}, \mp 1}(t_1)}; \bar{\rho}_{\textrm{B}}\big).\]
Importantly, this is true for arbitrary Gaussian dephasing noise.
{In} the specific case of a bosonic bath, $\rho_{\textrm{B}, \mp 1}(t_1)$ can be further recognized to have the structure of (multi-mode) thermal coherent states~\cite{TCS}, and regular coherent states in the limiting case of zero-temperature $-$ as is assumed here in Secs.\,\ref{ssec:case_studies_single_reset} and~\ref{ssec:case_studies_multiple_resets}.

Before moving to the case of multiple resets, we can use these exact results to discuss the (update of the) correlation functions of the full, physical post-reset bath state. Since $\rho_{\textrm{B}, \mp 1}(t_1)$ give rise to exactly opposite means,
the updated bath state has, as $\rho_{\textrm{B}}(0) = \bar{\rho}_{\textrm{B}}$, zero mean,
\begin{equation*}
    \expval{B(t_1 + \tau)}_{\rho_{\textrm{B}}(t_1^{+})} \equiv 0 \qq*{,} \forall \tau \in \mathbb{R}.
\end{equation*}
Crucially, however, $\rho_{\textrm{B}}(t_1^{+})$ is no longer stationary, unless $\mu(t_{1} ; \tau)$ is constant in $\tau$, and no longer Gaussian, unless $\mu(t_{1} ; \tau) = 0, \forall \tau \in \mathbb{R}$, which is highly non-generic. 
Note that, without the decomposition Eq.~\eqref{eq:sum_of_gaussians_single_reset} into Gaussian components, one would then require all-order cumulants to characterize the statistics of the post-reset state $\rho_{\textrm{B}}(t_1^{+})$.
For the leading-order non-zero cumulants, we have the following exact expressions:
\begin{subequations}
\begin{align*}
    C^{-}_{\!\rho_{\textrm{B}} (t_1^{+})} (\tau_{2}, \tau_{1})
                &\equiv \expval{ \comm{B(t_1 + \tau_{2})}{B(t_1 + \tau_{1})}}_{\rho_{\textrm{B}}(t_1^{+})}  \nonumber \\
                &= C^{-}_{\bar{\rho}_{\textrm{B}}}(\tau_{2} - \tau_{1}), \\
    C^{+}_{\! \rho_{\textrm{B}} (t_1^{+})}  (\tau_{2}, \! \tau_{1})
                &\equiv \expval{\acomm{B(t_1 + \tau_{2})}{B(t_1 + \tau_{1})}}_{\rho_{\textrm{B}}(t_1^{+})}  \nonumber \\
                &= C^{+}_{\bar{\rho}_{\textrm{B}}} \! (\tau_{2}  - \tau_{1}) + 2 \mu(t_{1} ; \tau_{2}) \mu(t_{1} ; \tau_{1}).
\end{align*}
\end{subequations}
The classical correlation function thus directly reveals that the updated noise-process is not stationary.
Interestingly, only the classical correlation function is updated, but the update itself depends solely on the original {\em quantum} correlation function, and, via Eq.\,\eqref{eq:updated_mean_single_reset}, it explicitly encodes the past control history.

Similar conclusions may be drawn for the general case of multiple reset operations, in which case $\lrt =\ts$.
By introducing the compact notation $\boldsymbol{z_{K}} \equiv (z_{1}, \ldots, z_{K}),$ with $z_{k}  \in \{-1,1\}, k = 1, \ldots, K$, as shown in Appendix~\ref{app:sec:deriv_bath_update} we find that the bath state following the last reset can still be expressed as a uniform mixture of Gaussian components:
\begin{equation}
\label{eq:decomposition_gaussians_multiple_resets}
   { \rho_{\textrm{B}}(\ts^{+})} =    \rho_{\textrm{B}}(\lrt^{+}) 
    = \frac{1}{2^K} \!\!\!\!\sum_{\boldsymbol{z_{K}} \in \{-1, 1\}^{K}} \!\! \rho_{\textrm{B}, \boldsymbol{z_{K}}}(\lrt),
\end{equation}
where all $\rho_{\textrm{B}, \boldsymbol{z_{K}}}(\lrt)$ have the same centralized second-order moment as $\bar{\rho}_{\textrm{B}}$ and, in analogy to Eq.\,\eqref{eq:def_opposite_means_single_reset}, we may write
\begin{equation}
\label{eq:means}
    \expval{B(\lrt + \tau)}_{\rho_{\textrm{B}, \boldsymbol{z_{K}}}(\lrt)} = \sum_{k = 1}^{K} z_{k} \,\mu_{[k]}({\lrt; \tau }),
\end{equation}
where 
\begin{equation}
\label{eq:def_mean_k_time_dom}
    \mu_{[k]}( { \lrt; \tau } ) \equiv - i \int_{t_{k-1}}^{t_{k}} \! \!  \dd s \, y_{\textrm{hist}}(s) \, C^{-}_{\bar{\rho}_{\textrm{B}}}(\lrt + \tau - s).
\end{equation}
Just as for the single-reset case, one deduces that the updated bath state $\rho_{\textrm{B}}(\lrt^{+})$ is zero-mean, non-Gaussian, has an unchanged quantum correlation function,
\[C^{-}_{\rho_{\textrm{B}}(\lrt^{+})}(\tau_{2}, \tau_{1}) \equiv C^{-}_{\bar{\rho}_{\textrm{B}}}(\tau_{2} - \tau_{1}) ,\]
and is explicitly non-stationary, as manifested in its updated classical correlation function:
\begin{align}
\label{eq:update_class_corr_func_multiple_resets}
    C^{+}_{\rho_{\textrm{B}}(\lrt^{+})}(\tau_{2}, \tau_{1}) &= C^{+}_{\bar{\rho}_{\textrm{B}}}(\tau_{2} - \tau_{1}) \nonumber \\
                    &+ 2 \sum_{k = 1}^{K} \mu_{[k]}( {\lrt; \tau_{2} ) \mu_{[k]}(\lrt; \tau_{1} ).}
\end{align}

All the different scenarios for the control-history-dependent gate fidelity studied in Secs.\,\ref{sec:single_reset}-\ref{sec:multiple_resets} can now be re-interpreted and explained based on the \emph{non-zero means $\mu_{[k]}(\lrt ; \tau)$ of the bath-state components}, which encode the control history.
Indeed, following an independent derivation for the qubit fidelity at time $\lrt + \taug$, now starting from the initial joint state $\ketbra{+} \otimes \rho_{\textrm{B}}(\lrt^{+})$, as shown in Appendix \ref{app:sec:deriv_bath_update}, we can derive the following alternative expressions directly:
\begin{subequations}
\begin{align}
\hspace*{-2mm}\chi_{\textrm{c}} &= \! \! \int_{0}^{\taug} \! \! \! \! \dd \tau_{2} \! \! \int_{0}^{\taug} \! \! \! \! \dd \tau_{1} \,
                         y_{\textrm{G}}(\tau_{2}) y_{\textrm{G}}(\tau_{1})
                         \Big[ C^{+}_{\rho_{\textrm{B}, \boldsymbol{z_{K}}}(\lrt)} (\tau_{2}, \tau_{1}) \nonumber\\
    & - 2 \expval{B(\lrt +  \tau_{2})}_{\rho_{\textrm{B}, \boldsymbol{z_{K}}} \! (\lrt)} \!
        \expval{B(\lrt +  \tau_{1})}_{\rho_{\textrm{B}, \boldsymbol{z_{K}}} \! (\lrt)} \!  \Big],
\label{eq:chi_from_centralized_mean}\\
    \hspace*{-2mm}\theta_{\textrm{q}, [k]} & \equiv 2 \int_{0}^{\taug} \! \dd \tau \, y_{\textrm{G}}(\tau) \, \mu_{[k]}( \lrt; \tau ).
\label{eq:thetas_from_means}
\end{align}
\end{subequations}
Note that these expressions are written directly in terms of the statistics of the updated bath state components, with no explicit reference to the initial bath states $\bar{\rho_{\textrm{B}}}$.
In turn, we have described the statistics of the updated bath state components in terms of those of the original bath state $-$ and importantly, the control history.
If we substitute Eq.\,\eqref{eq:means}, Eq.~\eqref{eq:def_mean_k_time_dom} into Eq.\,\eqref{eq:chi_from_centralized_mean}, respectively Eq.\,\eqref{eq:thetas_from_means}, we thus obtain back the original expressions in
Eq.\,\eqref{eq:timedom_def_chi}, respectively Eq.\,\eqref{eq:timedom_def_thetas}.
The expressions in Eq.\,\eqref{eq:chi_from_centralized_mean}, Eq.\,\eqref{eq:thetas_from_means} shed a different light on the dynamics, however, as now we can directly interpret the quantum phases as being acquired through the non-zero-mean (Gaussian) components of the physical quantum state of the bath, which are in turn determined by past control operations applied to the system.
In particular, one can see explicitly from Eq.\,\eqref{eq:thetas_from_means} that, if the bath state was reset after every qubit reset, one would have that $\mu_{[k]}(\lrt; \tau ) \equiv 0$, and consequently all the quantum phases would vanish.

In the remainder of this section, we specifically focus on explaining the results of Sec.\,\ref{sub:idling} (and Fig.\,\ref{fig:rethermalization} in particular), which are indirect evidence of approximate re-equilibration under prolonged high-fidelity DD.
For completeness, the case of periodic control repetition is also included in Appendix~\ref{app:sec:dressing_bath_spectra}.

\subsection{Approximate re-equilibration of quantum bath statistics}
\label{ssec:rethermalization}

\begin{figure}[t!]
\includegraphics[width=1.0\linewidth]{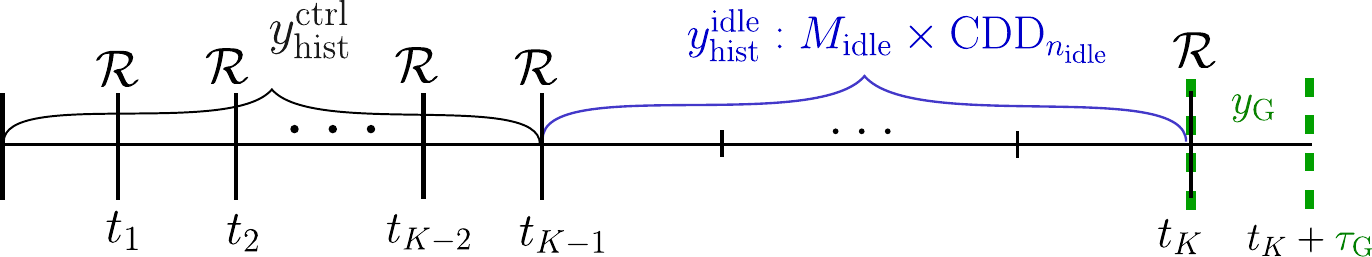}
\vspace*{-3mm}
\caption{Schematics for the re-equilibration analysis of the quantum bath statistics. The prolonged high-fidelity idling period has duration $t_{K} - t_{K - 1}\equiv T_{\textrm{idle}}$.}
\label{fig:retherm_theory}
\end{figure}

Consider $K - 1$ time intervals of general, possibly low-fidelity control, characterized by a switching function $y_{\textrm{hist}}(t)\equiv y_{\textrm{hist}}^{\textrm{ctrl}}(t)$, and separated by $K - 1$ reset operations applied at times $t_{k}, k = 1, \ldots, K - 1$.
The high-fidelity idling sequence of length $T_{\textrm{idle}}$ is, in turn, characterized by a switching function $y_{\textrm{hist}}(t)\equiv y_{\textrm{hist}}^{\textrm{idle}}(t)$ (see Fig.\,\ref{fig:retherm_theory} for a graphical illustration).
We will now argue that, under the same assumptions on the LF-noise as before, the means of the updated bath state after the last reset at time $t_K$,   $\expval{B(\lrt + \tau)}_{\rho_{\textrm{B}, \boldsymbol{z_{K}}}(\lrt)}$, converge to a value that is determined by the mean of the idling period alone, in the asymptotic limit where $T_{\textrm{idle}} \rightarrow\infty$. {Thus, the means converge to a value that is} ultimately limited by the {\em finite} control resources $\tau_0>0$.

By using Eq.\,\eqref{eq:means}, we can write the mean corresponding to $\rho_{\textrm{B}, \boldsymbol{z_{K}}}(t_{K})$ in the decomposition of the bath state after the $K$-th reset at $t_{K} = t_{K - 1} + T_{\textrm{idle}}$ as
\begin{align*}
 \expval{B(\lrt + \tau)}_{\rho_{\textrm{B}, \boldsymbol{z_{K}}}(\lrt)}  =  \expval{B(t_{K - 1} + T_{\textrm{idle}} + \tau)}_{\rho_{\textrm{B}, \boldsymbol{z_{K}}}(t_{K})} \\ 
                \equiv \sum_{k = 1}^{K - 1} z_{k} \,\mu_{\textrm{hist},k }^{\textrm{ctrl}}(t_K; \tau) + z_{K} \,\mu_{\textrm{hist}}^{\textrm{idle}}(t_K; \tau) ,
\end{align*}
where the two mean contributions are given by Eq.\,\eqref{eq:def_mean_k_time_dom}, in terms of the appropriate switching functions. By first focusing on $\mu_{k, \textrm{ctrl}}(\tau)$, we may write it as an overlap integral in the frequency domain:
\begin{eqnarray*}
\mu_{\textrm{hist}, k}^{\textrm{ctrl}}(t_K; \tau) &=& - i \int_{t_{k-1}}^{t_{k}} \! \! \! y_{\textrm{hist}}^{\textrm{ctrl}}(t) \,C^{-}(t_{K - 1} + T_{\textrm{idle}} + \tau - t)\\
&=& \!\frac{1}{2 \pi} \!\int_{- \infty}^{\infty} \!\!\!e^{i (t_{K - 1} - t_{k - 1} + T_{\textrm{idle}} + \tau) \omega} \\
& \mbox{} & \hspace*{5mm} e^{-i t_{k{-1}} \omega} F_{\textrm{hist}}^{\textrm{ctrl}}(-\omega ; t_{k-1}, t_{k}) S^{-}(\omega) \, \dd \omega.
\end{eqnarray*}
Under the stated LF-assumptions on the noise, we may then define the LF-expansion
\begin{align*}
e^{i \omega t_{k{-1}}} F_{\textrm{hist}}^{\textrm{ctrl}}(\omega ; t_{k-1}, t_{k})
        &\sim \, \tilde{F}_{\textrm{Re}} \, \omega^{\alpha_{\textrm{p}}^{\textrm{Re}}} {(t_{k} - t_{k-1})}^{\alpha_{\textrm{p}}^{\textrm{Re}} + 1} \\
        &+ i     \tilde{F}_{\textrm{Im}} \, \omega^{\alpha_{\textrm{p}}^{\textrm{Im}}} {(t_{k} - t_{k-1})}^{\alpha_{\textrm{p}}^{\textrm{Im}} + 1},       
\end{align*}
where $\tilde{F}_{\mathrm{Re / Im}} \in \mathbb{R}$. By following a reasoning analogous to the one in Theorem 1, one can then show that, provided that $s$ is not an odd integer, we obtain the asymptotic expansions 
\begin{align*}
    \abs{\mu_{\textrm{hist},k}^{\textrm{ctrl}}(t_K; \tau)} & \stackrel{T_{\textrm{idle}} \rightarrow \infty}{\sim} d_{\textrm{Re}} \LFstrength \LFwidth^{{\alpha_{\textrm{p}}^{\textrm{Re}}}} {\qty(t_{k} - t_{k-
    1})}^{{\alpha_{\textrm{p}}^{\textrm{Re}}} + 1}
                                        \\
                                      & \times \frac{\abs{\cos(\frac{\pi (s + \alpha_{\textrm{p}}^{\textrm{Re}})}{2})}}
             {{(\LFwidth \abs{t_{K - 1} - t_{k-1} + T_{\textrm{idle}} + \tau})}^{1 + s + \alpha_{\textrm{p}}^{\textrm{Re}}}} \nonumber\\
                                      &  +  d_{\textrm{Im}}
                                     \LFstrength \LFwidth^{{\alpha_{\textrm{p}}^{\textrm{Im}}}} {\qty(t_{k} - t_{k- 1})}^{{\alpha_{\textrm{p}}^{\textrm{Im}}} + 1}
                                         \nonumber\\
                                      & \times \frac{\abs{\sin(\frac{\pi (s + \alpha_{\textrm{p}}^{\textrm{Im}})}{2}) }}
             {{(\LFwidth \abs{t_{K - 1} - t_{k-1} + T_{\textrm{idle}} + \tau})}^{1 + s + \alpha_{\textrm{p}}^{\textrm{Im}}}} \nonumber ,
\end{align*}
with
\begin{subequations}
\begin{align*}
    d_{\textrm{Re}} \equiv \frac{2 \Gamma\qty(1 + s + \alpha_{\textrm{p}}^{\textrm{Re}})}
                     {\pi \Gamma\qty(\frac{1 + s}{2})} |\tilde{F}_{\textrm{Re}}|, \;
    d_{\textrm{Im}} \equiv \frac{2 \Gamma\qty(1 + s + \alpha_{\textrm{p}}^{\textrm{Im}})}
                     {\pi \Gamma\qty(\frac{1 + s}{2})} |\tilde{F}_{\textrm{Im}}| .
\end{align*}
\end{subequations}
Hence, the contributions before the idling sequence decay to zero in a power-law, on a timescale $\frac{1}{\LFwidth}$. 
For transient values of $\LFwidth T_{\textrm{idle}}$, however, this contribution can  \emph{a priori} be of the same order as for $T_{\textrm{idle}} = 0$.
Importantly, when $s$ is an odd integer, $\mu_{\textrm{hist},k}^{\textrm{ctrl}}(t_K;\tau)$ can still decay super-polynomially in $T_{\textrm{idle}}$, since $\alpha_{\textrm{p}}^{\mathrm{Re}}$ and $\alpha_{\textrm{p}}^{\mathrm{Im}}$ have opposite parity.
In such a case, the bath statistics would only re-equilibrate faster.

Turning to the second term, $\mu_{\textrm{hist}}^{\textrm{idle}}(t_K = t_{K - 1} + T_{\textrm{idling}} ; \tau)$,
one expects that by choosing a sufficiently high-order DD scheme applied in the fast control limit, 
$\LFwidth \tau_{0} \ll 1$, $\mu_{\textrm{hist}}^{\textrm{idle}}(t_K ; \tau)$
can be made negligibly small
for a value of the minimal timescale $T_{\textrm{idle}}$ needed 
to make the first term, $\mu_{\textrm{hist},k}^{\textrm{ctrl}}(t_K = t_{K - 1} + T_{\textrm{idle}}; \tau)$, decay. 
While obtaining a quantitative upper bound to $\sup_{\tau > 0} |\mu_{\textrm{hist}}^{\textrm{idle}}(t_K; \tau)|$
for fixed (but large enough) $T_{\textrm{idling}}$
is an interesting problem which we leave to future analysis,
the existence of such a bound is expected from the fact that a rigorous non-zero bound is known
for the minimum decay $\chi_c$ achievable by any DD sequence, subject to a timing constraint $\tau_0>0$~\cite{limits}.

{\section{Summary and Outlook}}
\label{sec:discussion}

In this work, we have shown that temporally correlated, nonclassical noise introduces an additional mechanism for bath-mediated error propagation, even when the system-side error propagation is fully removed through perfect reset operations.
Such a bath-side error propagation is generally detrimental to the performance of DES, which becomes dependent upon the control history through an intricate interplay between the properties of the underlying nonclassical noise spectrum as well as the details of the implemented quantum protocol.
In the context of a minimal single-qubit model, we have  quantitatively characterized the impact of such a control-history dependence on the fidelity of a DD-protected identity gate under temporally correlated Gaussian dephasing noise with both low- and high- frequency components, and dephasing-preserving unitary control subject to a finite timing constraint. 

We find that significant quantum effects, resulting in transient gate-fidelity oscillation and eventual gate-performance saturation, can arise even in parameter regimes where decay effects only contributed by classical noise are suppressed to high fidelity.
The presence of narrow high-frequency noise peaks is especially harmful in conjunction with digital control timing, due to the possible onset of resonant effects which can further contribute to the emergence of erratic gate-fidelity behavior.
Physically, these effects are shown to arise from the non-negligible ``back-action'' the quantum bath can suffer from the qubit and the non-trivial way in which its statistical properties are modified, when temporal correlations remain significant across multiple circuit locations.
While similar back-action effects have attracted renewed attention recently~\cite{Liu, Dykman, ClerkQuenches, ClerkQ, Luke, QBathSteering}, our work takes a first step toward elucidating their implications for DES and fault-tolerant QEC.

With the above in mind, we have seen that prolonged periods of high-fidelity qubit-idling can successfully decouple system and bath, and allow for approximately re-equilibration of the bath statistics.
Although this approximate reset procedure can be performed at will in principle, and it approximately restores the notion of a constant gate error, it comes at the expense of having to idle the computational degrees of freedom during an increasingly long time. Besides preventing the implementation of other {logical} operations (or otherwise imposing additional compatibility constraints on their execution~\cite{viola1999universal}), such extended idling could be problematic if temporally uncorrelated 
noise (not considered here) is also present -- as it would always be the case to some extent in practice. Since this could cause significant decoherence, additional tradeoffs would then have to be addressed for {DES to remain beneficial in the presence of both noise sources}.

{As we acknowledged, the present analysis is idealized in several ways. Nonetheless,} it serves as a caution that our current understanding about the interplay between DES and QEC, which largely ignores the possibility of back-action from the {driven} computational degrees of freedom onto a quantum bath, may fail to capture important aspects that temporal correlations can engender in the resulting error-propagation dynamics. 
{This initial analysis lends itself to a number of open questions and generalizations.  
From a characterization standpoint, it underscores the importance of further developing protocols that can unambiguously detect the presence of nonclassical noise. As it turns out, ``generalized Ramsey protocols'' may be devised to this end, which are less demanding than full QNS and, unlike existing ones \cite{Sen}, rely solely on unitary control resources. While we plan to report on that in a separate paper~\cite{free_evolution_paper}, it is interesting to note that the effect of arbitrary (instantaneous) off-axis gates can still be analyzed in terms of the same decay factor $\chi_{\textrm{c}}$ and quantum phases $\theta_{\textrm{q}, [k]}$ considered here. In terms of error mitigation, a direction for future work is to assess whether the knowledge of the applied control history, possibly combined with prior knowledge on the original noise statistics, obtained via QNS, can be leveraged to design improved, adaptive control schemes to make the bath statistics re-equilibrate faster, or with less taxing control resources. Such an approach would broadly connect to the idea of ``bath engineering'' -- although, in a non-Markovian setting and \emph{indirectly}, since the bath would be steered through the control applied to the system, akin to ``indirect controllability'' \cite{Domenico}.}

Most importantly, it is crucial to determine the extent to which similar limitations to combining DES and QES as identified in this single-qubit study will emerge in a proper QEC setting, where only the logical (not the full physical) state is projected onto the code space at each round. A first next step in this direction would be to revisit the setting of a three-qubit QEC code under bosonic dephasing considered in~\cite{novais2006decoherence}, and then move to one comprising at least two logical qubits, where protected two-qubit gates would also be needed. 
We expect that the presence of a quantum bath will still result in additional constraints that would need to be accounted for to properly choreograph DES and QEC into a layered quantum computing architecture. 
For instance, carefully designed evolution blocks may need to be included, that allow for the relevant temporally correlated
bath to be re-equilibrated, without reintroducing Markovian noise that may also generically be present.
Additionally, we expect that noise {\em spatial correlations}, which could analogously entail a component enabled solely by the nonclassical nature of the bath, would need to be treated on equal footing, as the system size increases. With quantum information and QEC protocols continuing to grow in complexity, our work highlights the importance of additional theoretical and experimental investigations into the {\em combined} effect of classical and quantum error mechanisms, in concert with DES and QEC taken together.

\bigskip

\section*{ACKNOWLEDGEMENTS}

It is a pleasure to thank Nikolay Gnezdilov, {Andy Goldschmidt}, Francisco Riberi, Francesco Ticozzi, Tommaso Grigoletto and Lev-Arcady Sellem for valuable discussions and comments, and Kaveh Khodjasteh for a critical reading of the manuscript.
This work was supported by the U.S.\ Army Research Office under grant No.\,W911NF2210004.
Work at Dartmouth was also supported in part by U.S.\ Army Research Office MURI Grant No.\,W911NF1810218.

\onecolumngrid

\section*{Appendix}
\appendix

\section{Derivation of general system observables and gate fidelity}
\label{app:sec:deriv_x_exp_vals}

Recall that the relevant gate fidelity [Eq.\,\eqref{eq:def_gate_fidelity_single_reset}] is defined as
\begin{align}
\label{eq:app:rep_def_gate_fidelity_multiple_resets}
    \mathcal{F}_{\textrm{G}}(\lrt) &= \mel{+}{\rho_S(\lrt + \taug)}{+} 
= \frac{1}{2} \!\!\sum_{z, z' \in \{-1, 1\}} \!\!\mel{z}{\rho_S(\lrt + \taug)}{z'} \nonumber\\
& = \frac{1}{2} \Big ( 1+ \!\mel{z=+1}{\rho_S(\lrt + \taug)}{z=-1} + \text{H.c.}\Big), 
\end{align} 
where, as noted in the main text, $\ket{z = -1} = \ket{g}$, $\ket{z = +1} =\ket{e}$ denotes the qubit eigenbasis. The remainder of this section is dedicated to calculating the above coherence matrix element, $\mel{+1}{\rho_S(\lrt + \taug)}{-1} = \tfrac{1}{2} \big(\langle \sigma_x (\lrt+\taug) \rangle  + i \langle \sigma_y (\lrt + \taug) \rangle \big) $ step by step.

Keeping the switching function $y(t), t \in [0, \lrt + \taug)$ fully general, and recalling that the reset operations ${\cal R}$ are assumed to be instantaneous, we can write the joint system-bath state
at the final time $\lrt + \taug$ as 
\begin{equation*}
    \rho_{\textrm{SB}}(t_K + \taug) = U(\lrt, \lrt + \taug) \, \reset\qty(\rho_{\textrm{SB}}(t_{K}^{-})) \, 
U^\dagger(\lrt, \lrt + \taug),
\end{equation*}
where the consecutive pre-reset states are given by the recursion 
\begin{equation}
    \rho_{\textrm{SB}}(t_{k}^{-}) = U(t_{k-1}, t_{k}) \, \reset\qty(\rho_{\textrm{SB}}(t_{k - 1}^{-})) \, U^\dagger(t_{k-1}, t_{k}), \quad k = 1, \ldots, K, 
\end{equation}
where the joint unitaries are defined as in Eq.\,\eqref{eq:def_joint_unitary} of the main text, and 
the initial joint system-bath state is factorized into $\rho_{\textrm{SB}}(0) = \ketbra{+} \otimes \bar{\rho}_B$. We can again represent the $\ketbra{+}$-state and also the reset operation $\reset$ in the $\sigma_z$-basis, 
\[\reset\qty(\rho_{\textrm{SB}}(t_{k}^{-})) = \frac{1}{2} \sum_{z_{k + 1}, z'_{k + 1}} \!\ketbra{z_{k + 1}}{z'_{k + 1}}
                        \otimes \sum_{z_{k}} \mel{z_{k}}{\rho_{\textrm{SB}}(t_{k}^{-})}{z_{k}}, \quad k = 1, \ldots, K,\]
where from here on out the summation limits $z \in \{-1, 1\}$ are tacitly assumed.
It is then easy to verify that
\begin{align}
    \mel{+1}{\rho_{S}(\lrt + \taug)}{-1}
        &=\frac{1}{2} \hspace*{-3mm} \sum_{z_{K + 1}, z'_{K + 1} ; z_K } \hspace*{-3mm} 
               \mel{+1}{\Tr_{\textrm{B}}[U(\lrt, \lrt + \taug) \qty(\ketbra{z_{K + 1}}{z'_{K + 1}} \!\otimes \!
               \mel{z_{K}}{\rho_{\textrm{SB}}(t^{-}_{K-1})}{z_{K}}) U^\dagger(\lrt, \lrt + \taug)]}{-1} \nonumber \\
        &= \frac{1}{2} \sum_{z_{K}}
               \Tr_{\textrm{B}}[\mel{+1}{U(\lrt, \lrt + \taug)}{+1} 
               \mel{z_{K}}{\rho_{\textrm{SB}}(t^{-}_{K-1})}{z_{K}}) \mel{-1}{U^\dagger(\lrt, \lrt + \taug)}{-1}] \nonumber \\
        &\ldots \nonumber\\
        &= \frac{1}{2^{K}} \sum_{z_{1}, \ldots, z_{K}} \Tr_{\textrm{B}}[
            \mel{+1}{U(\lrt, \lrt + \taug)}{+1}
            \mel{z_{K}}{U(t_{K - 1}, \lrt)}{z_{K}} \ldots \mel{z_{1}}{U(t_{0}, t_{1})}{z_{1}} \nonumber\\
            & \hspace*{-1cm} \hphantom{+\frac{1}{2^{K}} \sum_{z_{1}, \ldots, z_{K}} \Tr_{\textrm{B}}[} \; \;
            \mel{z_{1}}{\rho_{\textrm{S}(0)}}{z_{1}} \bar{\rho}_{\textrm{B}}
            \mel{z_{1}}{U^\dagger(t_{0}, t_{1})}{z_{1}} \ldots \mel{z_{K}}{U^\dagger(t_{K - 1}, t_{K})}{z_{K}}
            \mel{-1}{U^\dagger(\lrt, \lrt + \taug)}{-1}
            ] \nonumber \\
        &= \frac{1}{2^{K}} \sum_{z_{1}, \ldots, z_{K}} \biggl \langle
            \mel{z_{1}}{U^\dagger(t_{0}, t_{1})}{z_{1}} \ldots \mel{z_{K}}{U^\dagger(t_{K - 1}, t_{K})}{z_{K}} \mel{-1}{U^\dagger(\lrt, \lrt + \taug)}{-1} \nonumber\\
            & \hphantom{= \frac{1}{2^{K}} \sum_{z_{1}, \ldots, z_{K}}} \; \; \; \,
          \mel{+1}{U(\lrt, \lrt + \taug)}{+1} \mel{z_{K}}{U(t_{K - 1}, \lrt)}{z_{K}} \ldots \mel{z_{1}}{U(t_{0}, t_{1})}{z_{1}}
            \biggl\rangle_{\bar{\rho}_B}  . 
            \label{eq:app:last_line_coherence_element_expansion}
\end{align}
Here, we have used that the projectors $\ketbra{z}$ commute with all unitaries $\forall z \in \{-1, 1\}$, ensuring that no cross terms of the form $\mel{z_{k'}}{U(t_{k-1}, t_{k})}{z_{k}}, z_{k} \neq z_{k}'$ survive, and in the last line, we used the cyclicity of the trace.

At this point, it is useful to introduce an indexed family of switching functions that encode the specific values $z_{k}$,
\begin{equation}
\label{eq:app:generalized_switching_functions}
    y_{z, \boldsymbol{z_K}}(t) =   y_{\mp 1, \boldsymbol{z_K}}(t) \equiv 
    \begin{cases}
        \ z_k  \, y(t) &\qq*{,} t \in [t_{k-1}, t_{k}), \, k = 1, \ldots, K,\\
        \,\mp \, y(t) &\qq*{,} t \in [\lrt, \lrt + \taug).
    \end{cases}
\end{equation}
With this definition, we can write
\begin{equation*}
      \mel{z}{U(\lrt, \lrt + \taug)}{z} \mel{z_{K}}{U(t_{K - 1}, \lrt)}{z_{K}} \ldots \mel{z_{1}}{U(t_{0}, t_{1})}{z_{1}}
    = \mathcal{T}_+ \exp(- i \int_{0}^{\lrt + \taug} y_{z, \boldsymbol{z_{K}}}(s) B(s) \dd s).
\end{equation*}
Employing this relationship in Eq.\,\eqref{eq:app:last_line_coherence_element_expansion} then allows us to more compactly write the desired coherence element as 
\begin{equation}
\label{eq:general_sum_of_Ms}
    \mel{z=+1}{\rho_S(\lrt + \taug)}{z=-1} = \frac{1}{2^K} \!\!\sum_{z_1, \ldots, z_K } \!E_{\boldsymbol{z_K}},
\end{equation}
with 
\begin{equation*}
E_{\boldsymbol{z_K}} \equiv \expval{\mathcal{T}_- \exp(i \int_{0}^{\lrt + \taug} y_{-1, \boldsymbol{z_{K}}}(s) B(s) \dd s)
\mathcal{T}_+ \exp(- i \int_{0}^{\lrt + \taug} y_{+1, \boldsymbol{z_{K}}}(s) B(s) \dd s) }_{\bar{\rho}_B} .
\end{equation*}
Finally, defining
\begin{equation*}
    \bar{H}_{\boldsymbol{z_{K}}}(t) \equiv 
    \begin{cases}
        y_{-1,\boldsymbol{z_{K}}}(\lrt + \taug - t) B(\lrt + \taug + t) &\qq*{,} t \in [0, \lrt + \taug)\\
        y_{+1,\boldsymbol{z_{K}}}(\lrt + \taug + t) B(\lrt + \taug - t) &\qq*{,} t \in [- \lrt - \taug, 0),
    \end{cases}
\end{equation*}
we can write
\begin{equation*}
   E_{\boldsymbol{z_K}} \equiv \expval{\mathcal{T}_+ \exp(- i \int_{-\lrt - \taug}^{\lrt + \taug} \bar{H}_{\boldsymbol{z_{K}}}(s)\dd s)}_{\bar{\rho}_B}.
\end{equation*}

We can now utilize generalized cumulants to calculate $E_{\boldsymbol{z_K}}$. Since the noise process $(B(t), \bar{\rho}_{\textrm{B}})$ is Gaussian, stationary, and zero-mean, one only needs to calculate the second-order cumulant, namely, 
\[E_{\boldsymbol{z_K}} = e^{- \frac{1}{2} \mathcal{C}_{\boldsymbol{z_K}}^{(2)}(\lrt + \taug)},\]
where
\begin{align}
\label{eq:general_off_diagonal_cumulant}
    \frac{1}{2} \mathcal{C}_{\boldsymbol{z_K}}^{(2)}(\lrt + \taug)
       &= \int_{- \lrt + \taug}^{\lrt + \taug} \!\!\dd s \int_{-\lrt - \taug}^{s}  \!\!\!\dd s' \expval{\bar{H}(s) \bar{H}(s')}_{\bar{\rho}_{\textrm{B}}} \nonumber\\
       &= \int_{\lrt}^{\lrt + \taug}\!\! \dd s  \int_{\lrt}^{\lrt + \taug} \!\!\dd s' y(s) y(s') C^{+}(s, s')
            + 2 \int_{\lrt}^{\lrt + \taug} \!\!\dd s \int_0^{\lrt} \!\!\dd s' y(s) y_{\boldsymbol{z_K}}(s') C^{-}(s, s'). 
\end{align}
Here, the last line is obtained after splitting up the integration domain appropriately. Remembering that $C^{+}$ is real and $C^{-}$ is purely imaginary, we define the classical decay factor
\begin{equation}
\label{eq:app:timedom_def_chi}
\chi_c (\lrt, \lrt + \taug)
\equiv \int_{\lrt}^{\lrt + \taug} \!\! \dd s \int_{\lrt}^{\lrt + \taug} \!\! \dd s'
y(s) y(s') \; C^+(s - s') ,
\end{equation}
and the quantum phase
\begin{equation}
\label{eq:app:theta_zK_unraveled}
\theta_{q, \boldsymbol{z_{K}}}(0, \lrt; \lrt, \lrt + \taug) \equiv - 2 i \int_{\lrt}^{\lrt + \taug} \!\!\dd s \int_0^{\lrt} \!\!\dd s'
y(s) y_{\boldsymbol{z_K}}(s') C^{-}(s, s')
\end{equation}
to obtain
\begin{equation*}
    \frac{1}{2} \mathcal{C}_{\boldsymbol{z_K}}^{(2)}(\lrt + \taug) = \chi_{c}(\lrt, \lrt + \taug)
    + i \theta_{q, \boldsymbol{z_{K}}}(0, \lrt; \lrt, \lrt + \taug).
\end{equation*}
First, note that the first term does not depend on the multi-index $\boldsymbol{z_K}$, and hence $\exp(- \chi_c(\lrt, \lrt + \taug))$ factors out in Eq.\,\eqref{eq:general_sum_of_Ms}. To simplify the remaining sum in Eq.~\eqref{eq:general_sum_of_Ms}, it is instructive to split up the second integral in Eq.~\eqref{eq:general_off_diagonal_cumulant} into
\[\int_0^{t_{K}} \dd s = \sum_{k = 1}^K \int_{t_{k-1}}^{t_k} \dd s.\]
This yields
\begin{equation}
\label{eq:decomp_thetas_zk}
    \theta_{q, \boldsymbol{z_{K}}}(0, \lrt; \lrt, \lrt + \taug) =
    \sum_{k = 1}^{K} z_{k} \theta_{q,[k]}(t_{k-1}, t_{k}; \lrt, \lrt + \taug),
\end{equation}
where $\theta_{q,[k]}(t_{k-1})$ is defined in Eq.~\eqref{eq:timedom_def_single_theta}.  
Then, performing the sum over $z_{1}, \ldots, z_{K} \in \{-1, 1\}$ in~Eq.\,\eqref{eq:general_sum_of_Ms}, we obtain
\begin{align*}
\mel{+1}{\rho_S(\lrt + \taug)}{-1} = \frac{1}{2} \,e^{- \chi_c} \prod_{k = 1}^{K} \cos(\theta_{q,[k]}),
\end{align*}
with 
\[ \theta_{q,[k]} \equiv - 2 i \int_{\lrt}^{ \lrt + \taug} \! \! \dd s \int_{t_{k-1}}^{t_{k}} \! \! \dd s'
                                     y(s) \, y(s') \, C^-(s - s'). \]
Finally, inserting this into Eq.\,\eqref{eq:app:rep_def_gate_fidelity_multiple_resets} yields the stated gate fidelity in Eq.\,\eqref{eq:general_fidelity_multiple_resets} in the main text,
if one considers splitting up the general switching function as
\begin{equation*}
    y(t) \equiv \begin{cases}
                    y_{\textrm{hist}}(t) &\qq*{,} t \in [0, \lrt),           \\
                    y_{\textrm{G}}(t - \lrt)    &\qq*{,} t \in [\lrt, \lrt + \taug).
                \end{cases}
\end{equation*}
In the special case $K = 1$, Eq.\,\eqref{eq:general_fidelity_single_reset} is recovered.

\section{Derivation of updated bath correlation functions}
\label{app:sec:deriv_bath_update}

In this section, we set out to derive the main results reported in Sec.\,\ref{ssec:decomposition_gaussian_components}.
While the method of propagating quantum bath statistics to a later time after interaction with the system can be formulated for more general open quantum-system dynamics~\cite{bath_update_paper}, in order to make the presentation self-contained, the necessary ingredients are developed in the derivation in this section.

\subsection{Decomposition of evolved bath state into Gaussian components}

As a first step, let us work towards a representation of the bath state after the $K$-th reset operation, in terms of bath-only unitaries acting on the initial bath state,
\begin{align}
    \rho_{\textrm{B}}(\lrt^{+})
             &= \Tr_{S} \qty[U(t_{K-1}, \lrt) \qty(\ketbra{+} \otimes \rho_{\textrm{B}}(t_{K-1}^{+})) U^\dagger(t_{K-1}, \lrt)] \nonumber\\
             &= \sum_{{z_{K}}} \mel{z_{K}}{U(t_{K-1}, \lrt)}{z_{K}} \braket{z_{K}}{+} \braket{+}{z_{K}}
             \rho_{\textrm{B}}(t_{K-1}^{+}) \mel{z_{K}}{U^\dagger(t_{K-1}, \lrt)}{z_{K}} \nonumber\\
             &= \frac{1}{2} \sum_{{z_{K}}} \mel{z_{K}}{U(t_{K-1}, \lrt)}{z_{K}}
            \rho_{\textrm{B}}(t_{K-1}^{+}) \mel{z_{K}}{U^\dagger(t_{K-1}, \lrt)}{z_{K}}.
\label{eq:bath_pov_starting_point_app}
\end{align}
Following through with this recursion, we get
\begin{equation}
\label{eq:gaussian_decomposition_appendix}
    \rho_{\textrm{B}}(\lrt^{+}) = \frac{1}{2^K} \sum_{\boldsymbol{z_{K}} \in \{-1,1\}^{K}}
    \rho_{\textrm{B}, \boldsymbol{z_{K}}}(\lrt),
\end{equation}
with
\begin{equation*}
    \rho_{\textrm{B}, \boldsymbol{z_{K}}}(\lrt)
        \equiv \mel{z_{K}}{U(t_{K-1}, \lrt)}{z_{K}} \cdots \mel{z_{1}}{U(0, t_{1})}{z_{1}} \bar{\rho}_{\textrm{B}}
               \mel{z_{1}}{U^\dagger(t_{0}, t_{1})}{z_{1}} \cdots \mel{z_{K}}{U(t_{K-1}, t_{K})}{z_{K}}.
\end{equation*}
We thus see that $\rho_{\textrm{B}}(\lrt^{+})$ is the uniform convex combination of $2^K$ different bona fide bath states, as stated in Eq.\,\eqref{eq:decomposition_gaussians_multiple_resets} in the main text.  It is straightforward to see that all $ \rho_{\textrm{B}, \boldsymbol{z_{K}}}(\lrt)$ are Gaussian, by writing
\[    \mel{z_{K}}{U(t_{K-1}, \lrt)}{z_{K}} \cdots \mel{z_{1}}{U(0, t_{1})}{z_{1}}
    = \mathcal{T}_{+} \exp(- i \int_{0}^{\lrt} y_{\boldsymbol{z_{K}}}(t) B(t) \, \dd t),\]
with   $ y_{\boldsymbol{z_{K}}}(t) \equiv z_{k} y_{\textrm{hist}}(t),$ for $t \in [t_{k-1}, t_{k}).$
We are thus left with calculating the first- and second- order moments of the Gaussian component $\rho_{\textrm{B}, \boldsymbol{z_{K}}}(\lrt)$. For the first moment, we can write 
\begin{align*}
    \expval{B(\lrt +\tau)}_{\rho_{\textrm{B}, \boldsymbol{z_{K}}}(\lrt)}
        &\equiv \Tr_{\textrm{B}} \qty[ B(\lrt +\tau)  \rho_{\textrm{B}, \boldsymbol{z_{K}}}(\lrt) ] \nonumber \\
        &= \Tr_{\textrm{B}} \qty[\mathcal{T}_{-} \exp(i \int_{0}^{\lrt} y_{\boldsymbol{z_{K}}}(t) B(t) \, \dd t)
               \, B(\lrt + \tau) \,\mathcal{T}_{+}\exp(- i \int_{0}^{\lrt} y_{\boldsymbol{z_{K}}}(t) B(t) \, \dd t)
                                                                                         \bar{\rho}_{\textrm{B}}].
\end{align*}
The method to calculate this quantity exactly (by virtue of $\bar{\rho}_{\textrm{B}}$ being a Gaussian state), is to substitute 
\[B(\lrt + \tau) = i \dv{\lambda} \qty(e^{- i \lambda B(\lrt + \tau)}) \bigg|_{\lambda = 0},\]
in such a way that 
\begin{equation}
\label{eq:updated_mean_from_single_T+_exp}
    \expval{B(\lrt +\tau)}_{\rho_{\textrm{B}, \boldsymbol{z_{K}}}(\lrt)}
        = i \dv{\lambda} \expval{\mathcal{T}_{+} \exp(- i \int_{- \lrt - \frac{\lambda}{2}}^{\lrt + \frac{\lambda}{2}}
                                      B_{\lambda, \boldsymbol{z_{K}}}(t) \, \dd t)}_{\bar{\rho}_{\textrm{B}}} \Bigg|_{\lambda = 0},
\end{equation}
with
\begin{equation*}
    B_{\lambda, \boldsymbol{z_{K}}}(t)
        \equiv \begin{cases}
                              -  y_{\boldsymbol{z_{K}}}(\lrt + \frac{\lambda}{2} - t) B(\lrt + \frac{\lambda}{2} - t)
                                                            &\qq*{,} t \in [ \frac{\lambda}{2} , \lrt + \frac{\lambda}{2} ,),\\
                    \hphantom{-  y_{\boldsymbol{z_{K}}}(\lrt + \frac{\lambda}{2} - t)} B(\lrt + \tau)
                                                            &\qq*{,} t \in [ - \frac{\lambda}{2} , \frac{\lambda}{2}),\\
                    \hphantom{-} y_{\boldsymbol{z_{K}}}(\lrt + \frac{\lambda}{2} + t) B(\lrt + \frac{\lambda}{2} + t)
                                                            &\qq*{,} t \in [- \lrt - \frac{\lambda}{2} , - \frac{\lambda}{2}),
                \end{cases}
\end{equation*}
Relying on $(B_{\lambda, \boldsymbol{z_{K}}}(t), \bar{\rho}_{\textrm{B}})$ being a (zero-mean) Gaussian noise process,
we can again utilize generalized cumulants (and simple time-reparametrizations) to obtain
\[ \expval{\mathcal{T}_{+} \exp(- i \int_{- \lrt - \frac{\lambda}{2}}^{\lrt + \frac{\lambda}{2}}
   B_{\lambda, \boldsymbol{z_{K}}}(t) \, \dd t)} = e^{- \frac{1}{2} \mathcal{C}^{(2)}_{\lambda, \boldsymbol{z_{K}}}(0, \lrt)},\]
with
\begin{align*}
    \frac{1}{2} \mathcal{C}^{(2)}_{\lambda, \boldsymbol{z_{K}}}(0, \lrt)
                &= \frac{\lambda^2}{2} \expval{B^{2}(\lrt + \tau)}_{\bar{\rho}_B}
                + \lambda \int_{0}^{\lrt} y_{\boldsymbol{z_{K}}}(t)
                                           \expval{\comm{B(\lrt + \tau)}{B(t)}}_{\bar{\rho}_{\textrm{B}}} \, \dd t.
\end{align*}
Substituting this in Eq.\,\eqref{eq:updated_mean_from_single_T+_exp}, we then finally obtain
\begin{equation}
\label{eq:final_formula_updated_mean}
    \expval{B(\lrt +\tau)}_{\rho_{\textrm{B}, \boldsymbol{z_{K}}}(\lrt)}
            = - i \int_{0}^{\lrt} y_{\boldsymbol{z_{K}}}(t) C_{\bar{\rho}_{\textrm{B}}}^{-}(\lrt + \tau - t) \, \dd t.
\end{equation}

The second-order moment of $\rho_{\textrm{B}, \boldsymbol{z_{K}}}(\lrt)$ can be calculated in a completely analogous way. Explicitly, we have 
\begin{align*}
     &\expval{B(\lrt + \tau_{2}) B(\lrt + \tau_{1})}_{\rho_{\textrm{B}, \boldsymbol{z_{K}}}(\lrt)}\\
    =& \Tr_{\textrm{B}} \qty[\mathcal{T}_{-} \exp(i \int_{0}^{\lrt} y_{\boldsymbol{z_{K}}}(t) B(t) \, \dd t)
               B(\lrt + \tau_{2}) B(\lrt + \tau_{1}) \mathcal{T}_{+}
               \exp(- i \int_{0}^{\lrt} y_{\boldsymbol{z_{K}}}(t) B(t) \, \dd t) \bar{\rho}_{\textrm{B}}]\\
    =& - \pdv{}{\lambda_{2}}{\lambda_{1}} \expval{\mathcal{T}_{+}
                \exp(- i \int_{- \lrt - \lambda_{1}}^{\lrt + \lambda_{2}}
                         B_{\lambda_{2}, \lambda_{1}, \boldsymbol{z_{K}}}(t) \, \dd t)}_{\bar{\rho}_{\textrm{B}}}
                         \Bigg|_{\lambda_{2} = \lambda_{1} = 0},
\end{align*}
with
\[B_{\lambda_{2}, \lambda_{1}, \boldsymbol{z_{K}}}(t)
    = \begin{cases}
                  -  y_{\boldsymbol{z_{K}}}(\lrt + \lambda_{2} - t) B(\lrt + \lambda_{2} - t)
                                                &\qq*{,} t \in [ \lambda_{2} , \lrt + \lambda_{2} ,),\\
         B(\lrt + \tau_{2}) &\qq*{,} t \in [ 0, \lambda_{2}),\\
         B(\lrt + \tau_{1}) &\qq*{,} t \in [ - \lambda_{1} ,0 ),\\
        \hphantom{-} y_{\boldsymbol{z_{K}}}(\lrt + \lambda_{1} + t) B(\lrt + \lambda_{1} + t)
                                                &\qq*{,} t \in [- \lrt - \lambda_{1} , - \lambda_{1}),
      \end{cases}\]
Again, working this out by utilizing the second order of the generalized cumulant expansion, one obtains
\begin{align}
    \expval{B(\lrt +\tau_{2}) B(\lrt +\tau_{1})}_{\rho_{\textrm{B}, \boldsymbol{z_{K}}}(\lrt)}
            & = \expval{B(\tau_{2}) B(\tau_{1})}_{\bar{\rho}_{\textrm{B}}} \nonumber\\
            &- 
            \int_{0}^{\lrt} y_{\boldsymbol{z_{K}}}(t) C_{\bar{\rho}_{\textrm{B}}}^{-}(\lrt + \tau_{2} - t) \, \dd t 
               \int_{0}^{\lrt} y_{\boldsymbol{z_{K}}}(t) C_{\bar{\rho}_{\textrm{B}}}^{-}(\lrt + \tau_{1} - t) \, \dd t 
                           \nonumber\\
            & =\expval{B(\tau_{2}) B(\tau_{1})}_{\bar{\rho}_{\textrm{B}}}
             + \expval{B(\lrt +\tau_{2})}_{\rho_{\textrm{B}, \boldsymbol{z_{K}}}(\lrt)}
              \expval{B(\lrt +\tau_{1})}_{\rho_{\textrm{B}, \boldsymbol{z_{K}}}(\lrt)},
               \label{eq:result_same_centralized_means}
\end{align}
where, in the last line, we have used Eq.\,\eqref{eq:final_formula_updated_mean} for the updated mean.
Thus, we indeed see that $\rho_{\textrm{B}, \boldsymbol{z_{K}}}(\lrt)$ and $\bar{\rho}_{\textrm{B}}$ have the same centralized moments, as claimed in Sec.\,\ref{ssec:decomposition_gaussian_components}.

\subsection{Updated statistics for the full post-reset bath state}

Putting everything together, and performing the sum in Eq.\,\eqref{eq:gaussian_decomposition_appendix}, we then get
\begin{align*}
    \expval{B(\lrt + \tau)}_{\rho_{\textrm{B}}(\lrt^{+})}
                = - \frac{i}{2^{K}} \sum_{\boldsymbol{z_{K}} \in \{-1,1\}^{K}} \sum_{k = 1}^{K} z_{k}
                                \int_{t_{k-1}}^{t_{k}} y_{\textrm{hist}}(t) C^{-}(\lrt + \tau - t) \, \dd t
                = 0,
\end{align*}
whereas for the updated quantum and classical correlation functions, we obtain
\begin{subequations}
\begin{align*}
    C^{-}_{\rho_{\textrm{B}}(\lrt^{+})}(\tau_{2}, \tau_{1})
            &\equiv C^{-}_{\bar{\rho}_{\textrm{B}}}(\tau_{2}, \tau_{1}), \quad \forall \tau_{2}, \tau_{1},\\
    C^{+}_{\rho_{\textrm{B}}(\lrt^{+})}(\tau_{2}, \tau_{1})
            &= \frac{1}{2^{K}} \sum_{\boldsymbol{z_{K}} \in \{-1, 1\}^K}
               \Big(C^{+}_{\bar{\rho}_{\textrm{B}}}(\tau_{2}, \tau_{1}) - 2 \sum_{k_{1} = 1}^{K} \sum_{k_{2} = 1}^{K} 
               z_{k_{2}} z_{k_{1}} \mu_{[k_{2}]}(\lrt ; \tau_{2}) \mu_{[k_{1}]}(\lrt ; \tau_{1}) \Big)  \nonumber \\
            &=C^{+}_{\bar{\rho}_{\textrm{B}}}(\tau_{2}, \tau_{1}) - 2 \sum_{k = 1}^{K} \mu_{[k]}(\lrt ; \tau_{2}) \mu_{[k]}(\lrt ; \tau_{1}),
\end{align*}
\end{subequations}
as reported in Eq.\,\eqref{eq:update_class_corr_func_multiple_resets} in the main text.

\subsection{Alternative representation of post-reset qubit-only dynamics}
\label{sub:alternative}

Lastly, it is also interesting to calculate the qubit-only dynamics at the final time $t_K+\taug$ directly from the updated bath state at time $\rho_{\textrm{B}}(\lrt^{+})$.
Note that, for every Gaussian component $\rho_{\textrm{B}, \boldsymbol{z_{K}}}$ taken separately, the required calculation is analogous to the one given in Appendix~\ref{app:sec:deriv_x_exp_vals} starting from $\bar{\rho}_{\textrm{B}}$, so we merely provide the result of applying the generalized cumulant expansion to compute every term:
\begin{align}
    \mel{z=+1}{\rho_{\textrm{S}}(\lrt + \taug)}{z=-1}
        &= \frac{1}{2^{K}} \sum_{\boldsymbol{z_{K}} \in \{-1, 1\}}
            \expval{\mel{-1}{U^\dagger(\lrt, \lrt + \taug)}{-1} \mel{+1}{U(\lrt, \lrt + \taug)}{+1}
                   }_{\rho_{\textrm{B}, \boldsymbol{z_{K}}}(\lrt)}, \notag\\
        &= \frac{1}{2^{K}} \sum_{\boldsymbol{z_{K}} \in \{-1, 1\}}
                  e^{- i \mathcal{C}^{(1)}_{\boldsymbol{z_{K}}}(0, \taug)
                - \frac{1}{2} \mathcal{C}^{(2)}_{\boldsymbol{z_{K}}}(0, \taug)}.
\label{eq:coherence_element_from_updated_cumulants}
\end{align}
Since the sub-process ${\rho_{\textrm{B}, \boldsymbol{z_{K}}}(\lrt)}$ is no longer zero-mean, for the first cumulant we now have:
\begin{equation}
    \mathcal{C}^{(1)}_{\boldsymbol{z_{K}}}(0, \taug) = 2 \int_{0}^{\taug} \!\dd\tau \, y_{\textrm{G}}(\tau) 
            \expval{B(\lrt + \tau)}_{\rho_{\textrm{B}}, \boldsymbol{z_{K}}(\lrt)} 
           = \theta_{\textrm{q}, \boldsymbol{z_{K}}} (0, \lrt; \lrt , \lrt + \taug), 
\label{eq:thetas_zk_bath_pov_app}
\end{equation}
where, in obtaining the second equality, we have used the expression in Eq.\,\eqref{eq:final_formula_updated_mean}. The second cumulant reads 
\begin{subequations}
\begin{align}
    \mathcal{C}^{(2)}_{\boldsymbol{z_{K}}}(0, \taug) &= \int_{0}^{\taug} \dd \tau_{2} \int_{0}^{\taug} \!\dd \tau_{1}\,
            y_{\textrm{G}}(\tau_{2}) y_{\textrm{G}}(\tau_{1}) \qty(
                   C^{+}_{\rho_{\textrm{B}, \boldsymbol{z_{K}}}} (\tau_{2}, \tau_{1})
                 - 2 \expval{B(\lrt + \tau_{2})}_{\rho_{\textrm{B}}, \boldsymbol{z_{K}}(\lrt)}
                     \expval{B(\lrt + \tau_{1})}_{\rho_{\textrm{B}}, \boldsymbol{z_{K}}(\lrt)}) 
\nonumber \\
            &= \chi_{\textrm{c}}, \quad \forall \boldsymbol{z_{K}} \in \{-1, 1\}^{K},   
\label{eq:chi_bath_pov_app}
\end{align}
\end{subequations} 
where, in obtaining Eq.\,\eqref{eq:chi_bath_pov_app}, we have used the fact that $\rho_{\textrm{B}, \boldsymbol{z_{K}}}(\lrt)$ has the same centralized moments as $\bar{\rho}_{\textrm{B}}$, as established in Eq.\,\eqref{eq:result_same_centralized_means}.
We thus see that in Eqs.\,\eqref{eq:thetas_zk_bath_pov_app} and Eq.~\eqref{eq:chi_bath_pov_app} we have derived expressions for the quantum phase $\theta_{\textrm{q}, \boldsymbol{z_{K}}}$ and the classical decay factor $\chi_{\textrm{c}}$ that are equivalent, respectively, to those of Eqs.\,\eqref{eq:app:theta_zK_unraveled} and Eq.~\eqref{eq:app:timedom_def_chi}, but are directly expressed in terms of the statistics of the (Gaussian components of the) updated bath state $\rho_{\textrm{B}}(\lrt^{+})$. To arrive at Eq.\,\eqref{eq:thetas_from_means} for the phases $\theta_{\textrm{q}, [k]}$ cited in the main text, it suffices to use Eq.\,\eqref{eq:decomp_thetas_zk}.

\section{Dressing of bath noise spectra with past periodic control}
\label{app:sec:dressing_bath_spectra}

In this section, we set out to study the updated quantum bath statistics of Sec.\,\ref{ssec:decomposition_gaussian_components} for the case of periodic control repetition.
For this, we start by representing $\mu(t_{1} ; \tau)$ (single-reset case) and $\mu_{[k]}(\lrt ; \tau)$ (multiple-reset case) in the frequency domain, and considering the respective asymptotic regime.

\subsection{Single-reset case}

We start by expressing $\mu(t_{1} ; \tau)$ in Eq.\,\eqref{eq:updated_mean_single_reset} in the frequency domain,
\begin{align*}
    \mu({t_1; \tau}) = - i \int_{0}^{\ts} \! \! y_{\textrm{hist}}(t) \, C^{-}_{\bar{\rho}_B}(\ts + \tau - t) \, \dd t
                 = -\frac{i}{2 \pi} \int_{-\infty}^{\infty} e^{i \omega (\ts + \tau)} F_{\textrm{hist}}(- \omega;  0, \ts) S^{-}(\omega) \, \dd \omega.
\end{align*}
Now restricting to the case of periodic repetition, $t_1=\ts \equiv  M \taug$,
\[e^{i \omega \ts} F_{\textrm{hist}}(- \omega; 0, \ts) = e^{i (M + 1) \frac{\omega \taug}{2}}
\frac{\sin(M \frac{\omega \taug}{2})}{\sin(\frac{\omega \taug}{2})} F_{\textrm{G}}(- \omega ; \taug),\]
we obtain
\begin{equation}
\label{eq:periodic_mean_time_dom}
    \mu^{(M)}(\tau ; \taug) \equiv \mu(M \taug ; \tau) = \frac{1}{2 \pi} \int_{- \infty}^{\infty} e^{i \omega \tau} \mu^{(M)}(\omega ; \taug),
\end{equation}
with
\begin{subequations}
\begin{align}
   \mu^{(M)}(\omega ; \taug)
      &\equiv - \frac{i}{2} \qty(D^{(M)}(\omega \taug) - 1 + i \xi^{(M)}(\omega \taug)) F_{\textrm{G}}(- \omega ; \taug) S^{-}(\omega),
      \label{periodic_mean_kernel_version}\\
      &= - i \bigg(\sum_{m = 1}^{M} e^{i m \omega \taug}\bigg) F_{\textrm{G}}(- \omega ; \taug) S^{-}(\omega).
      \label{periodic_mean_fourier_version}
\end{align}
\end{subequations}
From Eq.~\eqref{periodic_mean_kernel_version}, we can deduce the asymptotic limit, as
\begin{subequations}
\begin{align*}
    \xi^{(M)}(\omega \taug) \rightarrow \mathcal{P}v_{\cot(\frac{\omega \taug}{2})},\qquad
      D^{(M)}(\omega \taug) \rightarrow \frac{2 \pi}{\taug} \Sh_{\frac{2 \pi}{\taug}}(\omega),
\end{align*}
\end{subequations}
with
\[\Sh_{\frac{2 \pi}{\taug}}(\omega) \equiv \sum_{\ell \in \mathbb{Z}} \delta\qty(\omega - \frac{2 \ell \pi}{\taug}).\]
Inserting this in Eq.~\eqref{eq:periodic_mean_time_dom}, we obtain
\begin{align*}
    \mu^{(\infty)}(\tau ; \taug) 
    = - \frac{i}{2 \taug} \sum_{\ell \in \mathbb{Z}} e^{2 i \ell \pi \frac{\tau}{\taug}}
                                    F_{\textrm{G}}\qty(- \frac{2 \ell \pi}{\taug} ; \taug) S^{-}\qty(\frac{2 \ell \pi}{\taug})
                                 + \frac{1}{4 \pi} \mathcal{P}v \int_{- \infty}^{\infty} \!\qty({\cot(\frac{\omega \taug}{2})} + i)
                                 F_{\textrm{G}}(- \omega ; \taug) S^{-}(\omega)\, \dd \omega.
\end{align*}
We thus see that the infinite periodic repetition singles out the IRF's, and additionally includes the (regularized) antisymmetric singularities of the $\mathcal{P}v_{\cot(\frac{\omega \taug}{2})}$-term, plus a last overall contribution.

From Eq.\,\eqref{periodic_mean_fourier_version}, we can in turn derive the speed of convergence, as well as the criterion for convergence, fully analogously to the derivation in Sec.\,\ref{ssec:exact_asymptotics_single_reset}.
We readily obtain that
\begin{align*}
    \abs{\mu^{(M)}(\tau ; \taug) - \mu^{(\infty)}(\tau ; \taug)}
        &\sim \tilde{F}_{\textrm{Re}} \Gamma(1 + s_{\textrm{p}}^\mathrm{Re}) \cos(\frac{\pi s_{\textrm{p}}^\mathrm{Re}}{2}) \bigg(\zeta(1 + s_{\textrm{p}}^\mathrm{Re})
        - \sum_{m = M + 1}^{\infty} \frac{1}{{m}^{1 + s_{\textrm{p}}^\mathrm{Re}}}\bigg)\nonumber\\
        &+ \tilde{F}_{\textrm{Im}} \Gamma(1 + s_{\textrm{p}}^\mathrm{Im}\cos(\frac{\pi s_{\textrm{p}}^\mathrm{Im}}{2}) \bigg(\zeta(1 + s_{\textrm{p}}^\mathrm{Im})
        - \sum_{m = M + 1}^{\infty} \frac{1}{{m}^{1 + s_{\textrm{p}}^\mathrm{Im}}}\bigg),
\end{align*}
where we have introduced 
\[ s_{\textrm{p}}^\mathrm{Re/Im} \equiv s + \alpha_{\textrm{p}}^\mathrm{Re/Im} > 0, \]
and we have exploited the LF-expansion
\begin{equation}
\label{eq:LF_expansion_re_im_parts}
    F_{\textrm{G}}(\omega ; \taug) \sim \tilde{F}_{\textrm{Re}} \omega^{\alpha_{\textrm{p}}^\mathrm{Re}}
                                   + i \tilde{F}_{\textrm{Im}} {\omega}^{\alpha_{\textrm{p}}^\mathrm{Im}}.
\end{equation}
Note that, due to the parity of the FF, $\alpha_{\textrm{p}, \mathrm{Re}}$ ($\alpha_{\textrm{p}, \mathrm{Im}}$) is an even (odd) integer.

\subsection{Multiple-reset case}

For the case of multiple resets, let us introduce 
\[\mu_{\tilde{k}}^{(M)}(\tau ; \taug) \equiv \mu_{[k = K - \tilde{k} + 1]}({K M \taug ; \tau}), \quad \tilde{k} = 1, \ldots, K,\]
We then get from the definition Eq.~\eqref{eq:def_mean_k_time_dom}
\begin{align*}
    \mu_{\tilde{k}}^{(M)}(\tau ; \taug) = \frac{1}{2 \pi} \int_{- \infty}^{\infty}
                                           e^{i (\tau + (\tilde{k} - 1) M \taug) \omega} \mu^{(M)}(\omega ; \taug) \, \dd \omega
                                        = - \frac{i}{2 \pi} \sum_{m = 1}^{M} \int_{- \infty}^{\infty} \dd \omega
                                           e^{i \omega \qty(\tau + (m + (\tilde{k} - 1) M) \taug)} F_{\textrm{G}}(- \omega ; \taug) S^{-}(\omega).
\end{align*}
Employing the same LF-expansion in Eq.\,\eqref{eq:LF_expansion_re_im_parts}, we can then upper-bound the means for asymptotically-large values of $\tilde{k}$:
\begin{align}
    \abs{\mu_{\tilde{k}}^{(M)}(\tau ; \taug)} &\leq \tilde{F}_{\textrm{Re}} \, \Gamma\qty(1 + s_{\textrm{p}}^{\mathrm{Re}}) \abs{\cos(\frac{\pi s_{\textrm{p}}^{\mathrm{Re}}}{2})}
                                                    \sum_{m = 1}^{M} \frac{1}{{(\tau + m \taug (\tilde{k} - 1) M \taug)}^{1 + s_{\textrm{p}}^{\mathrm{Re}}}} \nonumber\\
                                              &+ \tilde{F}_{\textrm{Im}} \, \Gamma\qty(1 + s_{\textrm{p}}^{\mathrm{Im}}) \abs{\sin(\frac{\pi s_{\textrm{p}}^{\mathrm{Im}}}{2})}
                                                    \sum_{m = 1}^{M} \frac{1}{{(\tau + m \taug (\tilde{k} - 1) M \taug)}^{1 + s_{\textrm{p}}^{\mathrm{Im}}}} \nonumber \\
                                              &\leq M \tilde{F}_{\textrm{Re}} \, \Gamma\qty(1 + s_{\textrm{p}}^{\mathrm{Re}}) \abs{\cos(\frac{\pi s_{\textrm{p}}^{\mathrm{Re}}}{2})}
                                                    \frac{1}{{(\tau + (\tilde{k} - 1) M \taug)}^{1 + s_{\textrm{p}}^{\mathrm{Re}}}} \nonumber \\
                                              &+    M \tilde{F}_{\textrm{Im}} \, \Gamma\qty(1 + s_{\textrm{p}}^{\mathrm{Im}}) \abs{\sin(\frac{\pi s_{\textrm{p}}^{\mathrm{Im}}}{2})}
                                                    \frac{1}{{(\tau + m \taug (\tilde{k} - 1) M \taug)}^{1 + s_{\textrm{p}}^{\mathrm{Im}}}},\label{eq:decay_law_mutli_reset_mean_app}
\end{align}
for $\tilde{k} \rightarrow \infty$, and for a fixed value of $\tau$. We have again excluded the case where the Ohmicity parameter $s$ is an odd integer.

This decay formula for $|{\mu_{\tilde{k}}^{(M)}(\tau ; \taug|}$ can then be utilized to show point-wise convergence of any multi-time correlator of the asymptotic bath state $\rho_{\textrm{B}}(t_{K}^{+})$ for the case of periodic control repetition ($t_{K} = K M \taug$).
This can easily be seen for the classical correlation function by rewriting Eq.\,\eqref{eq:update_class_corr_func_multiple_resets} as
\begin{align*}
    C^{+}_{\rho_{\textrm{B}}(\lrt^{+})}(\tau_{2}, \tau_{1}) = C^{+}_{\bar{\rho}_{\textrm{B}}}(\tau_{2} - \tau_{1}) 
    + 2 \sum_{\tilde{k} = 1}^{K} \mu_{\tilde{k}}^{(M)}(\tau_{2} ; \taug) \mu_{\tilde{k}}^{(M)}(\tau_{1} ; \taug),
\end{align*}
and utilizing the fact that, for fixed $\tau_{1}, \tau_{2}$, (and $M, \taug$), $\mu_{\tilde{k}}^{(M)}(\tau_{1,2} ; \taug)$ decay according to Eq.\,\eqref{eq:decay_law_mutli_reset_mean_app} for $\tilde{k} \rightarrow \infty$.
Point-wise convergence of any higher-order statistics of the total bath state in the multiple-reset case under periodic control can be shown in an analogous way.

\twocolumngrid

\bibliography{./bibliography}

\end{document}